\documentclass[twocolumn,floatfix,prl,aps,a4paper,superscriptaddress]{revtex4-2}

\usepackage{microtype}
\usepackage[utf8]{inputenx}
\usepackage{physics}                            % Per i simboli di fisica
\usepackage{bm}									% Per i simboli in grassetto
\usepackage{enumerate}							% Per personalizzare gli elenchi numerati
\usepackage{amsthm}								% Per teoremi e definizioni
\usepackage{comment}							% Per i commenti su più righe
\usepackage{t1enc}								% Per le virgolette caporali
\usepackage{mathtools}							% Per i sistemi di equazioni
\usepackage{tensor}								% Per gli indici alti e bassi
\usepackage{listings}							% Per le linee di codice
\usepackage[usenames,dvipsnames]{xcolor}		% Per parole o blocchi di testo colorati
\usepackage{hyperref}
\hypersetup{colorlinks=true}
\usepackage[capitalize]{cleveref}               % Clever referencing
\usepackage{graphicx,color}
\usepackage{xstring}                            % For clever citations
\usepackage{lipsum}
\usepackage{bbold}                              % Per il fottuto simbolo dell'identita
\usepackage{tikz}
\usetikzlibrary{shadings}
\usepackage[caption=false]{subfig}

\newcommand{\cor}[1]{\mathcal{#1}}									% Calligrafico
\newcommand{\dslash}[1]{\frac{\dd[d]{#1}}{(2\pi)^d}}                % Momentum integral
\def \r {_{\vb r}}

\def \q {_{\vb q}}

\def \y {^{(y)}}

\newcommand{\n}{\nonumber}
\newcommand{\ccite}[1]{\IfSubStr{#1}{,}{Refs.~}{Ref.~}\cite{#1}}

\newcommand{\rev}[1]{#1}
\newcommand{\revv}[1]{#1}
\newcommand{\revvv}[1]{#1}
\newcommand{\gdisc}{w^{(1)}}
\newcommand{\gcont}{w^{(1)}}
\newcommand{\prof}{w^{(0)}}

\begin{document}
%
%
% \title{\rev{Universal scale-free decay of 
% %spatial 
% tracer-bath
% correlations in $d$-dimensional interacting particle systems}}

\title{Universal scale-free decay of tracer-bath correlations in $d$-dimensional interacting particle systems}

\author{Davide Venturelli}
\affiliation{Sorbonne Universit\'e, CNRS, Laboratoire de Physique Th\'eorique de la Mati\`ere Condens\'ee (LPTMC), 4 Place Jussieu, 75005 Paris, France}

\author{Pierre Illien}
\affiliation{Sorbonne Universit\'e, CNRS, Laboratoire PHENIX (Physico-Chimie des Electrolytes et Nanosyst\`emes Interfaciaux), 4 Place Jussieu, 75005 Paris, France}

\author{Aur\'elien Grabsch}
\affiliation{Sorbonne Universit\'e, CNRS, Laboratoire de Physique Th\'eorique de la Mati\`ere Condens\'ee (LPTMC), 4 Place Jussieu, 75005 Paris, France}

\author{Olivier B\'enichou}
\affiliation{Sorbonne Universit\'e, CNRS, Laboratoire de Physique Th\'eorique de la Mati\`ere Condens\'ee (LPTMC), 4 Place Jussieu, 75005 Paris, France}

\date{ \today }

% \begin{abstract}
% \rev{Quantifying the correlations between the position of a tagged tracer and the density of surrounding bath particles is crucial for
% %key to 
% understanding tracer diffusion in interacting particle systems, and for characterizing the response properties of the bath}.
% We address this problem analytically for both hard-core and soft-core interactions, using minimal yet paradigmatic models in $d$ spatial dimensions.
% In both cases, we derive analytical expressions for the spatial correlation profiles in the reference frame of the tracer.
% We reveal unexpected universal features in their large-distance behavior, characterized by power-law tails with exponents that depend solely on the spatial dimensionality of the system.
% Beyond these simple models, we demonstrate the robustness of our results across different regimes using particle-based numerical simulations.
% \end{abstract}

\begin{abstract}
Quantifying the correlations between the position of a tagged tracer and the density of surrounding bath particles is crucial for understanding tracer diffusion in interacting particle systems, and for characterizing the response properties of the bath.
We address this problem analytically for both hard-core and soft-core interactions, using minimal yet paradigmatic models in $d$ spatial dimensions.
In both cases, we derive analytical expressions for the spatial correlation profiles in the reference frame of the tracer.
We reveal unexpected universal features in their large-distance behavior, characterized by power-law tails with exponents that depend solely on the spatial dimensionality of the system.
Beyond these simple models, we demonstrate the robustness of our results across different regimes using particle-based numerical simulations.
\end{abstract}

\maketitle

\textit{Introduction.---}
Assessing the statistics of the displacement of a tagged tracer particle evolving in a complex medium subject to thermal fluctuations is a paradigmatic problem in statistical physics, which found in the last decades numerous applications to microrheology~\cite{Squires2010,Chen2003,Lau2003,Puertas2014,Wilson2009,Wilson2011} and biophysics~\cite{Guo2014,Parry2014} experiments.
A basic but powerful approach toward describing tracer diffusion in complex media is to neglect the effect that the tracer exerts on the surrounding 
%particle 
bath, which is assumed to remain in equilibrium at all times, and can thus generically be treated as a source of noise~\cite{Marconi2008}. 
This is exemplified by the celebrated Langevin description of Brownian motion~\cite{Langevin_1908}, which assumes a large mass disparity between tracer and bath particles~\cite{Mazur1970,Lebowitz1963}.
However, such effective description naturally breaks down if the tracer particle (TP) and the bath particles have comparable sizes, because then correlations between the tracer position and the density of the surrounding bath 
%particles 
assume a relevant role, and can no longer be neglected.
Characterizing these correlations is in general a complex many-body problem.
However, their knowledge gives full access not only to the TP fluctuations, but also to the response properties of the medium itself,
which makes the pursuit of these correlations (at least in approximate forms)
valuable.

An ideal and fertile arena for seeking analytical predictions for these correlations 
%are
is provided by
lattice gases, which
are arguably among the most 
%paradigmatic
emblematic
models to study transport properties in interacting particle systems~\cite{Chou2011,Mallick2015}.
They consist of particles that can jump from one site of the lattice to another, with the constraint that each site can host at most one particle --- this minimally models hard-core
or excluded-volume interactions. 
The one-dimensional case, corresponding to the single-file geometry, exemplifies spectacularly how the correlations with the surrounding bath particles induce subdiffusive behavior of the TP displacement~\cite{Spohn1991}.
These correlations have recently been computed exactly for the symmetric exclusion process~\cite{Grabsch2022}
and related (integrable) one-dimensional models~\cite{Grabsch2023,Mallick_2022,Rizkallah2023}.
In higher dimensions, approximate
semi-analytical
solutions for these correlations have been key to obtaining the variance of the TP position, particularly in non-equilibrium settings~\cite{Benichou_2014_microscopic,Illien2015_distribution,Illien2018,Benichou2018}.

\begin{figure}
\subfloat{
    \centering
\begin{tikzpicture}[scale=0.9]

    \def\psize{0.1} % Particle diameter
    \def\pcolor{MidnightBlue} % Particle diameter
    \def\arrowcolor{OliveGreen} % Arrow color

        % Draw rounded arrows pointing to neighboring sites of (3,2)
        \draw[->, thick, \arrowcolor,>=latex] (3,2) .. controls (3.2,2) and (3.2,2.2) .. (4,2); % Right arrow
        \draw[->, thick, \arrowcolor,>=latex] (3,2) .. controls (3,2.2) and (3.2,2.2) .. (3,3); % Up arrow
        \draw[->, thick, \arrowcolor,>=latex] (3,2) .. controls (2.8,2) and (2.8,2.2) .. (2,2); % Left arrow
        \draw[->, thick, \arrowcolor,>=latex] (3,2) .. controls (3,1.8) and (3.2,1.8) .. (3,1); % Down arrow

        % Optional: Draw the grid for reference
        \draw[gray, thin] (-0.2,-0.2) grid (4.2,4.2);
        % Draw the square lattice
        \foreach \x in {0, 1, 2, 3, 4} {
            \foreach \y in {0, 1, 2, 3, 4} {
                % Fill certain vertices (manually select)
                \ifnum\x=0 \ifnum\y=0 \fill[\pcolor] (\x,\y) circle (\psize); \fi \fi
                \ifnum\x=0 \ifnum\y=1 \fill[\pcolor] (\x,\y) circle (\psize); \fi \fi
                \ifnum\x=1 \ifnum\y=1 \fill[\pcolor] (\x,\y) circle (\psize); \fi \fi
                \ifnum\x=1 \ifnum\y=3 \fill[\pcolor] (\x,\y) circle (\psize); \fi \fi
                \ifnum\x=2 \ifnum\y=0 \fill[\pcolor] (\x,\y) circle (\psize); \fi \fi
                \ifnum\x=2 \ifnum\y=4 \fill[\pcolor] (\x,\y) circle (\psize); \fi \fi
                \ifnum\x=3 \ifnum\y=2 \fill[\pcolor] (\x,\y) circle (\psize); \fi \fi
                \ifnum\x=4 \ifnum\y=1 \fill[\pcolor] (\x,\y) circle (\psize); \fi \fi
                \ifnum\x=4 \ifnum\y=4 \fill[\pcolor] (\x,\y) circle (\psize); \fi \fi
                \ifnum\x=0 \ifnum\y=4 \fill[\pcolor] (\x,\y) circle (\psize); \fi \fi
                \ifnum\x=3 \ifnum\y=0 \fill[\pcolor] (\x,\y) circle (\psize); \fi \fi
            }
        }

        % Draw the tracer particle in red
        \fill[BrickRed] (1,2) circle (\psize); % Tracer particle
        
        % Label the tracer particle
        \node at (1,2) [above left] {$\vb{X}(t)$};

    \end{tikzpicture}
}
\hspace{0.2cm}
\subfloat{
    \centering
\begin{tikzpicture}[scale=0.4]
% Parameters
\def\particlesize{0.8} % Particle diameter

% Draw regular particles with radial shading (maximum intensity at the center)
\node[circle, shading=radial, outer color=MidnightBlue!1, inner color=MidnightBlue!100, minimum size=\particlesize cm] at (0,0) {};    % Particle 1
\node[circle, shading=radial, outer color=MidnightBlue!1, inner color=MidnightBlue!100, minimum size=\particlesize cm] at (4,-2) {};  % Particle 2
\node[circle, shading=radial, outer color=MidnightBlue!1, inner color=MidnightBlue!100, minimum size=\particlesize cm] at (2,-3.5) {};  % Particle 3
\node[circle, shading=radial, outer color=MidnightBlue!1, inner color=MidnightBlue!100, minimum size=\particlesize cm] at (6,-4.5) {};  % Particle 4
\node[circle, shading=radial, outer color=MidnightBlue!1, inner color=MidnightBlue!100, minimum size=\particlesize cm] at (3,3) {};  % Particle 5
\node[circle, shading=radial, outer color=MidnightBlue!1, inner color=MidnightBlue!100, minimum size=\particlesize cm] at (6.5,2) {};  % Particle 6

% Draw tracer particle with radial shading in red
\node[circle, shading=radial, outer color=BrickRed!1, inner color=BrickRed!100, minimum size=\particlesize cm] at (3,0) {};  % Tracer Particle

% Label the tracer particle
\node at (3,0) [above right] {$\vb{X}(t)$};

\end{tikzpicture}
}
\put(-245,105){(a)} 
\put(-100,105){(b)} 
    \caption{Schematic representation of the two models considered in this work, namely \textbf{(a)} a hard-core $d$-dimensional lattice gas, where each particle can jump to one of its neighbouring sites with equal rates, only if the target site is empty; and \textbf{(b)} a $d$-dimensional system of Brownian particles interacting via soft pairwise potentials (see the main text). 
    In both cases, we single out the position $\vb X(t)$ of a tagged tracer particle (red). 
    }
    \label{fig:sketch}
\end{figure}

On the other hand, in continuum space,
particles are often assumed to evolve 
according to a system of coupled Langevin equations, featuring two-body interaction potentials
--- which provide a more realistic modeling of their interactions (beyond excluded volume). 
In the overdamped case,
the exact 
evolution equation
of the coarse-grained particle density was derived in seminal works by Dean and Kawasaki~\cite{Dean1996,Kawasaki1998},
yet the presence of nonlinear multiplicative noise has so far hindered all efforts toward its exact solution.
Conversely, the correction to the diffusion coefficient of a driven TP induced by the bath, as well as the average density profile in the frame of the TP, have only more recently been addressed in~\cite{Demery2014} by 
linearizing the coarse-grained equations around a fixed background density. 
However, this approach typically fails when attempting to interpolate between the Dean-Kawasaki theory and the hard-core particle limit by simply increasing the strength of the two-body potentials~\footnote{For instance, this approach is unable to reproduce the subdiffusive behavior expected for hard-core particles in one dimension.}, which prevents the description of proper hardcore-like repulsive interactions.
Besides, computing the correlations between the TP position and the coarse-grained bath density 
beyond the average density profile 
has never been attempted (to the best of our knowledge) within this setting. 

Due to the key role played by these correlations in determining the TP statistics, their use in $d$-dimensional lattice models has been so far mainly instrumental in the description of the TP variance. 
In this work, we focus instead on the stationary TP-bath correlations in their own right, with the aim of analyzing their spatial properties.
\rev{Indeed, these profiles quantify the variation of the bath density distribution due to a 
%positive 
fluctuation of the TP position, thus allowing to probe the bath's response even far from the TP.}

\smallskip

\revv{\textit{Tracer-bath correlations. ---} We start by defining the TP-bath correlation functions, which are the focus of this Letter, and highlight their importance and intrinsic dynamical nature. 
Although we will consider them in both discrete and continuous settings, we introduce them here in the continuous case for clarity. 
We denote by $\{\vb X_i(t)\}$ the positions of the particles at time $t$. We single out particle $i=0$, referred to hereafter as the TP, and denote its displacement by $\vb X(t) \equiv \vb X_0(t) - \vb X_0(0) \equiv \vb X_0(t)$, adopting the convention $\vb X_0(0)=0$. 
Let $\rho(\vb x,t)$ denote the density of the other particles ($i \neq 0$), referred to as the bath particles.}

\revv{We first note that $\vb X(t)$ and $\rho(\vb x,t)$ are coupled. Qualitatively, any large displacement $\vb X(t)$ of the TP in a given direction requires the reorganization of a large number of bath particles, encoded in the density $\rho(\vb X(t)+ \vb r,t)$ around the tracer. In turn, the dynamics of this reorganization governs that of the TP. The two random variables, $\vb X(t)$ and $\rho(\vb X(t)+ \vb r,t)$, are thus coupled, and characterizing this coupling is essential to understanding the transport properties of the tracer and quantifying the perturbation induced by its displacement.}

\revv{Quantitatively, this coupling is naturally described by the covariance 
$
\mathrm{Cov}[\vb X(t), \rho(\vb X(t)+ \vb r,t)] \equiv \langle \vb X(t) \rho(\vb X(t)+ \vb r,t)\rangle - \langle \vb X(t) \rangle \langle \rho(\vb X(t)+ \vb r,t) \rangle = \langle \vb X(t) \rho(\vb X(t)+ \vb r,t)\rangle,
$
since $\langle \vb X(t) \rangle = 0$. 
We emphasize that these correlation functions are intrinsically dynamical, because they involve the \textit{displacement} $\vb X(t)$ of the TP rather than its absolute position. Unlike equilibrium quantities, these correlation functions cannot, in practice, be computed from the equilibrium Gibbs measure~\cite{McDonald_book}.}

\revv{Several important questions arise. 
(i)~What is the sign of this coupling? Without loss of generality, consider the projection onto the direction $\vu e_1$, namely
$
\langle \vb X(t) \rho(\vb X(t)+ \vb r,t)\rangle \cdot \vu e_1 = \langle X_t \rho(\vb X(t)+ \vb r,t)\rangle,
$
where $X_t = \vb X(t) \cdot \vu e_1$. 
A positive sign indicates that an increase in $X_t$ is correlated with an increase in the density at a relative position $\vb r$ from the tracer, while a negative sign indicates anti-correlation.
(ii)~What is the range of this coupling? 
In particular, can short-range interactions 
lead to long-range reorganization of the bath particles, or do they affect only a finite region of space around the TP? 
(iii)~What is the dynamics of this coupling? 
Specifically, is this reorganization efficient enough to result in a stationary correlation profile? 
(iv)~What are the relevant parameters controlling this coupling? 
Does it depend on the nature of the interaction, or on the efficiency of homogenization by diffusion (related to the spatial dimension)?}

\revv{In our Letter we provide
explicit answers to these key questions by determining and analyzing the TP-bath correlation profiles.
To this end,
we first derive 
analytic expressions for these quantities both for a lattice gas model 
with \textit{hard-core} interactions,
and for a system of 
\textit{soft}
interacting Brownian particles. 
We then
find that a power-law behavior at large distances, with a simple algebraic decay exponent depending solely on the spatial dimensionality, encompasses all the interacting particle systems mentioned above --- in spite of the intrinsically distinct nature of their interactions, which otherwise prevents their description on an equal footing. 
Using numerical simulations, we argue that this behavior remains robust
beyond the assumptions underlying our analytic derivation,
and for Brownian suspensions with strongly repulsive Lennard-Jones-type potentials.}

\begin{figure}
    \includegraphics[width=0.482\textwidth]{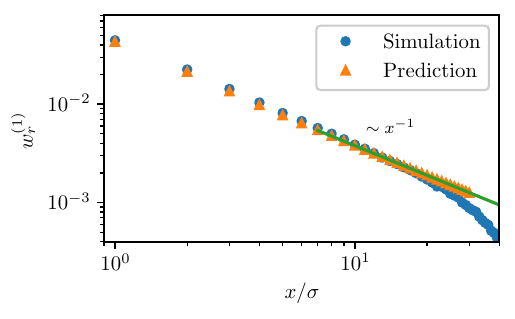}
    %\put(-195,39)
    \put(-194,39)
    {\includegraphics[scale=0.22]{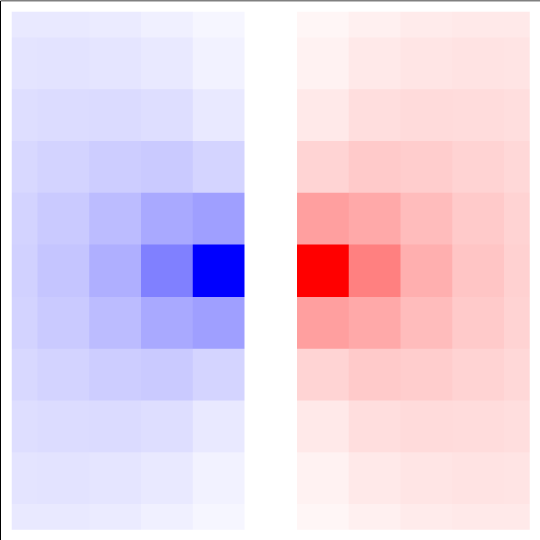}}
    \caption{Stationary tracer-bath correlation profile \rev{$\gdisc\r$} [see \cref{eq:def_g_discrete,eq:gr_discrete_sol}] for the hard-core lattice gas model, as measured from simulations in $d=2$. Its large-distance tails are characterized by the algebraic behavior~\eqref{eq:gr_discrete_asymptotics}.
    In the simulation we used $L=100$ and $\bar\rho=0.8$. Inset: color map indicating the spatial symmetries of \rev{$\gdisc\r$} on a square lattice (red means positive while blue means negative). 
    }
    \label{fig:gr_discrete}
\end{figure}

\smallskip
\textit{Hard-core lattice gas.---} 
We first consider particles evolving on an infinite $d$-dimensional cubic lattice, with spacing $\sigma$.
Initially, each site is occupied with probability $\bar{\rho}$.
Particles then perform symmetric random walks with nearest-neighbor jumping rate $1/(2d\tau)$ [so that their bare diffusion coefficient is $\sigma^2/(2\tau)$], with the constraint that the target site must be empty [see \cref{fig:sketch}(a)]. 
\rev{The state of the system at a given time $t$ is specified by the tracer position $\vb X(t)$,
and by the set of occupations $\rho_{\vb r}(t)=\{0,1\}$ for each site $\vb r$.}
The joint probability distribution $P(\vb X,\rho\r,t)$ satisfies a master equation
$\partial_t P = \mathcal{L}P$,
where the well-known form of $\mathcal L$
is reported in~\cite{sm}.
Multiplying its two sides by 
$\mathrm e^{\bm \lambda\cdot  \vb X}$
and averaging with respect to both $\rho\r$ and $\vb X$ then yields an evolution equation for the moment generating function 
$\Psi(\bm \lambda, t)\equiv \ln \expval*{\mathrm e^{\bm \lambda\cdot  \vb X(t)}}$ 
of the tracer position as~\cite{Illien2015_distribution}
%\dav{See Eqs.~(56)--(58) in~\cite{Illien2015_distribution}}
\begin{equation}
    \partial_t\Psi(\bm\lambda,t) = 
    \frac{1}{2d \,\tau}
    \sum_\mu \left( e^{\sigma \bm \lambda \cdot \vu e_\mu}-1 \right)\left[  1- w_{\vu{e}_\mu}(\bm\lambda,t) \right].
    \label{eq:CGF_discrete}
\end{equation}
Here the sum runs over $\mu \in \{\pm 1,\dots, \pm d\}$, 
we called $\vu{e}_\mu$ the unit vectors along the Cartesian directions, 
and finally
the generalized profile
\rev{$   w\r(\bm \lambda,t) =   \expval{\rho_{\vb X +\vb r}\, \mathrm{e}^{\bm \lambda \cdot \vb X}}/\expval{\mathrm e^{\bm \lambda \cdot \vb X}}$}
encodes all correlations between the tracer position and the occupations. 
Equation~\eqref{eq:CGF_discrete} demonstrates how these correlations completely control the statistical properties of the tracer position.
The first among these cross correlations is
\rev{\begin{align}
    \gdisc_{\vb r} &\equiv \eval{\dv{w\r}{\lambda_1}}_{\bm \lambda=\bm 0}= \big\langle\left(X_t-\expval{X_t} \right) \left( \rho_{\vb X+\vb r} -\expval{\rho_{\vb X+\vb r}} \right) \big\rangle \n\\
    &= \expval{X_t\,\rho_{\vb X+\vb r}} , 
    \label{eq:def_g_discrete}
\end{align}}
where we indicated without loss of generality $X_t\equiv \vb X\cdot \vu e_1$ \rev{[since $\gdisc\r$ can only depend on the relative orientations of $\vb X$ and $\vb r$]},
and where
in the last step
we used that $\expval{X_t}=0$ and \rev{$\expval{\rho_{\vb X+\vb r}}=\bar{\rho}$}.
This is the simplest and most physically significant
among the TP-bath correlation profiles, and
it completely characterizes the variance of the TP [see \cref{eq:CGF_discrete}]. In the following, we 
aim at analyzing its spatial properties in the stationary limit attained at long times, and for an infinite system. 
To this end, we note that an evolution equation for \rev{$\gdisc\r(t)$} [akin to \cref{eq:CGF_discrete} for $\Psi(\bm \lambda,t)$] can be obtained starting from the master equation, but it naturally involves higher-order cross-correlation functions~\cite{sm}. 
\rev{This hierarchy can however be closed by approximating $\expval*{\rho_{\vb X+\vb r}\rho_{\vb X+\vb r'}}\simeq \expval*{\rho_{\vb X+\vb r}}\expval*{\rho_{\vb X+\vb r'}}$, and
$
    \expval*{\delta X_t \,\rho_{\vb X+\vb r}\,\rho_{\vb X+\vb r'}}\simeq \expval*{\rho_{\vb X+\vb r}}\expval*{\delta X_t\,\rho_{\vb X+\vb r'}}+\expval*{\delta X_t\,\rho_{\vb X+\vb r}}\expval*{\rho_{\vb X+\vb r'}}$,
where $\delta x \equiv x-\expval*{x}$.}
Such \textit{decoupling} approximation~\cite{Benichou2013,Benichou_2014_microscopic,Illien2015_distribution}
goes beyond the simple mean-field [above we only discarded terms of $\order{\delta x^2}$ and $\order{\delta x^3}$, respectively], and has been shown in~\cite{Illien2018}
to become exact both in the dense 
limit, and in the dilute limit with fixed bath particles.
In the same work,
this method was used to derive a closed set of self-consistent equations satisfied by \rev{$\gdisc\r$} for the case of a driven tracer, which were then solved numerically and used to 
%inspect 
accurately predict
the diffusion coefficient of the tracer. 
Here we focus instead on the unbiased case, for which we show in~\cite{sm} that such self-consistent equations can in fact be solved \textit{analytically}.
In the stationary limit attained by \rev{$\gdisc\r(t)$} if $d\geq 2$, the result reads simply
\rev{
\begin{equation}
    \gdisc\r = \frac{ \sigma \bar{\rho} (1-\bar{\rho}) }{(2-\bar{\rho})d-(2-3\bar{\rho}) \cor I_d(\vu e_1)}\cor I_d(\vb r),
    \label{eq:gr_discrete_sol}
\end{equation}}
with
$
    \cor I_d(\vb r) = \int_{-\pi}^\pi \dslash{q} \frac{\sin (q_1) \sin(\vb q\cdot \vb r) }{1-\frac{1}{d}\sum_{j=1}^d \cos(q_j) } .
$
Its 
asymptotic behavior for large positive $x\equiv \vb r\cdot \vu{e}_1$ can then be inspected by standard techniques~\cite{sm}, yielding 
\begin{equation}
    \gdisc\r \sim 
    \frac{ \sigma \bar{\rho} (1-\bar{\rho}) }{(2-\bar{\rho})d-(2-3\bar{\rho}) \cor I_d(\vu e_1)}
    %\frac{2\sigma \bar{\rho} (1-\bar{\rho}) }{1-(1/d-4\bar{\rho}) } 
    \frac{\Gamma (d/2)\, d}{\pi^{d/2}}x^{1-d}.
    \label{eq:gr_discrete_asymptotics}
\end{equation}
\revv{Note that $\gdisc_{-\vb r}=-\gdisc\r$ by symmetry, see \cref{eq:def_g_discrete}, so that in particular it vanishes for $\vb r \perp \vu{e}_1$, see the inset of \cref{fig:gr_discrete}.
Additionally, $\gdisc\r > 0$ for $\vb r \cdot \vu e_1 > 0$, indicating that in a realization where $X_t > 0$ (corresponding to a net displacement to the ``right''), the density of particles around the tracer will show an accumulation of particles in the same direction (``in front'' of the tracer), and a depletion of particles in the opposite direction ($\vb r \cdot \vu e_1 < 0$, ``behind'').}
%Note that $g_{-\vb r}=-g\r$ by symmetry, see \cref{eq:def_g_discrete}.
This behavior is tested in \cref{fig:gr_discrete} against numerical simulations, showing excellent agreement 
[while we stress that the prediction~\eqref{eq:gr_discrete_sol} is expected to become exact both in the dense and dilute limits].
The simple algebraic decay \rev{$\gdisc\r\sim x^{1-d}$}
is our first main result. 
Below we consider a system of Brownian particles featuring a substantially distinct type of interaction, and construct an analogous tracer-bath correlation profile, which we then characterize.

\smallskip
\textit{Soft interacting Brownian particles.---} As a second paradigmatic model, we now consider a system of $(N+1)$ particles at positions $\vb{X}_i(t)\in \mathbb{R}^d$ [as in \cref{fig:sketch}(b)], each evolving according to the overdamped Langevin dynamics
\begin{equation}
    \partial_t \vb{X}_i(t) =- \mu \sum_{j\neq i} \nabla_{\vb{X}_i} U\left(\vb{X}_i(t)-\vb{X}_j(t)\right)+\bm\eta_i(t).
    \label{eq:langevin}
\end{equation}
Here the noise terms are Gaussian with zero mean and correlations
$
    \expval{ \bm \eta_i(t) \bm\eta_j^{\mathrm{T}}(t')}=2\mu T \delta_{i,j} \delta (t-t') \mathbb{1}
$, where we set the Boltzmann constant $k_B=1$,
while $U(\vb x)=U(\abs{\vb x})$ is a pairwise interaction potential. This type of system
lends itself to analytical treatment within the Dean-Kawasaki framework~\cite{Dean1996,Kawasaki1998},
upon linearization of the effective coarse-grained dynamics.
Following~\cite{Dean1996,Demery2014}, we first derive the coupled evolution equations for the tracer position $\vb X(t)\equiv\vb X_{i=0}(t)$, and the fluctuating density $\rho(\vb x,t)=\sum_{i=1}^N \delta \left( \vb x-\vb X_i(t) \right)$ of the other particles. We then linearize the latter around a fixed background density $\bar\rho$ as $\rho(\vb x,t) = \bar\rho +  \bar\rho^{1/2} \phi(\vb x,t)$, according to the prescription $ h \phi(\vb x,t) \ll 1$, where $h\equiv 1/\bar\rho^{1/2}$. This yields~\cite{sm}
\begin{align}
    \partial_t \vb{X}(t)&=-
    %\bar\rho^{-1/2} 
    h
    \mu \nabla_{\vb{X}} \cor H[\phi,\vb{X}]+\bm\eta_0(t), \label{eq:tracer}\\
    \partial_t \phi(\vb x,t) &= \mu \div \left[ \grad \fdv{\mathcal H[\phi,\vb{X}]}{\phi(\vb x,t)}+\bm \xi (\vb x,t) \right],\label{eq:field}
\end{align}
where
$    \left\langle \bm \xi(\vb x,t) \bm\xi^{\mathrm{T}}\left(\vb x',t^{\prime}\right)\right\rangle=2\mu T\delta (\vb x-\vb x') \delta (t-t' )\mathbb{1} $,
and with the pseudo-Hamiltonian
\begin{align}
    \cor H[\phi, \vb{X}]=&  \,
    %\frac{T}{2} \int \dd{\vb x} \phi^2(\vb x) +  \frac 12 \int \dd{\vb x} \dd{\vb y} \phi(\vb x) u(\vb x-\vb y)\phi(\vb y)\n\\
    \frac{1}{2}\int \dd{\vb x} \dd{\vb y} \phi(\vb x)\left[T\delta(\vb x-\vb y)+ u(\vb x-\vb y)\right]\phi(\vb y)\n\\
    &+ 
    %\bar\rho^{-1/2}  
    h
    \int \dd{\vb y}\phi(\vb y) u(\vb y-\vb{X}),
    \label{eq:pseudo-H}
\end{align}
where we rescaled $u(\vb x) = \bar\rho\, U(\vb x)$.

\begin{figure*}
\centering
\includegraphics[width=\columnwidth]{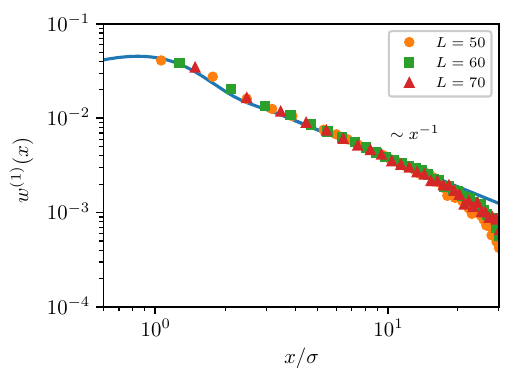}
\put(-195,40){\includegraphics[scale=0.425]{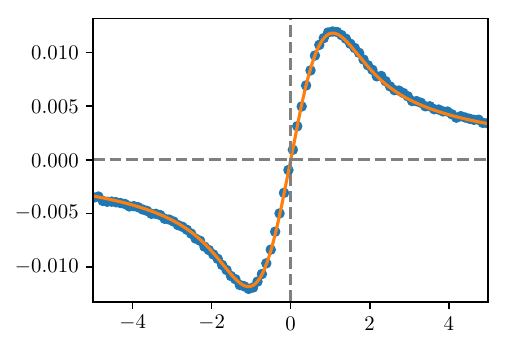}}
  \put(-195,162){(a)}
\hspace{10pt}
\includegraphics[width=\columnwidth]{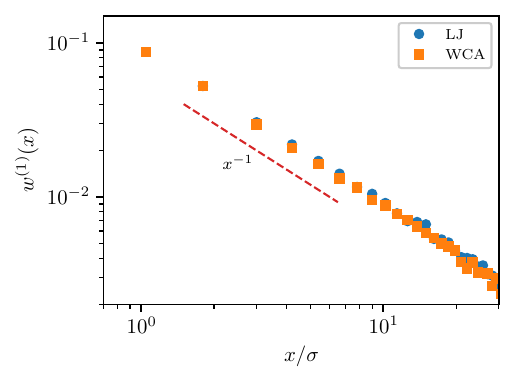}
  \put(-195,162){(b)}
    \caption{\textbf{(a)} Stationary correlation profile $\gcont(x)$ for a system of Brownian particles interacting via Gaussian pairwise potentials, i.e.~$U(r) = \epsilon \, \exp[-(r/\sigma)^2]$, in $d=2$.
    Its large-distance behavior predicted in \cref{eq:gr_continuum_asymptotics} is compared to the result of 
    numerical
    simulations, obtained with $T=0.8$, $\bar\rho=0.5$, and for various system sizes (the other parameters were set to unity). 
    The solid line corresponds to the analytic expression~\eqref{eq:gr_continuum_sol}, computed along
    $x=\vb r\cdot \vu{e}_1$.
    Inset: check of the accuracy of the prediction at short distances, in a system of size $L=20$ with 
    %$\bar\rho=0.8$ and
    $\epsilon=0.1$.
    \rev{In panel \textbf{(b)} we replaced the Gaussian potential by Lennard-Jones and WCA potentials, showing the robustness of the large-distance algebraic decay
    (the dashed line is a guide to the eye)
    for potentials not described by the Dean-Kawasaki theory.
    \revv{This simulation was performed in $d=2$ with $L=120$, see~\cite{sm} for additional details.}
    }
    %of the simulation).
    }
\label{fig:gr_continuum}
\end{figure*}

Using Stratonovich calculus, we first derive the relation~\cite{sm}
\begin{align}
    \partial_t \Psi(\bm \lambda,t) = \lambda^2\mu T  -h \mu  \bm \lambda \cdot \int \dslash{q} \mathrm i \vb q \, u\q  w\q(\bm \lambda,t),
    \label{eq:cgf-continuum}
\end{align}
where 
$\Psi(\bm \lambda, t)\equiv \ln \expval*{\mathrm e^{\bm \lambda\cdot  \vb X(t)}}$ 
as above, and
where we denoted
$u\q = \int \dd[d]{x} \mathrm e^{-\mathrm i\vb{q}\cdot \vb{x}} u(\vb x)$
the spatial Fourier transform of the interaction potential (which we assume henceforth to exist).
The cumulants of the tracer position are thus dictated by the correlations with the particle bath, encoded in the generalized correlation profile $ w(\vb x,\bm \lambda,t) \equiv \expval{  \phi(\vb x +\vb X(t),t)\, \mathrm e^{ \bm \lambda \cdot \vb X(t)}} / \expval{\mathrm e^{ \bm \lambda \cdot \vb X(t)}}$, in complete analogy with \cref{eq:CGF_discrete} for the discrete case.
\revv{We emphasize that these correlation profiles are in general
%\textit{nonequilibrium} 
dynamical
observables, and as such they are not directly reducible to the usual pair correlation functions predicted and measured for simple liquids~\cite{McDonald_book,Evans1993,Evans1994,LeotedeCarvalho1994,Dijkstra2000,Hopkins2005,Stopper2019}.}
Again, the first and most relevant among these cross correlations is
\rev{\begin{equation}
    \gcont(\vb x,t) \equiv \vu{e}_1 \cdot \expval{\vb{X}(t) \, \phi(\vb x +\vb{X}(t),t)  },
    \label{eq:def_g_continuum}
\end{equation}}
i.e.~the continuous counterpart of \rev{$\gdisc\r$} introduced in \cref{eq:def_g_discrete} for the lattice gas model. In the following, we set out to compute its stationary value \rev{$\gcont(\vb x)$} attained in the long-time limit. 
To this end, 
we treat perturbatively the term proportional to $h$ in~\cref{eq:pseudo-H}, which controls the interaction between the tracer and the coarse-grained bath density. 
This corresponds to an expansion around the \textit{soft} interaction limit, 
as we argue below and in~\cite{sm}.
Working in the Fourier domain, we focus on~\footnote{Since the Gaussian white noises $\bm \eta_0$ and $\bm \xi$ in \cref{eq:tracer,eq:field} are additive, we adopt the Stratonovich calculus throughout without loss of generality.}
\rev{
\begin{align}
    \partial_t \gcont\q(t) &= \Big\lbrace \expval{\dot{\vb X}(t) \phi_{\vb q}(t)\,\mathrm e^{\mathrm i \vb q \cdot \vb X(t)}} + \expval{\vb X (t) \dot \phi_{\vb q}(t)\,\mathrm e^{\mathrm i \vb q \cdot \vb X(t)} } \n\\
    &+\mathrm i \expval{[\vb q \cdot \dot{\vb X}(t)] \,\vb X(t) \,\phi_{\vb q}(t)\,\mathrm e^{\mathrm i \vb q \cdot \vb X(t)} } \Big\rbrace \cdot \vu{e}_1 , \label{appeq:gq_steady}
\end{align}}
and we replace
$\dot{\vb X}(t)$ and $\dot{\phi}_{\vb q}(t)$ on the right-hand side by using the equations of motion~\eqref{eq:tracer}~and~\eqref{eq:field}. This produces several expectation values, which we compute in~\cite{sm} within perturbation theory --- in particular, the terms involving the Gaussian noises $\bm \eta_0(t)$ and $\bm \xi(\vb x,t)$
%$\nu\q(t)$ 
can be evaluated within Stratonovich calculus using the Furutsu-Novikov-Donker formula~\cite{Novikov_1965,Luczka_2005,Venturelli_2022,Venturelli_2022_2parts,Venturelli_2023,venturelli2024heat}. In the stationary limit, we thus find
\rev{
\begin{equation}
    \gcont(\vb x) = -hT \int \dslash{q}\frac{ \mathrm e^{\mathrm i \vb q\cdot \vb x} \, \mathrm i q_1 u\q  }{q^2(T+u\q) (2T+u\q)} +\order{h^2}. \label{eq:gr_continuum_sol}
\end{equation}}

Note that the calculation delineated above can be promptly extended to the case in which the tracer is biased by an external force $\vb f$, which is obtained by adding a term $\vb f \delta_{i,0}$ to the right-hand side of \cref{eq:tracer}. 
In particular, 
in~\cite{sm} we recover by the same method the average density profile \rev{$\prof(\vb x) =\expval{  \phi(\vb x +\vb X)}$} in the frame of the tracer, which had been previously derived in~\cite{Demery2014} by distinct techniques. 

The large-distance behavior of \rev{$\gcont(\vb x)$} in \cref{eq:gr_continuum_sol} can be extracted without the need to specify the interaction potential $u\q$.
Indeed, in~\cite{sm} we analyze the singular behavior of \rev{$\gcont_{\vb q}$} around $\vb q \sim \bm 0$ and show that, for large
(positive)
$x=\vb x\cdot \vu{e}_1$ and in $d\geq 2$,
\rev{
\begin{equation}
    \gcont(x) \sim \frac{hT u_{\vb q=\bm 0}  }{\Omega_d(T+u_{\vb q=\bm 0})(2T+u_{\vb q=\bm 0})} 
    %\T{sgn}(x) \abs{x}^{1-d},
    x^{1-d},
    \label{eq:gr_continuum_asymptotics}
\end{equation}}
where $\Omega_d$ is the $d$-dimensional solid angle.
Remarkably, \rev{$\gcont(x)$} displays the same
algebraic decay exponent as the one found in \cref{eq:gr_discrete_asymptotics} for the hard-core lattice gas, in spite of the 
inherently distinct
nature of the inter-particle interactions in these two models.
Besides, also the prefactor in \cref{eq:gr_continuum_asymptotics} is largely insensitive to the spatial details of the interaction potential, as it only depends on 
its typical energy scale via
$u_{\vb q=\bm 0}$~\cite{sm}. 
In \cref{fig:gr_continuum}(a) we test this prediction for a system of Brownian particles in $d=2$ interacting via Gaussian pairwise potentials, finding excellent agreement \revvv{[an analogous test for a smaller system in $d=3$ is reported in~\cite{sm}]}. 

\smallskip
\textit{Robustness of the algebraic behavior.---} 
First, since in our calculation we neglected perturbative terms of $\order{h^2}$ or higher, it is natural to wonder whether the qualitative features of \rev{$\gcont(\vb x)$} presented above remain robust away from the limit in which $h= 1/\bar\rho^{1/2}$ is small. 
We verified numerically that this is indeed the case
[see e.g.~\cref{fig:gr_continuum}(a), for which $\bar\rho=0.5$]; 
even
in regimes where
discrepancies do show up at short distances, by contrast they hardly affect the large-distance tail~\eqref{eq:gr_continuum_asymptotics} of the correlation profile. 
In fact, in~\cite{sm} we verify that the validity of perturbation theory is rather linked to the interaction energy being small compared to typical thermal fluctuations, as hinted above and in~\cite{Demery2014}.

Furthermore, 
the expression~\eqref{eq:gr_continuum_sol} relies on the existence of the Fourier transform $u\q$ of the interaction potential. 
%$U(\vb x)$. 
This,
together with the use of perturbation theory,
in principle restricts the applicability of the theory to soft interactions, and excludes strong short-distance repulsion --- such as the one typically modeled by Lennard-Jones (LJ) potentials in the description of fluids~\cite{McDonald_book,mcquarrie2000statistical}.
To test the robustness of the asymptotic algebraic decay predicted for the correlation profile \rev{$\gcont(\vb x)$} in \cref{eq:gr_continuum_asymptotics},
we have thus simulated LJ fluids and measured \rev{$\gcont(\vb x)$} numerically. 
The results in \cref{fig:gr_continuum}(b) indeed confirm the persistence of this large-distance algebraic behavior.
In the simulations we used both LJ and Weeks-Chandler-Andersen (WCA) potentials, the latter corresponding to retaining only the repulsive part of the LJ interaction, and shifting it vertically so that the minimum potential energy is zero~\cite{sm}. Remarkably, the two curves shown in \cref{fig:gr_continuum}(b) 
appear to coincide, provided the typical length and energy scales of the two potentials are chosen to be equal.
This is consistent with the prefactor in \cref{eq:gr_continuum_asymptotics} depending only on
$u_{\vb q=\bm 0}$, but it is nontrivial in the present context, since the LJ and WCA potentials 
do not admit a Fourier transform.

\smallskip
\textit{Conclusion.---}
In summary, we have characterized the
cross-correlation profile between the TP position and the density of surrounding particles, for several paradigmatic interacting particle models. 
Remarkably,
its large-distance algebraic behavior \rev{$\gcont(x)\sim x^{1-d}$} turns out to be simple and robust upon changing the details of the pairwise interaction.
\revv{This decay is faster in higher dimensions, which traces back to the efficiency of the homogenization by the diffusion of the bath particles in high dimensions. Note that, in confined quasi-$1d$ systems such as strips, diffusion is even more efficient, so that the decay is expected to be faster than algebraic (see also~\cite{Benichou2016} for a similar effect on the mean density around a driven tracer).}

Our findings pave the way towards the analysis of higher-order correlation profiles, which will be the subject of future work.  
We expect these and related observables 
to be accessible 
using the approach developed in this work, 
where the preliminary calculation of the stationary TP-bath correlation profiles allows one to access the statistical properties of the TP without resorting to the 
path-integral formalism developed in~\cite{demerypath}. For instance, it would be interesting to inspect if the crossover from algebraic to exponential decay of the average density profile \rev{$\prof(\vb x)$}, observed in~\cite{Benichou2016} upon spatial confinement of the system, carries over also to the continuum case~\cite{Venturelli_2022_confined}.
Finally, our analysis naturally calls for a unified description of such seemingly different interacting particle systems, which would shed light on the physical origin of the universal behavior that we unveiled in this work.

\smallskip
\begin{acknowledgments}
DV thanks Luca Capizzi and Pietro Luigi Muzzeddu for stimulating discussions.
\end{acknowledgments}

\bibliography{references}% Produces the bibliography via BibTeX.

\begin{thebibliography}{54}%
\makeatletter
\providecommand \@ifxundefined [1]{%
 \@ifx{#1\undefined}
}%
\providecommand \@ifnum [1]{%
 \ifnum #1\expandafter \@firstoftwo
 \else \expandafter \@secondoftwo
 \fi
}%
\providecommand \@ifx [1]{%
 \ifx #1\expandafter \@firstoftwo
 \else \expandafter \@secondoftwo
 \fi
}%
\providecommand \natexlab [1]{#1}%
\providecommand \enquote  [1]{``#1''}%
\providecommand \bibnamefont  [1]{#1}%
\providecommand \bibfnamefont [1]{#1}%
\providecommand \citenamefont [1]{#1}%
\providecommand \href@noop [0]{\@secondoftwo}%
\providecommand \href [0]{\begingroup \@sanitize@url \@href}%
\providecommand \@href[1]{\@@startlink{#1}\@@href}%
\providecommand \@@href[1]{\endgroup#1\@@endlink}%
\providecommand \@sanitize@url [0]{\catcode `\\12\catcode `\$12\catcode `\&12\catcode `\#12\catcode `\^12\catcode `\_12\catcode `\%12\relax}%
\providecommand \@@startlink[1]{}%
\providecommand \@@endlink[0]{}%
\providecommand \url  [0]{\begingroup\@sanitize@url \@url }%
\providecommand \@url [1]{\endgroup\@href {#1}{\urlprefix }}%
\providecommand \urlprefix  [0]{URL }%
\providecommand \Eprint [0]{\href }%
\providecommand \doibase [0]{https://doi.org/}%
\providecommand \selectlanguage [0]{\@gobble}%
\providecommand \bibinfo  [0]{\@secondoftwo}%
\providecommand \bibfield  [0]{\@secondoftwo}%
\providecommand \translation [1]{[#1]}%
\providecommand \BibitemOpen [0]{}%
\providecommand \bibitemStop [0]{}%
\providecommand \bibitemNoStop [0]{.\EOS\space}%
\providecommand \EOS [0]{\spacefactor3000\relax}%
\providecommand \BibitemShut  [1]{\csname bibitem#1\endcsname}%
\let\auto@bib@innerbib\@empty
%</preamble>
\bibitem [{\citenamefont {Squires}\ and\ \citenamefont {Mason}(2010)}]{Squires2010}%
  \BibitemOpen
  \bibfield  {author} {\bibinfo {author} {\bibfnamefont {T.~M.}\ \bibnamefont {Squires}}\ and\ \bibinfo {author} {\bibfnamefont {T.~G.}\ \bibnamefont {Mason}},\ }\bibfield  {title} {\bibinfo {title} {Fluid {Mechanics} of {Microrheology}},\ }\href {https://doi.org/10.1146/annurev-fluid-121108-145608} {\bibfield  {journal} {\bibinfo  {journal} {Annu. Rev. Fluid Mech.}\ }\textbf {\bibinfo {volume} {42}},\ \bibinfo {pages} {413} (\bibinfo {year} {2010})}\BibitemShut {NoStop}%
\bibitem [{\citenamefont {Chen}\ \emph {et~al.}(2003)\citenamefont {Chen}, \citenamefont {Weeks}, \citenamefont {Crocker}, \citenamefont {Islam}, \citenamefont {Verma}, \citenamefont {Gruber}, \citenamefont {Levine}, \citenamefont {Lubensky},\ and\ \citenamefont {Yodh}}]{Chen2003}%
  \BibitemOpen
  \bibfield  {author} {\bibinfo {author} {\bibfnamefont {D.~T.}\ \bibnamefont {Chen}}, \bibinfo {author} {\bibfnamefont {E.~R.}\ \bibnamefont {Weeks}}, \bibinfo {author} {\bibfnamefont {J.~C.}\ \bibnamefont {Crocker}}, \bibinfo {author} {\bibfnamefont {M.~F.}\ \bibnamefont {Islam}}, \bibinfo {author} {\bibfnamefont {R.}~\bibnamefont {Verma}}, \bibinfo {author} {\bibfnamefont {J.}~\bibnamefont {Gruber}}, \bibinfo {author} {\bibfnamefont {A.~J.}\ \bibnamefont {Levine}}, \bibinfo {author} {\bibfnamefont {T.~C.}\ \bibnamefont {Lubensky}},\ and\ \bibinfo {author} {\bibfnamefont {a.~G.}\ \bibnamefont {Yodh}},\ }\bibfield  {title} {\bibinfo {title} {Rheological {Microscopy}: {Local} {Mechanical} {Properties} from {Microrheology}},\ }\href {https://doi.org/10.1103/PhysRevLett.90.108301} {\bibfield  {journal} {\bibinfo  {journal} {Phys. Rev. Lett.}\ }\textbf {\bibinfo {volume} {90}},\ \bibinfo {pages} {108301} (\bibinfo {year} {2003})}\BibitemShut {NoStop}%
\bibitem [{\citenamefont {Lau}\ \emph {et~al.}(2003)\citenamefont {Lau}, \citenamefont {Hoffman}, \citenamefont {Davies}, \citenamefont {Crocker},\ and\ \citenamefont {Lubensky}}]{Lau2003}%
  \BibitemOpen
  \bibfield  {author} {\bibinfo {author} {\bibfnamefont {A.~W.~C.}\ \bibnamefont {Lau}}, \bibinfo {author} {\bibfnamefont {B.~D.}\ \bibnamefont {Hoffman}}, \bibinfo {author} {\bibfnamefont {A.}~\bibnamefont {Davies}}, \bibinfo {author} {\bibfnamefont {J.~C.}\ \bibnamefont {Crocker}},\ and\ \bibinfo {author} {\bibfnamefont {T.~C.}\ \bibnamefont {Lubensky}},\ }\bibfield  {title} {\bibinfo {title} {Microrheology, {Stress} {Fluctuations}, and {Active} {Behavior} of {Living} {Cells}.},\ }\href {https://doi.org/10.1103/PhysRevLett.91.198101} {\bibfield  {journal} {\bibinfo  {journal} {Phys. Rev. Lett.}\ }\textbf {\bibinfo {volume} {91}},\ \bibinfo {pages} {198101} (\bibinfo {year} {2003})}\BibitemShut {NoStop}%
\bibitem [{\citenamefont {Puertas}\ and\ \citenamefont {Voigtmann}(2014)}]{Puertas2014}%
  \BibitemOpen
  \bibfield  {author} {\bibinfo {author} {\bibfnamefont {A.~M.}\ \bibnamefont {Puertas}}\ and\ \bibinfo {author} {\bibfnamefont {T.}~\bibnamefont {Voigtmann}},\ }\bibfield  {title} {\bibinfo {title} {Microrheology of colloidal systems},\ }\href {https://doi.org/10.1088/0953-8984/26/24/243101} {\bibfield  {journal} {\bibinfo  {journal} {J. Phys. Condens. Mat.}\ }\textbf {\bibinfo {volume} {26}},\ \bibinfo {pages} {243101} (\bibinfo {year} {2014})}\BibitemShut {NoStop}%
\bibitem [{\citenamefont {Wilson}\ \emph {et~al.}(2009)\citenamefont {Wilson}, \citenamefont {Harrison}, \citenamefont {Schofield}, \citenamefont {Arlt},\ and\ \citenamefont {Poon}}]{Wilson2009}%
  \BibitemOpen
  \bibfield  {author} {\bibinfo {author} {\bibfnamefont {L.~G.}\ \bibnamefont {Wilson}}, \bibinfo {author} {\bibfnamefont {a.~W.}\ \bibnamefont {Harrison}}, \bibinfo {author} {\bibfnamefont {a.~B.}\ \bibnamefont {Schofield}}, \bibinfo {author} {\bibfnamefont {J.}~\bibnamefont {Arlt}},\ and\ \bibinfo {author} {\bibfnamefont {W.~C.~K.}\ \bibnamefont {Poon}},\ }\bibfield  {title} {\bibinfo {title} {Passive and active microrheology of hard-sphere colloids},\ }\href {https://doi.org/10.1021/jp8079028} {\bibfield  {journal} {\bibinfo  {journal} {J. Phys. Chem. B}\ }\textbf {\bibinfo {volume} {113}},\ \bibinfo {pages} {3806} (\bibinfo {year} {2009})}\BibitemShut {NoStop}%
\bibitem [{\citenamefont {Wilson}\ \emph {et~al.}(2011)\citenamefont {Wilson}, \citenamefont {Harrison}, \citenamefont {Poon},\ and\ \citenamefont {Puertas}}]{Wilson2011}%
  \BibitemOpen
  \bibfield  {author} {\bibinfo {author} {\bibfnamefont {L.~G.}\ \bibnamefont {Wilson}}, \bibinfo {author} {\bibfnamefont {A.~W.}\ \bibnamefont {Harrison}}, \bibinfo {author} {\bibfnamefont {W.~C.~K.}\ \bibnamefont {Poon}},\ and\ \bibinfo {author} {\bibfnamefont {A.~M.}\ \bibnamefont {Puertas}},\ }\bibfield  {title} {\bibinfo {title} {Microrheology and the fluctuation theorem in dense colloids},\ }\href {https://doi.org/10.1209/0295-5075/93/58007} {\bibfield  {journal} {\bibinfo  {journal} {EPL}\ }\textbf {\bibinfo {volume} {93}},\ \bibinfo {pages} {58007} (\bibinfo {year} {2011})}\BibitemShut {NoStop}%
\bibitem [{\citenamefont {Guo}\ \emph {et~al.}(2014)\citenamefont {Guo}, \citenamefont {Ehrlicher}, \citenamefont {Jensen}, \citenamefont {Renz}, \citenamefont {Moore}, \citenamefont {Goldman}, \citenamefont {Lippincott-Schwartz}, \citenamefont {Mackintosh},\ and\ \citenamefont {Weitz}}]{Guo2014}%
  \BibitemOpen
  \bibfield  {author} {\bibinfo {author} {\bibfnamefont {M.}~\bibnamefont {Guo}}, \bibinfo {author} {\bibfnamefont {A.~J.}\ \bibnamefont {Ehrlicher}}, \bibinfo {author} {\bibfnamefont {M.~H.}\ \bibnamefont {Jensen}}, \bibinfo {author} {\bibfnamefont {M.}~\bibnamefont {Renz}}, \bibinfo {author} {\bibfnamefont {J.~R.}\ \bibnamefont {Moore}}, \bibinfo {author} {\bibfnamefont {R.~D.}\ \bibnamefont {Goldman}}, \bibinfo {author} {\bibfnamefont {J.}~\bibnamefont {Lippincott-Schwartz}}, \bibinfo {author} {\bibfnamefont {F.~C.}\ \bibnamefont {Mackintosh}},\ and\ \bibinfo {author} {\bibfnamefont {D.~A.}\ \bibnamefont {Weitz}},\ }\bibfield  {title} {\bibinfo {title} {Probing the stochastic, motor-driven properties of the cytoplasm using force spectrum microscopy},\ }\href {https://doi.org/10.1016/j.cell.2014.06.051} {\bibfield  {journal} {\bibinfo  {journal} {Cell}\ }\textbf {\bibinfo {volume} {158}},\ \bibinfo {pages} {822} (\bibinfo {year} {2014})}\BibitemShut {NoStop}%
\bibitem [{\citenamefont {Parry}\ \emph {et~al.}(2014)\citenamefont {Parry}, \citenamefont {Surovtsev}, \citenamefont {Cabeen}, \citenamefont {O'Hern}, \citenamefont {Dufresne},\ and\ \citenamefont {Jacobs-Wagner}}]{Parry2014}%
  \BibitemOpen
  \bibfield  {author} {\bibinfo {author} {\bibfnamefont {B.~R.}\ \bibnamefont {Parry}}, \bibinfo {author} {\bibfnamefont {I.~V.}\ \bibnamefont {Surovtsev}}, \bibinfo {author} {\bibfnamefont {M.~T.}\ \bibnamefont {Cabeen}}, \bibinfo {author} {\bibfnamefont {C.~S.}\ \bibnamefont {O'Hern}}, \bibinfo {author} {\bibfnamefont {E.~R.}\ \bibnamefont {Dufresne}},\ and\ \bibinfo {author} {\bibfnamefont {C.}~\bibnamefont {Jacobs-Wagner}},\ }\bibfield  {title} {\bibinfo {title} {The bacterial cytoplasm has glass-like properties and is fluidized by metabolic activity},\ }\href {https://doi.org/10.1016/j.cell.2013.11.028} {\bibfield  {journal} {\bibinfo  {journal} {Cell}\ }\textbf {\bibinfo {volume} {156}},\ \bibinfo {pages} {183} (\bibinfo {year} {2014})}\BibitemShut {NoStop}%
\bibitem [{\citenamefont {Marconi}\ \emph {et~al.}(2008)\citenamefont {Marconi}, \citenamefont {Puglisi}, \citenamefont {Rondoni},\ and\ \citenamefont {Vulpiani}}]{Marconi2008}%
  \BibitemOpen
  \bibfield  {author} {\bibinfo {author} {\bibfnamefont {U.~M.~B.}\ \bibnamefont {Marconi}}, \bibinfo {author} {\bibfnamefont {A.}~\bibnamefont {Puglisi}}, \bibinfo {author} {\bibfnamefont {L.}~\bibnamefont {Rondoni}},\ and\ \bibinfo {author} {\bibfnamefont {A.}~\bibnamefont {Vulpiani}},\ }\bibfield  {title} {\bibinfo {title} {Fluctuation–dissipation: Response theory in statistical physics},\ }\href {https://doi.org/10.1016/j.physrep.2008.02.002} {\bibfield  {journal} {\bibinfo  {journal} {Phys. Rep.}\ }\textbf {\bibinfo {volume} {461}},\ \bibinfo {pages} {111–195} (\bibinfo {year} {2008})}\BibitemShut {NoStop}%
\bibitem [{\citenamefont {Langevin}(1908)}]{Langevin_1908}%
  \BibitemOpen
  \bibfield  {author} {\bibinfo {author} {\bibfnamefont {P.}~\bibnamefont {Langevin}},\ }\bibfield  {title} {\bibinfo {title} {Sur la th{\'e}orie du mouvement {Brownien}},\ }\href@noop {} {\bibfield  {journal} {\bibinfo  {journal} {Compt. Rendus}\ }\textbf {\bibinfo {volume} {146}},\ \bibinfo {pages} {530} (\bibinfo {year} {1908})}\BibitemShut {NoStop}%
\bibitem [{\citenamefont {Mazur}\ and\ \citenamefont {Oppenheim}(1970)}]{Mazur1970}%
  \BibitemOpen
  \bibfield  {author} {\bibinfo {author} {\bibfnamefont {P.}~\bibnamefont {Mazur}}\ and\ \bibinfo {author} {\bibfnamefont {I.}~\bibnamefont {Oppenheim}},\ }\bibfield  {title} {\bibinfo {title} {Molecular theory of {Brownian} motion},\ }\href {https://doi.org/10.1016/0031-8914(70)90005-4} {\bibfield  {journal} {\bibinfo  {journal} {Physica}\ }\textbf {\bibinfo {volume} {50}},\ \bibinfo {pages} {241} (\bibinfo {year} {1970})}\BibitemShut {NoStop}%
\bibitem [{\citenamefont {Lebowitz}\ and\ \citenamefont {Rubin}(1963)}]{Lebowitz1963}%
  \BibitemOpen
  \bibfield  {author} {\bibinfo {author} {\bibfnamefont {J.~L.}\ \bibnamefont {Lebowitz}}\ and\ \bibinfo {author} {\bibfnamefont {E.}~\bibnamefont {Rubin}},\ }\bibfield  {title} {\bibinfo {title} {Dynamical study of {Brownian} motion},\ }\href {https://doi.org/10.1103/PhysRev.131.2381} {\bibfield  {journal} {\bibinfo  {journal} {Phys. Rev.}\ }\textbf {\bibinfo {volume} {131}},\ \bibinfo {pages} {2381} (\bibinfo {year} {1963})}\BibitemShut {NoStop}%
\bibitem [{\citenamefont {Chou}\ \emph {et~al.}(2011)\citenamefont {Chou}, \citenamefont {Mallick},\ and\ \citenamefont {Zia}}]{Chou2011}%
  \BibitemOpen
  \bibfield  {author} {\bibinfo {author} {\bibfnamefont {T.}~\bibnamefont {Chou}}, \bibinfo {author} {\bibfnamefont {K.}~\bibnamefont {Mallick}},\ and\ \bibinfo {author} {\bibfnamefont {R.~K.~P.}\ \bibnamefont {Zia}},\ }\bibfield  {title} {\bibinfo {title} {Non-equilibrium statistical mechanics: {From} a paradigmatic model to biological transport},\ }\href {https://doi.org/10.1088/0034-4885/74/11/116601} {\bibfield  {journal} {\bibinfo  {journal} {Rep. Prog. Phys.}\ }\textbf {\bibinfo {volume} {74}},\ \bibinfo {pages} {116601} (\bibinfo {year} {2011})}\BibitemShut {NoStop}%
\bibitem [{\citenamefont {Mallick}(2015)}]{Mallick2015}%
  \BibitemOpen
  \bibfield  {author} {\bibinfo {author} {\bibfnamefont {K.}~\bibnamefont {Mallick}},\ }\bibfield  {title} {\bibinfo {title} {The exclusion process: {A} paradigm for non-equilibrium behaviour},\ }\href {https://doi.org/10.1016/j.physa.2014.07.046} {\bibfield  {journal} {\bibinfo  {journal} {Physica A}\ }\textbf {\bibinfo {volume} {418}},\ \bibinfo {pages} {17} (\bibinfo {year} {2015})}\BibitemShut {NoStop}%
\bibitem [{\citenamefont {Spohn}(1991)}]{Spohn1991}%
  \BibitemOpen
  \bibfield  {author} {\bibinfo {author} {\bibfnamefont {H.}~\bibnamefont {Spohn}},\ }\href@noop {} {\emph {\bibinfo {title} {Large scale dynamics of interacting particles}}}\ (\bibinfo  {publisher} {Springer Berlin, Heidelberg},\ \bibinfo {year} {1991})\BibitemShut {NoStop}%
\bibitem [{\citenamefont {Grabsch}\ \emph {et~al.}(2022)\citenamefont {Grabsch}, \citenamefont {Poncet}, \citenamefont {Rizkallah}, \citenamefont {Illien},\ and\ \citenamefont {B{\'e}nichou}}]{Grabsch2022}%
  \BibitemOpen
  \bibfield  {author} {\bibinfo {author} {\bibfnamefont {A.}~\bibnamefont {Grabsch}}, \bibinfo {author} {\bibfnamefont {A.}~\bibnamefont {Poncet}}, \bibinfo {author} {\bibfnamefont {P.}~\bibnamefont {Rizkallah}}, \bibinfo {author} {\bibfnamefont {P.}~\bibnamefont {Illien}},\ and\ \bibinfo {author} {\bibfnamefont {O.}~\bibnamefont {B{\'e}nichou}},\ }\bibfield  {title} {\bibinfo {title} {Exact closure and solution for spatial correlations in single-file diffusion},\ }\href {https://doi.org/10.1126/sciadv.abm5043} {\bibfield  {journal} {\bibinfo  {journal} {Sci. Adv.}\ }\textbf {\bibinfo {volume} {8}},\ \bibinfo {pages} {eabm5043} (\bibinfo {year} {2022})}\BibitemShut {NoStop}%
\bibitem [{\citenamefont {Grabsch}\ \emph {et~al.}(2023)\citenamefont {Grabsch}, \citenamefont {Rizkallah}, \citenamefont {Poncet}, \citenamefont {Illien},\ and\ \citenamefont {B\'enichou}}]{Grabsch2023}%
  \BibitemOpen
  \bibfield  {author} {\bibinfo {author} {\bibfnamefont {A.}~\bibnamefont {Grabsch}}, \bibinfo {author} {\bibfnamefont {P.}~\bibnamefont {Rizkallah}}, \bibinfo {author} {\bibfnamefont {A.}~\bibnamefont {Poncet}}, \bibinfo {author} {\bibfnamefont {P.}~\bibnamefont {Illien}},\ and\ \bibinfo {author} {\bibfnamefont {O.}~\bibnamefont {B\'enichou}},\ }\bibfield  {title} {\bibinfo {title} {Exact spatial correlations in single-file diffusion},\ }\href {https://doi.org/10.1103/PhysRevE.107.044131} {\bibfield  {journal} {\bibinfo  {journal} {Phys. Rev. E}\ }\textbf {\bibinfo {volume} {107}},\ \bibinfo {pages} {044131} (\bibinfo {year} {2023})}\BibitemShut {NoStop}%
\bibitem [{\citenamefont {Mallick}\ \emph {et~al.}(2022)\citenamefont {Mallick}, \citenamefont {Moriya},\ and\ \citenamefont {Sasamoto}}]{Mallick_2022}%
  \BibitemOpen
  \bibfield  {author} {\bibinfo {author} {\bibfnamefont {K.}~\bibnamefont {Mallick}}, \bibinfo {author} {\bibfnamefont {H.}~\bibnamefont {Moriya}},\ and\ \bibinfo {author} {\bibfnamefont {T.}~\bibnamefont {Sasamoto}},\ }\bibfield  {title} {\bibinfo {title} {Exact solution of the macroscopic fluctuation theory for the symmetric exclusion process},\ }\href {https://doi.org/10.1103/PhysRevLett.129.040601} {\bibfield  {journal} {\bibinfo  {journal} {Phys. Rev. Lett.}\ }\textbf {\bibinfo {volume} {129}},\ \bibinfo {pages} {040601} (\bibinfo {year} {2022})}\BibitemShut {NoStop}%
\bibitem [{\citenamefont {Rizkallah}\ \emph {et~al.}(2023)\citenamefont {Rizkallah}, \citenamefont {Grabsch}, \citenamefont {Illien},\ and\ \citenamefont {Bénichou}}]{Rizkallah2023}%
  \BibitemOpen
  \bibfield  {author} {\bibinfo {author} {\bibfnamefont {P.}~\bibnamefont {Rizkallah}}, \bibinfo {author} {\bibfnamefont {A.}~\bibnamefont {Grabsch}}, \bibinfo {author} {\bibfnamefont {P.}~\bibnamefont {Illien}},\ and\ \bibinfo {author} {\bibfnamefont {O.}~\bibnamefont {Bénichou}},\ }\bibfield  {title} {\bibinfo {title} {Duality relations in single-file diffusion},\ }\href {https://doi.org/10.1088/1742-5468/aca8fb} {\bibfield  {journal} {\bibinfo  {journal} {J. Stat. Mech.}\ }\textbf {\bibinfo {volume} {2023}},\ \bibinfo {pages} {013202} (\bibinfo {year} {2023})}\BibitemShut {NoStop}%
\bibitem [{\citenamefont {B\'enichou}\ \emph {et~al.}(2014)\citenamefont {B\'enichou}, \citenamefont {Illien}, \citenamefont {Oshanin}, \citenamefont {Sarracino},\ and\ \citenamefont {Voituriez}}]{Benichou_2014_microscopic}%
  \BibitemOpen
  \bibfield  {author} {\bibinfo {author} {\bibfnamefont {O.}~\bibnamefont {B\'enichou}}, \bibinfo {author} {\bibfnamefont {P.}~\bibnamefont {Illien}}, \bibinfo {author} {\bibfnamefont {G.}~\bibnamefont {Oshanin}}, \bibinfo {author} {\bibfnamefont {A.}~\bibnamefont {Sarracino}},\ and\ \bibinfo {author} {\bibfnamefont {R.}~\bibnamefont {Voituriez}},\ }\bibfield  {title} {\bibinfo {title} {Microscopic theory for negative differential mobility in crowded environments},\ }\href {https://doi.org/10.1103/PhysRevLett.113.268002} {\bibfield  {journal} {\bibinfo  {journal} {Phys. Rev. Lett.}\ }\textbf {\bibinfo {volume} {113}},\ \bibinfo {pages} {268002} (\bibinfo {year} {2014})}\BibitemShut {NoStop}%
\bibitem [{\citenamefont {Illien}\ \emph {et~al.}(2015)\citenamefont {Illien}, \citenamefont {Bénichou}, \citenamefont {Oshanin},\ and\ \citenamefont {Voituriez}}]{Illien2015_distribution}%
  \BibitemOpen
  \bibfield  {author} {\bibinfo {author} {\bibfnamefont {P.}~\bibnamefont {Illien}}, \bibinfo {author} {\bibfnamefont {O.}~\bibnamefont {Bénichou}}, \bibinfo {author} {\bibfnamefont {G.}~\bibnamefont {Oshanin}},\ and\ \bibinfo {author} {\bibfnamefont {R.}~\bibnamefont {Voituriez}},\ }\bibfield  {title} {\bibinfo {title} {Distribution of the position of a driven tracer in a hardcore lattice gas},\ }\href {https://doi.org/10.1088/1742-5468/2015/11/p11016} {\bibfield  {journal} {\bibinfo  {journal} {J. Stat. Mech.}\ }\textbf {\bibinfo {volume} {2015}},\ \bibinfo {pages} {P11016} (\bibinfo {year} {2015})}\BibitemShut {NoStop}%
\bibitem [{\citenamefont {Illien}\ \emph {et~al.}(2018)\citenamefont {Illien}, \citenamefont {B\'enichou}, \citenamefont {Oshanin}, \citenamefont {Sarracino},\ and\ \citenamefont {Voituriez}}]{Illien2018}%
  \BibitemOpen
  \bibfield  {author} {\bibinfo {author} {\bibfnamefont {P.}~\bibnamefont {Illien}}, \bibinfo {author} {\bibfnamefont {O.}~\bibnamefont {B\'enichou}}, \bibinfo {author} {\bibfnamefont {G.}~\bibnamefont {Oshanin}}, \bibinfo {author} {\bibfnamefont {A.}~\bibnamefont {Sarracino}},\ and\ \bibinfo {author} {\bibfnamefont {R.}~\bibnamefont {Voituriez}},\ }\bibfield  {title} {\bibinfo {title} {Nonequilibrium fluctuations and enhanced diffusion of a driven particle in a dense environment},\ }\href {https://doi.org/10.1103/PhysRevLett.120.200606} {\bibfield  {journal} {\bibinfo  {journal} {Phys. Rev. Lett.}\ }\textbf {\bibinfo {volume} {120}},\ \bibinfo {pages} {200606} (\bibinfo {year} {2018})}\BibitemShut {NoStop}%
\bibitem [{\citenamefont {Bénichou}\ \emph {et~al.}(2018)\citenamefont {Bénichou}, \citenamefont {Illien}, \citenamefont {Oshanin}, \citenamefont {Sarracino},\ and\ \citenamefont {Voituriez}}]{Benichou2018}%
  \BibitemOpen
  \bibfield  {author} {\bibinfo {author} {\bibfnamefont {O.}~\bibnamefont {Bénichou}}, \bibinfo {author} {\bibfnamefont {P.}~\bibnamefont {Illien}}, \bibinfo {author} {\bibfnamefont {G.}~\bibnamefont {Oshanin}}, \bibinfo {author} {\bibfnamefont {A.}~\bibnamefont {Sarracino}},\ and\ \bibinfo {author} {\bibfnamefont {R.}~\bibnamefont {Voituriez}},\ }\bibfield  {title} {\bibinfo {title} {Tracer diffusion in crowded narrow channels},\ }\href {https://doi.org/10.1088/1361-648x/aae13a} {\bibfield  {journal} {\bibinfo  {journal} {J. Phys. Condens. Mat.}\ }\textbf {\bibinfo {volume} {30}},\ \bibinfo {pages} {443001} (\bibinfo {year} {2018})}\BibitemShut {NoStop}%
\bibitem [{\citenamefont {Dean}(1996)}]{Dean1996}%
  \BibitemOpen
  \bibfield  {author} {\bibinfo {author} {\bibfnamefont {D.~S.}\ \bibnamefont {Dean}},\ }\bibfield  {title} {\bibinfo {title} {Langevin equation for the density of a system of interacting {Langevin} processes},\ }\href {https://doi.org/10.1088/0305-4470/29/24/001} {\bibfield  {journal} {\bibinfo  {journal} {J. Phys. A: Math. Gen.}\ }\textbf {\bibinfo {volume} {29}},\ \bibinfo {pages} {L613–L617} (\bibinfo {year} {1996})}\BibitemShut {NoStop}%
\bibitem [{\citenamefont {Kawasaki}(1998)}]{Kawasaki1998}%
  \BibitemOpen
  \bibfield  {author} {\bibinfo {author} {\bibfnamefont {K.}~\bibnamefont {Kawasaki}},\ }\bibfield  {title} {\bibinfo {title} {Microscopic analyses of the dynamical density functional equation of dense fluids},\ }\href {https://doi.org/10.1023/b:joss.0000033240.66359.6c} {\bibfield  {journal} {\bibinfo  {journal} {J. Stat. Phys.}\ }\textbf {\bibinfo {volume} {93}},\ \bibinfo {pages} {527–546} (\bibinfo {year} {1998})}\BibitemShut {NoStop}%
\bibitem [{\citenamefont {Démery}\ \emph {et~al.}(2014)\citenamefont {Démery}, \citenamefont {Bénichou},\ and\ \citenamefont {Jacquin}}]{Demery2014}%
  \BibitemOpen
  \bibfield  {author} {\bibinfo {author} {\bibfnamefont {V.}~\bibnamefont {Démery}}, \bibinfo {author} {\bibfnamefont {O.}~\bibnamefont {Bénichou}},\ and\ \bibinfo {author} {\bibfnamefont {H.}~\bibnamefont {Jacquin}},\ }\bibfield  {title} {\bibinfo {title} {Generalized {Langevin} equations for a driven tracer in dense soft colloids: construction and applications},\ }\href {https://doi.org/10.1088/1367-2630/16/5/053032} {\bibfield  {journal} {\bibinfo  {journal} {New J. Phys.}\ }\textbf {\bibinfo {volume} {16}},\ \bibinfo {pages} {053032} (\bibinfo {year} {2014})}\BibitemShut {NoStop}%
\bibitem [{Note1()}]{Note1}%
  \BibitemOpen
  \bibinfo {note} {For instance, this approach is unable to reproduce the subdiffusive behavior expected for hard-core particles in one dimension.}\BibitemShut {Stop}%
\bibitem [{\citenamefont {Hansen}\ and\ \citenamefont {McDonald}(2013)}]{McDonald_book}%
  \BibitemOpen
  \bibfield  {author} {\bibinfo {author} {\bibfnamefont {J.-P.}\ \bibnamefont {Hansen}}\ and\ \bibinfo {author} {\bibfnamefont {I.~R.}\ \bibnamefont {McDonald}},\ }\href {https://doi.org/10.1016/c2010-0-66723-x} {\emph {\bibinfo {title} {Theory of Simple Liquids}}}\ (\bibinfo  {publisher} {Elsevier},\ \bibinfo {year} {2013})\BibitemShut {NoStop}%
\bibitem [{sm()}]{sm}%
  \BibitemOpen
  \href@noop {} {\bibinfo {title} {See the {S}upplemental {M}aterial at [{URL}], which includes {R}efs.~\cite{hughes1995random,Poncet_2021,table,Tauber,Benichou_2000,wellGauss,Thompson2022}.}}\BibitemShut {Stop}%
\bibitem [{\citenamefont {B\'enichou}\ \emph {et~al.}(2013)\citenamefont {B\'enichou}, \citenamefont {Illien}, \citenamefont {Oshanin},\ and\ \citenamefont {Voituriez}}]{Benichou2013}%
  \BibitemOpen
  \bibfield  {author} {\bibinfo {author} {\bibfnamefont {O.}~\bibnamefont {B\'enichou}}, \bibinfo {author} {\bibfnamefont {P.}~\bibnamefont {Illien}}, \bibinfo {author} {\bibfnamefont {G.}~\bibnamefont {Oshanin}},\ and\ \bibinfo {author} {\bibfnamefont {R.}~\bibnamefont {Voituriez}},\ }\bibfield  {title} {\bibinfo {title} {Fluctuations and correlations of a driven tracer in a hard-core lattice gas},\ }\href {https://doi.org/10.1103/PhysRevE.87.032164} {\bibfield  {journal} {\bibinfo  {journal} {Phys. Rev. E}\ }\textbf {\bibinfo {volume} {87}},\ \bibinfo {pages} {032164} (\bibinfo {year} {2013})}\BibitemShut {NoStop}%
\bibitem [{\citenamefont {Evans}\ \emph {et~al.}(1993)\citenamefont {Evans}, \citenamefont {Henderson}, \citenamefont {Hoyle}, \citenamefont {Parry},\ and\ \citenamefont {Sabeur}}]{Evans1993}%
  \BibitemOpen
  \bibfield  {author} {\bibinfo {author} {\bibfnamefont {R.}~\bibnamefont {Evans}}, \bibinfo {author} {\bibfnamefont {J.}~\bibnamefont {Henderson}}, \bibinfo {author} {\bibfnamefont {D.}~\bibnamefont {Hoyle}}, \bibinfo {author} {\bibfnamefont {A.}~\bibnamefont {Parry}},\ and\ \bibinfo {author} {\bibfnamefont {Z.}~\bibnamefont {Sabeur}},\ }\bibfield  {title} {\bibinfo {title} {Asymptotic decay of liquid structure: oscillatory liquid-vapour density profiles and the {Fisher-Widom} line},\ }\href {https://doi.org/10.1080/00268979300102621} {\bibfield  {journal} {\bibinfo  {journal} {Mol. Phys.}\ }\textbf {\bibinfo {volume} {80}},\ \bibinfo {pages} {755–775} (\bibinfo {year} {1993})}\BibitemShut {NoStop}%
\bibitem [{\citenamefont {Evans}\ \emph {et~al.}(1994)\citenamefont {Evans}, \citenamefont {Leote~de Carvalho}, \citenamefont {Henderson},\ and\ \citenamefont {Hoyle}}]{Evans1994}%
  \BibitemOpen
  \bibfield  {author} {\bibinfo {author} {\bibfnamefont {R.}~\bibnamefont {Evans}}, \bibinfo {author} {\bibfnamefont {R.~J.~F.}\ \bibnamefont {Leote~de Carvalho}}, \bibinfo {author} {\bibfnamefont {J.~R.}\ \bibnamefont {Henderson}},\ and\ \bibinfo {author} {\bibfnamefont {D.~C.}\ \bibnamefont {Hoyle}},\ }\bibfield  {title} {\bibinfo {title} {Asymptotic decay of correlations in liquids and their mixtures},\ }\href {https://doi.org/10.1063/1.466920} {\bibfield  {journal} {\bibinfo  {journal} {J. Chem. Phys.}\ }\textbf {\bibinfo {volume} {100}},\ \bibinfo {pages} {591–603} (\bibinfo {year} {1994})}\BibitemShut {NoStop}%
\bibitem [{\citenamefont {Leote~de Carvalho}\ and\ \citenamefont {Evans}(1994)}]{LeotedeCarvalho1994}%
  \BibitemOpen
  \bibfield  {author} {\bibinfo {author} {\bibfnamefont {R.}~\bibnamefont {Leote~de Carvalho}}\ and\ \bibinfo {author} {\bibfnamefont {R.}~\bibnamefont {Evans}},\ }\bibfield  {title} {\bibinfo {title} {The decay of correlations in ionic fluids},\ }\href {https://doi.org/10.1080/00268979400101491} {\bibfield  {journal} {\bibinfo  {journal} {Mol. Phys.}\ }\textbf {\bibinfo {volume} {83}},\ \bibinfo {pages} {619–654} (\bibinfo {year} {1994})}\BibitemShut {NoStop}%
\bibitem [{\citenamefont {Dijkstra}\ and\ \citenamefont {Evans}(2000)}]{Dijkstra2000}%
  \BibitemOpen
  \bibfield  {author} {\bibinfo {author} {\bibfnamefont {M.}~\bibnamefont {Dijkstra}}\ and\ \bibinfo {author} {\bibfnamefont {R.}~\bibnamefont {Evans}},\ }\bibfield  {title} {\bibinfo {title} {A simulation study of the decay of the pair correlation function in simple fluids},\ }\href {https://doi.org/10.1063/1.480598} {\bibfield  {journal} {\bibinfo  {journal} {J. Chem. Phys.}\ }\textbf {\bibinfo {volume} {112}},\ \bibinfo {pages} {1449–1456} (\bibinfo {year} {2000})}\BibitemShut {NoStop}%
\bibitem [{\citenamefont {Hopkins}\ \emph {et~al.}(2005)\citenamefont {Hopkins}, \citenamefont {Archer},\ and\ \citenamefont {Evans}}]{Hopkins2005}%
  \BibitemOpen
  \bibfield  {author} {\bibinfo {author} {\bibfnamefont {P.}~\bibnamefont {Hopkins}}, \bibinfo {author} {\bibfnamefont {A.~J.}\ \bibnamefont {Archer}},\ and\ \bibinfo {author} {\bibfnamefont {R.}~\bibnamefont {Evans}},\ }\bibfield  {title} {\bibinfo {title} {Asymptotic decay of pair correlations in a {Yukawa} fluid},\ }\href {https://doi.org/10.1103/PhysRevE.71.027401} {\bibfield  {journal} {\bibinfo  {journal} {Phys. Rev. E}\ }\textbf {\bibinfo {volume} {71}},\ \bibinfo {pages} {027401} (\bibinfo {year} {2005})}\BibitemShut {NoStop}%
\bibitem [{\citenamefont {Stopper}\ \emph {et~al.}(2019)\citenamefont {Stopper}, \citenamefont {Hansen-Goos}, \citenamefont {Roth},\ and\ \citenamefont {Evans}}]{Stopper2019}%
  \BibitemOpen
  \bibfield  {author} {\bibinfo {author} {\bibfnamefont {D.}~\bibnamefont {Stopper}}, \bibinfo {author} {\bibfnamefont {H.}~\bibnamefont {Hansen-Goos}}, \bibinfo {author} {\bibfnamefont {R.}~\bibnamefont {Roth}},\ and\ \bibinfo {author} {\bibfnamefont {R.}~\bibnamefont {Evans}},\ }\bibfield  {title} {\bibinfo {title} {On the decay of the pair correlation function and the line of vanishing excess isothermal compressibility in simple fluids},\ }\href {http://dx.doi.org/10.1063/1.5110044} {\bibfield  {journal} {\bibinfo  {journal} {J. Chem. Phys.}\ }\textbf {\bibinfo {volume} {151}},\ \bibinfo {pages} {014501} (\bibinfo {year} {2019})}\BibitemShut {NoStop}%
\bibitem [{Note2()}]{Note2}%
  \BibitemOpen
  \bibinfo {note} {Since the Gaussian white noises $\protect \bm {\eta }_0$ and $\protect \bm {\xi }$ in \protect \cref {eq:tracer,eq:field} are additive, we adopt the Stratonovich calculus throughout without loss of generality.}\BibitemShut {Stop}%
\bibitem [{\citenamefont {Novikov}(1965)}]{Novikov_1965}%
  \BibitemOpen
  \bibfield  {author} {\bibinfo {author} {\bibfnamefont {E.~A.}\ \bibnamefont {Novikov}},\ }\bibfield  {title} {\bibinfo {title} {Functionals and the random-force method in turbulence theory},\ }\href {http://www.jetp.ras.ru/cgi-bin/dn/e_020_05_1290.pdf} {\bibfield  {journal} {\bibinfo  {journal} {Sov. Phys. JETP}\ }\textbf {\bibinfo {volume} {20}},\ \bibinfo {pages} {1290} (\bibinfo {year} {1965})}\BibitemShut {NoStop}%
\bibitem [{\citenamefont {Łuczka}(2005)}]{Luczka_2005}%
  \BibitemOpen
  \bibfield  {author} {\bibinfo {author} {\bibfnamefont {J.}~\bibnamefont {Łuczka}},\ }\bibfield  {title} {\bibinfo {title} {{Non-Markovian stochastic processes: Colored noise}},\ }\href {https://doi.org/10.1063/1.1860471} {\bibfield  {journal} {\bibinfo  {journal} {Chaos}\ }\textbf {\bibinfo {volume} {15}},\ \bibinfo {pages} {026107} (\bibinfo {year} {2005})}\BibitemShut {NoStop}%
\bibitem [{\citenamefont {Venturelli}\ \emph {et~al.}(2022)\citenamefont {Venturelli}, \citenamefont {Ferraro},\ and\ \citenamefont {Gambassi}}]{Venturelli_2022}%
  \BibitemOpen
  \bibfield  {author} {\bibinfo {author} {\bibfnamefont {D.}~\bibnamefont {Venturelli}}, \bibinfo {author} {\bibfnamefont {F.}~\bibnamefont {Ferraro}},\ and\ \bibinfo {author} {\bibfnamefont {A.}~\bibnamefont {Gambassi}},\ }\bibfield  {title} {\bibinfo {title} {{Nonequilibrium relaxation of a trapped particle in a near-critical Gaussian field}},\ }\href {https://doi.org/10.1103/PhysRevE.105.054125} {\bibfield  {journal} {\bibinfo  {journal} {Phys. Rev. E}\ }\textbf {\bibinfo {volume} {105}},\ \bibinfo {pages} {054125} (\bibinfo {year} {2022})}\BibitemShut {NoStop}%
\bibitem [{\citenamefont {Venturelli}\ and\ \citenamefont {Gambassi}(2022)}]{Venturelli_2022_2parts}%
  \BibitemOpen
  \bibfield  {author} {\bibinfo {author} {\bibfnamefont {D.}~\bibnamefont {Venturelli}}\ and\ \bibinfo {author} {\bibfnamefont {A.}~\bibnamefont {Gambassi}},\ }\bibfield  {title} {\bibinfo {title} {{Inducing oscillations of trapped particles in a near-critical Gaussian field}},\ }\href {https://doi.org/10.1103/PhysRevE.106.044112} {\bibfield  {journal} {\bibinfo  {journal} {Phys. Rev. E}\ }\textbf {\bibinfo {volume} {106}},\ \bibinfo {pages} {044112} (\bibinfo {year} {2022})}\BibitemShut {NoStop}%
\bibitem [{\citenamefont {Venturelli}\ and\ \citenamefont {Gambassi}(2023)}]{Venturelli_2023}%
  \BibitemOpen
  \bibfield  {author} {\bibinfo {author} {\bibfnamefont {D.}~\bibnamefont {Venturelli}}\ and\ \bibinfo {author} {\bibfnamefont {A.}~\bibnamefont {Gambassi}},\ }\bibfield  {title} {\bibinfo {title} {Memory-induced oscillations of a driven particle in a dissipative correlated medium},\ }\href {https://doi.org/10.1088/1367-2630/acf240} {\bibfield  {journal} {\bibinfo  {journal} {New J. Phys.}\ }\textbf {\bibinfo {volume} {25}},\ \bibinfo {pages} {093025} (\bibinfo {year} {2023})}\BibitemShut {NoStop}%
\bibitem [{\citenamefont {Venturelli}\ \emph {et~al.}(2024)\citenamefont {Venturelli}, \citenamefont {Loos}, \citenamefont {Walter}, \citenamefont {Rold{\'a}n},\ and\ \citenamefont {Gambassi}}]{venturelli2024heat}%
  \BibitemOpen
  \bibfield  {author} {\bibinfo {author} {\bibfnamefont {D.}~\bibnamefont {Venturelli}}, \bibinfo {author} {\bibfnamefont {S.~A.~M.}\ \bibnamefont {Loos}}, \bibinfo {author} {\bibfnamefont {B.}~\bibnamefont {Walter}}, \bibinfo {author} {\bibfnamefont {{\'E}.}~\bibnamefont {Rold{\'a}n}},\ and\ \bibinfo {author} {\bibfnamefont {A.}~\bibnamefont {Gambassi}},\ }\bibfield  {title} {\bibinfo {title} {Stochastic thermodynamics of a probe in a fluctuating correlated field},\ }\href {https://doi.org/10.1209/0295-5075/ad3469} {\bibfield  {journal} {\bibinfo  {journal} {EPL}\ }\textbf {\bibinfo {volume} {146}},\ \bibinfo {pages} {27001} (\bibinfo {year} {2024})}\BibitemShut {NoStop}%
\bibitem [{\citenamefont {McQuarrie}(2000)}]{mcquarrie2000statistical}%
  \BibitemOpen
  \bibfield  {author} {\bibinfo {author} {\bibfnamefont {D.}~\bibnamefont {McQuarrie}},\ }\href {https://books.google.fr/books?id=itcpPnDnJM0C} {\emph {\bibinfo {title} {Statistical Mechanics}}}\ (\bibinfo  {publisher} {University Science Books},\ \bibinfo {year} {2000})\BibitemShut {NoStop}%
\bibitem [{\citenamefont {B\'enichou}\ \emph {et~al.}(2016)\citenamefont {B\'enichou}, \citenamefont {Illien}, \citenamefont {Oshanin}, \citenamefont {Sarracino},\ and\ \citenamefont {Voituriez}}]{Benichou2016}%
  \BibitemOpen
  \bibfield  {author} {\bibinfo {author} {\bibfnamefont {O.}~\bibnamefont {B\'enichou}}, \bibinfo {author} {\bibfnamefont {P.}~\bibnamefont {Illien}}, \bibinfo {author} {\bibfnamefont {G.}~\bibnamefont {Oshanin}}, \bibinfo {author} {\bibfnamefont {A.}~\bibnamefont {Sarracino}},\ and\ \bibinfo {author} {\bibfnamefont {R.}~\bibnamefont {Voituriez}},\ }\bibfield  {title} {\bibinfo {title} {Nonlinear response and emerging nonequilibrium microstructures for biased diffusion in confined crowded environments},\ }\href {https://doi.org/10.1103/PhysRevE.93.032128} {\bibfield  {journal} {\bibinfo  {journal} {Phys. Rev. E}\ }\textbf {\bibinfo {volume} {93}},\ \bibinfo {pages} {032128} (\bibinfo {year} {2016})}\BibitemShut {NoStop}%
\bibitem [{\citenamefont {D\'emery}\ and\ \citenamefont {Dean}(2011)}]{demerypath}%
  \BibitemOpen
  \bibfield  {author} {\bibinfo {author} {\bibfnamefont {V.}~\bibnamefont {D\'emery}}\ and\ \bibinfo {author} {\bibfnamefont {D.~S.}\ \bibnamefont {Dean}},\ }\bibfield  {title} {\bibinfo {title} {Perturbative path-integral study of active- and passive-tracer diffusion in fluctuating fields},\ }\href {https://doi.org/10.1103/PhysRevE.84.011148} {\bibfield  {journal} {\bibinfo  {journal} {Phys. Rev. E}\ }\textbf {\bibinfo {volume} {84}},\ \bibinfo {pages} {011148} (\bibinfo {year} {2011})}\BibitemShut {NoStop}%
\bibitem [{\citenamefont {Venturelli}\ and\ \citenamefont {Gross}(2022)}]{Venturelli_2022_confined}%
  \BibitemOpen
  \bibfield  {author} {\bibinfo {author} {\bibfnamefont {D.}~\bibnamefont {Venturelli}}\ and\ \bibinfo {author} {\bibfnamefont {M.}~\bibnamefont {Gross}},\ }\bibfield  {title} {\bibinfo {title} {Tracer particle in a confined correlated medium: an adiabatic elimination method},\ }\href {https://doi.org/10.1088/1742-5468/aca8fa} {\bibfield  {journal} {\bibinfo  {journal} {J. Stat. Mech.}\ }\textbf {\bibinfo {volume} {2022}},\ \bibinfo {pages} {123210} (\bibinfo {year} {2022})}\BibitemShut {NoStop}%
\bibitem [{\citenamefont {Hughes}(1995)}]{hughes1995random}%
  \BibitemOpen
  \bibfield  {author} {\bibinfo {author} {\bibfnamefont {B.}~\bibnamefont {Hughes}},\ }\href {https://books.google.fr/books?id=QhOen_t0LeQC} {\emph {\bibinfo {title} {Random Walks and Random Environments}}},\ \bibinfo {series} {{Oxford Science Publications}}, Vol.~\bibinfo {volume} {1}\ (\bibinfo  {publisher} {Clarendon Press},\ \bibinfo {year} {1995})\BibitemShut {NoStop}%
\bibitem [{\citenamefont {Poncet}\ \emph {et~al.}(2021)\citenamefont {Poncet}, \citenamefont {Grabsch}, \citenamefont {Illien},\ and\ \citenamefont {B\'enichou}}]{Poncet_2021}%
  \BibitemOpen
  \bibfield  {author} {\bibinfo {author} {\bibfnamefont {A.}~\bibnamefont {Poncet}}, \bibinfo {author} {\bibfnamefont {A.}~\bibnamefont {Grabsch}}, \bibinfo {author} {\bibfnamefont {P.}~\bibnamefont {Illien}},\ and\ \bibinfo {author} {\bibfnamefont {O.}~\bibnamefont {B\'enichou}},\ }\bibfield  {title} {\bibinfo {title} {Generalized correlation profiles in single-file systems},\ }\href {https://doi.org/10.1103/PhysRevLett.127.220601} {\bibfield  {journal} {\bibinfo  {journal} {Phys. Rev. Lett.}\ }\textbf {\bibinfo {volume} {127}},\ \bibinfo {pages} {220601} (\bibinfo {year} {2021})}\BibitemShut {NoStop}%
\bibitem [{\citenamefont {Gradshteyn}\ and\ \citenamefont {Ryzhik}(2007)}]{table}%
  \BibitemOpen
  \bibfield  {author} {\bibinfo {author} {\bibfnamefont {I.~S.}\ \bibnamefont {Gradshteyn}}\ and\ \bibinfo {author} {\bibfnamefont {I.~M.}\ \bibnamefont {Ryzhik}},\ }\href {https://www.sciencedirect.com/book/9780123736376/table-of-integrals-series-and-products} {\emph {\bibinfo {title} {Table of integrals, series, and products}}},\ \bibinfo {edition} {7th}\ ed.\ (\bibinfo  {publisher} {Elsevier/Academic Press, Amsterdam},\ \bibinfo {year} {2007})\BibitemShut {NoStop}%
\bibitem [{\citenamefont {Täuber}(2014)}]{Tauber}%
  \BibitemOpen
  \bibfield  {author} {\bibinfo {author} {\bibfnamefont {U.~C.}\ \bibnamefont {Täuber}},\ }\href {https://doi.org/10.1017/CBO9781139046213} {\emph {\bibinfo {title} {Critical Dynamics}}}\ (\bibinfo  {publisher} {Cambridge University Press},\ \bibinfo {year} {2014})\BibitemShut {NoStop}%
\bibitem [{\citenamefont {B\'enichou}\ \emph {et~al.}(2000)\citenamefont {B\'enichou}, \citenamefont {Cazabat}, \citenamefont {De~Coninck}, \citenamefont {Moreau},\ and\ \citenamefont {Oshanin}}]{Benichou_2000}%
  \BibitemOpen
  \bibfield  {author} {\bibinfo {author} {\bibfnamefont {O.}~\bibnamefont {B\'enichou}}, \bibinfo {author} {\bibfnamefont {A.~M.}\ \bibnamefont {Cazabat}}, \bibinfo {author} {\bibfnamefont {J.}~\bibnamefont {De~Coninck}}, \bibinfo {author} {\bibfnamefont {M.}~\bibnamefont {Moreau}},\ and\ \bibinfo {author} {\bibfnamefont {G.}~\bibnamefont {Oshanin}},\ }\bibfield  {title} {\bibinfo {title} {Stokes formula and density perturbances for driven tracer diffusion in an adsorbed monolayer},\ }\href {https://doi.org/10.1103/PhysRevLett.84.511} {\bibfield  {journal} {\bibinfo  {journal} {Phys. Rev. Lett.}\ }\textbf {\bibinfo {volume} {84}},\ \bibinfo {pages} {511} (\bibinfo {year} {2000})}\BibitemShut {NoStop}%
\bibitem [{\citenamefont {Basu}\ \emph {et~al.}(2022)\citenamefont {Basu}, \citenamefont {Démery},\ and\ \citenamefont {Gambassi}}]{wellGauss}%
  \BibitemOpen
  \bibfield  {author} {\bibinfo {author} {\bibfnamefont {U.}~\bibnamefont {Basu}}, \bibinfo {author} {\bibfnamefont {V.}~\bibnamefont {Démery}},\ and\ \bibinfo {author} {\bibfnamefont {A.}~\bibnamefont {Gambassi}},\ }\bibfield  {title} {\bibinfo {title} {{Dynamics of a colloidal particle coupled to a Gaussian field: from a confinement-dependent to a non-linear memory}},\ }\href {https://doi.org/10.21468/SciPostPhys.13.4.078} {\bibfield  {journal} {\bibinfo  {journal} {SciPost Phys.}\ }\textbf {\bibinfo {volume} {13}},\ \bibinfo {pages} {078} (\bibinfo {year} {2022})}\BibitemShut {NoStop}%
\bibitem [{\citenamefont {Thompson}\ \emph {et~al.}(2022)\citenamefont {Thompson}, \citenamefont {Aktulga}, \citenamefont {Berger}, \citenamefont {Bolintineanu}, \citenamefont {Brown}, \citenamefont {Crozier}, \citenamefont {in~'t Veld}, \citenamefont {Kohlmeyer}, \citenamefont {Moore}, \citenamefont {Nguyen}, \citenamefont {Shan}, \citenamefont {Stevens}, \citenamefont {Tranchida}, \citenamefont {Trott},\ and\ \citenamefont {Plimpton}}]{Thompson2022}%
  \BibitemOpen
  \bibfield  {author} {\bibinfo {author} {\bibfnamefont {A.~P.}\ \bibnamefont {Thompson}}, \bibinfo {author} {\bibfnamefont {H.~M.}\ \bibnamefont {Aktulga}}, \bibinfo {author} {\bibfnamefont {R.}~\bibnamefont {Berger}}, \bibinfo {author} {\bibfnamefont {D.~S.}\ \bibnamefont {Bolintineanu}}, \bibinfo {author} {\bibfnamefont {W.~M.}\ \bibnamefont {Brown}}, \bibinfo {author} {\bibfnamefont {P.~S.}\ \bibnamefont {Crozier}}, \bibinfo {author} {\bibfnamefont {P.~J.}\ \bibnamefont {in~'t Veld}}, \bibinfo {author} {\bibfnamefont {A.}~\bibnamefont {Kohlmeyer}}, \bibinfo {author} {\bibfnamefont {S.~G.}\ \bibnamefont {Moore}}, \bibinfo {author} {\bibfnamefont {T.~D.}\ \bibnamefont {Nguyen}}, \bibinfo {author} {\bibfnamefont {R.}~\bibnamefont {Shan}}, \bibinfo {author} {\bibfnamefont {M.~J.}\ \bibnamefont {Stevens}}, \bibinfo {author} {\bibfnamefont {J.}~\bibnamefont {Tranchida}}, \bibinfo {author} {\bibfnamefont {C.}~\bibnamefont {Trott}},\ and\ \bibinfo {author} {\bibfnamefont {S.~J.}\ \bibnamefont {Plimpton}},\
  }\bibfield  {title} {\bibinfo {title} {{LAMMPS} - a flexible simulation tool for particle-based materials modeling at the atomic, meso, and continuum scales},\ }\href {https://doi.org/10.1016/j.cpc.2021.108171} {\bibfield  {journal} {\bibinfo  {journal} {Comput. Phys. Commun.}\ }\textbf {\bibinfo {volume} {271}},\ \bibinfo {pages} {108171} (\bibinfo {year} {2022})}\BibitemShut {NoStop}%
\end{thebibliography}%


\begin{thebibliography}{27}%
\makeatletter
\providecommand \@ifxundefined [1]{%
 \@ifx{#1\undefined}
}%
\providecommand \@ifnum [1]{%
 \ifnum #1\expandafter \@firstoftwo
 \else \expandafter \@secondoftwo
 \fi
}%
\providecommand \@ifx [1]{%
 \ifx #1\expandafter \@firstoftwo
 \else \expandafter \@secondoftwo
 \fi
}%
\providecommand \natexlab [1]{#1}%
\providecommand \enquote  [1]{``#1''}%
\providecommand \bibnamefont  [1]{#1}%
\providecommand \bibfnamefont [1]{#1}%
\providecommand \citenamefont [1]{#1}%
\providecommand \href@noop [0]{\@secondoftwo}%
\providecommand \href [0]{\begingroup \@sanitize@url \@href}%
\providecommand \@href[1]{\@@startlink{#1}\@@href}%
\providecommand \@@href[1]{\endgroup#1\@@endlink}%
\providecommand \@sanitize@url [0]{\catcode `\\12\catcode `\$12\catcode `\&12\catcode `\#12\catcode `\^12\catcode `\_12\catcode `\%12\relax}%
\providecommand \@@startlink[1]{}%
\providecommand \@@endlink[0]{}%
\providecommand \url  [0]{\begingroup\@sanitize@url \@url }%
\providecommand \@url [1]{\endgroup\@href {#1}{\urlprefix }}%
\providecommand \urlprefix  [0]{URL }%
\providecommand \Eprint [0]{\href }%
\providecommand \doibase [0]{https://doi.org/}%
\providecommand \selectlanguage [0]{\@gobble}%
\providecommand \bibinfo  [0]{\@secondoftwo}%
\providecommand \bibfield  [0]{\@secondoftwo}%
\providecommand \translation [1]{[#1]}%
\providecommand \BibitemOpen [0]{}%
\providecommand \bibitemStop [0]{}%
\providecommand \bibitemNoStop [0]{.\EOS\space}%
\providecommand \EOS [0]{\spacefactor3000\relax}%
\providecommand \BibitemShut  [1]{\csname bibitem#1\endcsname}%
\let\auto@bib@innerbib\@empty
%</preamble>
\bibitem [{\citenamefont {Illien}\ \emph {et~al.}(2015)\citenamefont {Illien}, \citenamefont {Bénichou}, \citenamefont {Oshanin},\ and\ \citenamefont {Voituriez}}]{Illien2015_distribution}%
  \BibitemOpen
  \bibfield  {author} {\bibinfo {author} {\bibfnamefont {P.}~\bibnamefont {Illien}}, \bibinfo {author} {\bibfnamefont {O.}~\bibnamefont {Bénichou}}, \bibinfo {author} {\bibfnamefont {G.}~\bibnamefont {Oshanin}},\ and\ \bibinfo {author} {\bibfnamefont {R.}~\bibnamefont {Voituriez}},\ }\bibfield  {title} {\bibinfo {title} {Distribution of the position of a driven tracer in a hardcore lattice gas},\ }\href {https://doi.org/10.1088/1742-5468/2015/11/p11016} {\bibfield  {journal} {\bibinfo  {journal} {J. Stat. Mech.}\ }\textbf {\bibinfo {volume} {2015}},\ \bibinfo {pages} {P11016} (\bibinfo {year} {2015})}\BibitemShut {NoStop}%
\bibitem [{\citenamefont {Illien}\ \emph {et~al.}(2018)\citenamefont {Illien}, \citenamefont {B\'enichou}, \citenamefont {Oshanin}, \citenamefont {Sarracino},\ and\ \citenamefont {Voituriez}}]{Illien2018}%
  \BibitemOpen
  \bibfield  {author} {\bibinfo {author} {\bibfnamefont {P.}~\bibnamefont {Illien}}, \bibinfo {author} {\bibfnamefont {O.}~\bibnamefont {B\'enichou}}, \bibinfo {author} {\bibfnamefont {G.}~\bibnamefont {Oshanin}}, \bibinfo {author} {\bibfnamefont {A.}~\bibnamefont {Sarracino}},\ and\ \bibinfo {author} {\bibfnamefont {R.}~\bibnamefont {Voituriez}},\ }\bibfield  {title} {\bibinfo {title} {Nonequilibrium fluctuations and enhanced diffusion of a driven particle in a dense environment},\ }\href {https://doi.org/10.1103/PhysRevLett.120.200606} {\bibfield  {journal} {\bibinfo  {journal} {Phys. Rev. Lett.}\ }\textbf {\bibinfo {volume} {120}},\ \bibinfo {pages} {200606} (\bibinfo {year} {2018})}\BibitemShut {NoStop}%
\bibitem [{Note1()}]{Note1}%
  \BibitemOpen
  \bibinfo {note} {Unfortunately, the definition of $A_\mu $ given in Eq.~(8) of the Main Text of~\cite {Illien2018} differs from the one actually used in the calculation presented in the Supplemental Material. Here we adopt the latter for consistency.}\BibitemShut {Stop}%
\bibitem [{\citenamefont {Hughes}(1995)}]{hughes1995random}%
  \BibitemOpen
  \bibfield  {author} {\bibinfo {author} {\bibfnamefont {B.}~\bibnamefont {Hughes}},\ }\href {https://books.google.fr/books?id=QhOen_t0LeQC} {\emph {\bibinfo {title} {Random Walks and Random Environments}}},\ \bibinfo {series} {{Oxford Science Publications}}, Vol.~\bibinfo {volume} {1}\ (\bibinfo  {publisher} {Clarendon Press},\ \bibinfo {year} {1995})\BibitemShut {NoStop}%
\bibitem [{\citenamefont {Démery}\ \emph {et~al.}(2014)\citenamefont {Démery}, \citenamefont {Bénichou},\ and\ \citenamefont {Jacquin}}]{Demery2014}%
  \BibitemOpen
  \bibfield  {author} {\bibinfo {author} {\bibfnamefont {V.}~\bibnamefont {Démery}}, \bibinfo {author} {\bibfnamefont {O.}~\bibnamefont {Bénichou}},\ and\ \bibinfo {author} {\bibfnamefont {H.}~\bibnamefont {Jacquin}},\ }\bibfield  {title} {\bibinfo {title} {Generalized {Langevin} equations for a driven tracer in dense soft colloids: construction and applications},\ }\href {https://doi.org/10.1088/1367-2630/16/5/053032} {\bibfield  {journal} {\bibinfo  {journal} {New J. Phys.}\ }\textbf {\bibinfo {volume} {16}},\ \bibinfo {pages} {053032} (\bibinfo {year} {2014})}\BibitemShut {NoStop}%
\bibitem [{\citenamefont {Poncet}\ \emph {et~al.}(2021)\citenamefont {Poncet}, \citenamefont {Grabsch}, \citenamefont {Illien},\ and\ \citenamefont {B\'enichou}}]{Poncet_2021}%
  \BibitemOpen
  \bibfield  {author} {\bibinfo {author} {\bibfnamefont {A.}~\bibnamefont {Poncet}}, \bibinfo {author} {\bibfnamefont {A.}~\bibnamefont {Grabsch}}, \bibinfo {author} {\bibfnamefont {P.}~\bibnamefont {Illien}},\ and\ \bibinfo {author} {\bibfnamefont {O.}~\bibnamefont {B\'enichou}},\ }\bibfield  {title} {\bibinfo {title} {Generalized correlation profiles in single-file systems},\ }\href {https://doi.org/10.1103/PhysRevLett.127.220601} {\bibfield  {journal} {\bibinfo  {journal} {Phys. Rev. Lett.}\ }\textbf {\bibinfo {volume} {127}},\ \bibinfo {pages} {220601} (\bibinfo {year} {2021})}\BibitemShut {NoStop}%
\bibitem [{\citenamefont {Grabsch}\ \emph {et~al.}(2022)\citenamefont {Grabsch}, \citenamefont {Poncet}, \citenamefont {Rizkallah}, \citenamefont {Illien},\ and\ \citenamefont {B{\'e}nichou}}]{Grabsch2022}%
  \BibitemOpen
  \bibfield  {author} {\bibinfo {author} {\bibfnamefont {A.}~\bibnamefont {Grabsch}}, \bibinfo {author} {\bibfnamefont {A.}~\bibnamefont {Poncet}}, \bibinfo {author} {\bibfnamefont {P.}~\bibnamefont {Rizkallah}}, \bibinfo {author} {\bibfnamefont {P.}~\bibnamefont {Illien}},\ and\ \bibinfo {author} {\bibfnamefont {O.}~\bibnamefont {B{\'e}nichou}},\ }\bibfield  {title} {\bibinfo {title} {Exact closure and solution for spatial correlations in single-file diffusion},\ }\href {https://doi.org/10.1126/sciadv.abm5043} {\bibfield  {journal} {\bibinfo  {journal} {Sci. Adv.}\ }\textbf {\bibinfo {volume} {8}},\ \bibinfo {pages} {eabm5043} (\bibinfo {year} {2022})}\BibitemShut {NoStop}%
\bibitem [{\citenamefont {Gradshteyn}\ and\ \citenamefont {Ryzhik}(2007)}]{table}%
  \BibitemOpen
  \bibfield  {author} {\bibinfo {author} {\bibfnamefont {I.~S.}\ \bibnamefont {Gradshteyn}}\ and\ \bibinfo {author} {\bibfnamefont {I.~M.}\ \bibnamefont {Ryzhik}},\ }\href {https://www.sciencedirect.com/book/9780123736376/table-of-integrals-series-and-products} {\emph {\bibinfo {title} {Table of integrals, series, and products}}},\ \bibinfo {edition} {7th}\ ed.\ (\bibinfo  {publisher} {Elsevier/Academic Press, Amsterdam},\ \bibinfo {year} {2007})\BibitemShut {NoStop}%
\bibitem [{\citenamefont {Dean}(1996)}]{Dean1996}%
  \BibitemOpen
  \bibfield  {author} {\bibinfo {author} {\bibfnamefont {D.~S.}\ \bibnamefont {Dean}},\ }\bibfield  {title} {\bibinfo {title} {Langevin equation for the density of a system of interacting {Langevin} processes},\ }\href {https://doi.org/10.1088/0305-4470/29/24/001} {\bibfield  {journal} {\bibinfo  {journal} {J. Phys. A: Math. Gen.}\ }\textbf {\bibinfo {volume} {29}},\ \bibinfo {pages} {L613–L617} (\bibinfo {year} {1996})}\BibitemShut {NoStop}%
\bibitem [{\citenamefont {Hansen}\ and\ \citenamefont {McDonald}(2013)}]{McDonald_book}%
  \BibitemOpen
  \bibfield  {author} {\bibinfo {author} {\bibfnamefont {J.-P.}\ \bibnamefont {Hansen}}\ and\ \bibinfo {author} {\bibfnamefont {I.~R.}\ \bibnamefont {McDonald}},\ }\href {https://doi.org/10.1016/c2010-0-66723-x} {\emph {\bibinfo {title} {Theory of Simple Liquids}}}\ (\bibinfo  {publisher} {Elsevier},\ \bibinfo {year} {2013})\BibitemShut {NoStop}%
\bibitem [{\citenamefont {McQuarrie}(2000)}]{mcquarrie2000statistical}%
  \BibitemOpen
  \bibfield  {author} {\bibinfo {author} {\bibfnamefont {D.}~\bibnamefont {McQuarrie}},\ }\href {https://books.google.fr/books?id=itcpPnDnJM0C} {\emph {\bibinfo {title} {Statistical Mechanics}}}\ (\bibinfo  {publisher} {University Science Books},\ \bibinfo {year} {2000})\BibitemShut {NoStop}%
\bibitem [{\citenamefont {Venturelli}\ \emph {et~al.}(2022)\citenamefont {Venturelli}, \citenamefont {Ferraro},\ and\ \citenamefont {Gambassi}}]{Venturelli_2022}%
  \BibitemOpen
  \bibfield  {author} {\bibinfo {author} {\bibfnamefont {D.}~\bibnamefont {Venturelli}}, \bibinfo {author} {\bibfnamefont {F.}~\bibnamefont {Ferraro}},\ and\ \bibinfo {author} {\bibfnamefont {A.}~\bibnamefont {Gambassi}},\ }\bibfield  {title} {\bibinfo {title} {{Nonequilibrium relaxation of a trapped particle in a near-critical Gaussian field}},\ }\href {https://doi.org/10.1103/PhysRevE.105.054125} {\bibfield  {journal} {\bibinfo  {journal} {Phys. Rev. E}\ }\textbf {\bibinfo {volume} {105}},\ \bibinfo {pages} {054125} (\bibinfo {year} {2022})}\BibitemShut {NoStop}%
\bibitem [{\citenamefont {Evans}\ \emph {et~al.}(1993)\citenamefont {Evans}, \citenamefont {Henderson}, \citenamefont {Hoyle}, \citenamefont {Parry},\ and\ \citenamefont {Sabeur}}]{Evans1993}%
  \BibitemOpen
  \bibfield  {author} {\bibinfo {author} {\bibfnamefont {R.}~\bibnamefont {Evans}}, \bibinfo {author} {\bibfnamefont {J.}~\bibnamefont {Henderson}}, \bibinfo {author} {\bibfnamefont {D.}~\bibnamefont {Hoyle}}, \bibinfo {author} {\bibfnamefont {A.}~\bibnamefont {Parry}},\ and\ \bibinfo {author} {\bibfnamefont {Z.}~\bibnamefont {Sabeur}},\ }\bibfield  {title} {\bibinfo {title} {Asymptotic decay of liquid structure: oscillatory liquid-vapour density profiles and the {Fisher-Widom} line},\ }\href {https://doi.org/10.1080/00268979300102621} {\bibfield  {journal} {\bibinfo  {journal} {Mol. Phys.}\ }\textbf {\bibinfo {volume} {80}},\ \bibinfo {pages} {755–775} (\bibinfo {year} {1993})}\BibitemShut {NoStop}%
\bibitem [{\citenamefont {Evans}\ \emph {et~al.}(1994)\citenamefont {Evans}, \citenamefont {Leote~de Carvalho}, \citenamefont {Henderson},\ and\ \citenamefont {Hoyle}}]{Evans1994}%
  \BibitemOpen
  \bibfield  {author} {\bibinfo {author} {\bibfnamefont {R.}~\bibnamefont {Evans}}, \bibinfo {author} {\bibfnamefont {R.~J.~F.}\ \bibnamefont {Leote~de Carvalho}}, \bibinfo {author} {\bibfnamefont {J.~R.}\ \bibnamefont {Henderson}},\ and\ \bibinfo {author} {\bibfnamefont {D.~C.}\ \bibnamefont {Hoyle}},\ }\bibfield  {title} {\bibinfo {title} {Asymptotic decay of correlations in liquids and their mixtures},\ }\href {https://doi.org/10.1063/1.466920} {\bibfield  {journal} {\bibinfo  {journal} {J. Chem. Phys.}\ }\textbf {\bibinfo {volume} {100}},\ \bibinfo {pages} {591–603} (\bibinfo {year} {1994})}\BibitemShut {NoStop}%
\bibitem [{\citenamefont {Leote~de Carvalho}\ and\ \citenamefont {Evans}(1994)}]{LeotedeCarvalho1994}%
  \BibitemOpen
  \bibfield  {author} {\bibinfo {author} {\bibfnamefont {R.}~\bibnamefont {Leote~de Carvalho}}\ and\ \bibinfo {author} {\bibfnamefont {R.}~\bibnamefont {Evans}},\ }\bibfield  {title} {\bibinfo {title} {The decay of correlations in ionic fluids},\ }\href {https://doi.org/10.1080/00268979400101491} {\bibfield  {journal} {\bibinfo  {journal} {Mol. Phys.}\ }\textbf {\bibinfo {volume} {83}},\ \bibinfo {pages} {619–654} (\bibinfo {year} {1994})}\BibitemShut {NoStop}%
\bibitem [{\citenamefont {Dijkstra}\ and\ \citenamefont {Evans}(2000)}]{Dijkstra2000}%
  \BibitemOpen
  \bibfield  {author} {\bibinfo {author} {\bibfnamefont {M.}~\bibnamefont {Dijkstra}}\ and\ \bibinfo {author} {\bibfnamefont {R.}~\bibnamefont {Evans}},\ }\bibfield  {title} {\bibinfo {title} {A simulation study of the decay of the pair correlation function in simple fluids},\ }\href {https://doi.org/10.1063/1.480598} {\bibfield  {journal} {\bibinfo  {journal} {J. Chem. Phys.}\ }\textbf {\bibinfo {volume} {112}},\ \bibinfo {pages} {1449–1456} (\bibinfo {year} {2000})}\BibitemShut {NoStop}%
\bibitem [{\citenamefont {Hopkins}\ \emph {et~al.}(2005)\citenamefont {Hopkins}, \citenamefont {Archer},\ and\ \citenamefont {Evans}}]{Hopkins2005}%
  \BibitemOpen
  \bibfield  {author} {\bibinfo {author} {\bibfnamefont {P.}~\bibnamefont {Hopkins}}, \bibinfo {author} {\bibfnamefont {A.~J.}\ \bibnamefont {Archer}},\ and\ \bibinfo {author} {\bibfnamefont {R.}~\bibnamefont {Evans}},\ }\bibfield  {title} {\bibinfo {title} {Asymptotic decay of pair correlations in a {Yukawa} fluid},\ }\href {https://doi.org/10.1103/PhysRevE.71.027401} {\bibfield  {journal} {\bibinfo  {journal} {Phys. Rev. E}\ }\textbf {\bibinfo {volume} {71}},\ \bibinfo {pages} {027401} (\bibinfo {year} {2005})}\BibitemShut {NoStop}%
\bibitem [{\citenamefont {Stopper}\ \emph {et~al.}(2019)\citenamefont {Stopper}, \citenamefont {Hansen-Goos}, \citenamefont {Roth},\ and\ \citenamefont {Evans}}]{Stopper2019}%
  \BibitemOpen
  \bibfield  {author} {\bibinfo {author} {\bibfnamefont {D.}~\bibnamefont {Stopper}}, \bibinfo {author} {\bibfnamefont {H.}~\bibnamefont {Hansen-Goos}}, \bibinfo {author} {\bibfnamefont {R.}~\bibnamefont {Roth}},\ and\ \bibinfo {author} {\bibfnamefont {R.}~\bibnamefont {Evans}},\ }\bibfield  {title} {\bibinfo {title} {On the decay of the pair correlation function and the line of vanishing excess isothermal compressibility in simple fluids},\ }\href {http://dx.doi.org/10.1063/1.5110044} {\bibfield  {journal} {\bibinfo  {journal} {J. Chem. Phys.}\ }\textbf {\bibinfo {volume} {151}},\ \bibinfo {pages} {014501} (\bibinfo {year} {2019})}\BibitemShut {NoStop}%
\bibitem [{\citenamefont {Täuber}(2014)}]{Tauber}%
  \BibitemOpen
  \bibfield  {author} {\bibinfo {author} {\bibfnamefont {U.~C.}\ \bibnamefont {Täuber}},\ }\href {https://doi.org/10.1017/CBO9781139046213} {\emph {\bibinfo {title} {Critical Dynamics}}}\ (\bibinfo  {publisher} {Cambridge University Press},\ \bibinfo {year} {2014})\BibitemShut {NoStop}%
\bibitem [{\citenamefont {D\'emery}\ and\ \citenamefont {Dean}(2011)}]{demerypath}%
  \BibitemOpen
  \bibfield  {author} {\bibinfo {author} {\bibfnamefont {V.}~\bibnamefont {D\'emery}}\ and\ \bibinfo {author} {\bibfnamefont {D.~S.}\ \bibnamefont {Dean}},\ }\bibfield  {title} {\bibinfo {title} {Perturbative path-integral study of active- and passive-tracer diffusion in fluctuating fields},\ }\href {https://doi.org/10.1103/PhysRevE.84.011148} {\bibfield  {journal} {\bibinfo  {journal} {Phys. Rev. E}\ }\textbf {\bibinfo {volume} {84}},\ \bibinfo {pages} {011148} (\bibinfo {year} {2011})}\BibitemShut {NoStop}%
\bibitem [{\citenamefont {Novikov}(1965)}]{Novikov_1965}%
  \BibitemOpen
  \bibfield  {author} {\bibinfo {author} {\bibfnamefont {E.~A.}\ \bibnamefont {Novikov}},\ }\bibfield  {title} {\bibinfo {title} {Functionals and the random-force method in turbulence theory},\ }\href {http://www.jetp.ras.ru/cgi-bin/dn/e_020_05_1290.pdf} {\bibfield  {journal} {\bibinfo  {journal} {Sov. Phys. JETP}\ }\textbf {\bibinfo {volume} {20}},\ \bibinfo {pages} {1290} (\bibinfo {year} {1965})}\BibitemShut {NoStop}%
\bibitem [{\citenamefont {Venturelli}\ and\ \citenamefont {Gambassi}(2022)}]{Venturelli_2022_2parts}%
  \BibitemOpen
  \bibfield  {author} {\bibinfo {author} {\bibfnamefont {D.}~\bibnamefont {Venturelli}}\ and\ \bibinfo {author} {\bibfnamefont {A.}~\bibnamefont {Gambassi}},\ }\bibfield  {title} {\bibinfo {title} {{Inducing oscillations of trapped particles in a near-critical Gaussian field}},\ }\href {https://doi.org/10.1103/PhysRevE.106.044112} {\bibfield  {journal} {\bibinfo  {journal} {Phys. Rev. E}\ }\textbf {\bibinfo {volume} {106}},\ \bibinfo {pages} {044112} (\bibinfo {year} {2022})}\BibitemShut {NoStop}%
\bibitem [{\citenamefont {Venturelli}\ and\ \citenamefont {Gambassi}(2023)}]{Venturelli_2023}%
  \BibitemOpen
  \bibfield  {author} {\bibinfo {author} {\bibfnamefont {D.}~\bibnamefont {Venturelli}}\ and\ \bibinfo {author} {\bibfnamefont {A.}~\bibnamefont {Gambassi}},\ }\bibfield  {title} {\bibinfo {title} {Memory-induced oscillations of a driven particle in a dissipative correlated medium},\ }\href {https://doi.org/10.1088/1367-2630/acf240} {\bibfield  {journal} {\bibinfo  {journal} {New J. Phys.}\ }\textbf {\bibinfo {volume} {25}},\ \bibinfo {pages} {093025} (\bibinfo {year} {2023})}\BibitemShut {NoStop}%
\bibitem [{\citenamefont {Łuczka}(2005)}]{Luczka_2005}%
  \BibitemOpen
  \bibfield  {author} {\bibinfo {author} {\bibfnamefont {J.}~\bibnamefont {Łuczka}},\ }\bibfield  {title} {\bibinfo {title} {{Non-Markovian stochastic processes: Colored noise}},\ }\href {https://doi.org/10.1063/1.1860471} {\bibfield  {journal} {\bibinfo  {journal} {Chaos}\ }\textbf {\bibinfo {volume} {15}},\ \bibinfo {pages} {026107} (\bibinfo {year} {2005})}\BibitemShut {NoStop}%
\bibitem [{\citenamefont {B\'enichou}\ \emph {et~al.}(2000)\citenamefont {B\'enichou}, \citenamefont {Cazabat}, \citenamefont {De~Coninck}, \citenamefont {Moreau},\ and\ \citenamefont {Oshanin}}]{Benichou_2000}%
  \BibitemOpen
  \bibfield  {author} {\bibinfo {author} {\bibfnamefont {O.}~\bibnamefont {B\'enichou}}, \bibinfo {author} {\bibfnamefont {A.~M.}\ \bibnamefont {Cazabat}}, \bibinfo {author} {\bibfnamefont {J.}~\bibnamefont {De~Coninck}}, \bibinfo {author} {\bibfnamefont {M.}~\bibnamefont {Moreau}},\ and\ \bibinfo {author} {\bibfnamefont {G.}~\bibnamefont {Oshanin}},\ }\bibfield  {title} {\bibinfo {title} {Stokes formula and density perturbances for driven tracer diffusion in an adsorbed monolayer},\ }\href {https://doi.org/10.1103/PhysRevLett.84.511} {\bibfield  {journal} {\bibinfo  {journal} {Phys. Rev. Lett.}\ }\textbf {\bibinfo {volume} {84}},\ \bibinfo {pages} {511} (\bibinfo {year} {2000})}\BibitemShut {NoStop}%
\bibitem [{\citenamefont {Basu}\ \emph {et~al.}(2022)\citenamefont {Basu}, \citenamefont {Démery},\ and\ \citenamefont {Gambassi}}]{wellGauss}%
  \BibitemOpen
  \bibfield  {author} {\bibinfo {author} {\bibfnamefont {U.}~\bibnamefont {Basu}}, \bibinfo {author} {\bibfnamefont {V.}~\bibnamefont {Démery}},\ and\ \bibinfo {author} {\bibfnamefont {A.}~\bibnamefont {Gambassi}},\ }\bibfield  {title} {\bibinfo {title} {{Dynamics of a colloidal particle coupled to a Gaussian field: from a confinement-dependent to a non-linear memory}},\ }\href {https://doi.org/10.21468/SciPostPhys.13.4.078} {\bibfield  {journal} {\bibinfo  {journal} {SciPost Phys.}\ }\textbf {\bibinfo {volume} {13}},\ \bibinfo {pages} {078} (\bibinfo {year} {2022})}\BibitemShut {NoStop}%
\bibitem [{\citenamefont {Thompson}\ \emph {et~al.}(2022)\citenamefont {Thompson}, \citenamefont {Aktulga}, \citenamefont {Berger}, \citenamefont {Bolintineanu}, \citenamefont {Brown}, \citenamefont {Crozier}, \citenamefont {in~'t Veld}, \citenamefont {Kohlmeyer}, \citenamefont {Moore}, \citenamefont {Nguyen}, \citenamefont {Shan}, \citenamefont {Stevens}, \citenamefont {Tranchida}, \citenamefont {Trott},\ and\ \citenamefont {Plimpton}}]{Thompson2022}%
  \BibitemOpen
  \bibfield  {author} {\bibinfo {author} {\bibfnamefont {A.~P.}\ \bibnamefont {Thompson}}, \bibinfo {author} {\bibfnamefont {H.~M.}\ \bibnamefont {Aktulga}}, \bibinfo {author} {\bibfnamefont {R.}~\bibnamefont {Berger}}, \bibinfo {author} {\bibfnamefont {D.~S.}\ \bibnamefont {Bolintineanu}}, \bibinfo {author} {\bibfnamefont {W.~M.}\ \bibnamefont {Brown}}, \bibinfo {author} {\bibfnamefont {P.~S.}\ \bibnamefont {Crozier}}, \bibinfo {author} {\bibfnamefont {P.~J.}\ \bibnamefont {in~'t Veld}}, \bibinfo {author} {\bibfnamefont {A.}~\bibnamefont {Kohlmeyer}}, \bibinfo {author} {\bibfnamefont {S.~G.}\ \bibnamefont {Moore}}, \bibinfo {author} {\bibfnamefont {T.~D.}\ \bibnamefont {Nguyen}}, \bibinfo {author} {\bibfnamefont {R.}~\bibnamefont {Shan}}, \bibinfo {author} {\bibfnamefont {M.~J.}\ \bibnamefont {Stevens}}, \bibinfo {author} {\bibfnamefont {J.}~\bibnamefont {Tranchida}}, \bibinfo {author} {\bibfnamefont {C.}~\bibnamefont {Trott}},\ and\ \bibinfo {author} {\bibfnamefont {S.~J.}\ \bibnamefont {Plimpton}},\
  }\bibfield  {title} {\bibinfo {title} {{LAMMPS} - a flexible simulation tool for particle-based materials modeling at the atomic, meso, and continuum scales},\ }\href {https://doi.org/10.1016/j.cpc.2021.108171} {\bibfield  {journal} {\bibinfo  {journal} {Comput. Phys. Commun.}\ }\textbf {\bibinfo {volume} {271}},\ \bibinfo {pages} {108171} (\bibinfo {year} {2022})}\BibitemShut {NoStop}%
\end{thebibliography}%

% ------- SUPPLEMENTAL (for arXiv) ------
% \foreach \x in {1,...,21} % 21 is the number of pages in SM.pdf
% {
%     \clearpage
%     \includepdf[pages={\x,{}}]{\supplementfilename}
% }
% --- END ---

% If you instead include supplemental.tex on arXiv, then hyperref stops working (both within the main and the SM).
% This problem is unsolvable unless all the figures are in vector format (.eps), in which case the document gets automatically compiled with latex instead of pdflatex.
% In this latter case you don't need to do anything in particular, apart from loading on arXiv both the main.tex and supplemental.tex. The server will take care of everything automatically.

\end{document}

\title{Supplemental Material\\ for ``Universal scale-free decay of tracer-bath correlations in $d$-dimensional interacting particle systems''}

\author{Davide Venturelli}
\affiliation{Sorbonne Universit\'e, CNRS, Laboratoire de Physique Th\'eorique de la Mati\`ere Condens\'ee (LPTMC), 4 Place Jussieu, 75005 Paris, France}

\author{Pierre Illien}
\affiliation{Sorbonne Universit\'e, CNRS, Laboratoire PHENIX (Physico-Chimie des Electrolytes et Nanosyst\`emes Interfaciaux), 4 Place Jussieu, 75005 Paris, France}

\author{Aur\'elien Grabsch}
\affiliation{Sorbonne Universit\'e, CNRS, Laboratoire de Physique Th\'eorique de la Mati\`ere Condens\'ee (LPTMC), 4 Place Jussieu, 75005 Paris, France}

\author{Olivier B\'enichou}
\affiliation{Sorbonne Universit\'e, CNRS, Laboratoire de Physique Th\'eorique de la Mati\`ere Condens\'ee (LPTMC), 4 Place Jussieu, 75005 Paris, France}

\date{ \today }

\begin{abstract}
    In this Supplemental Material, we provide additional details on the derivation of the results presented in the Main Text, and on their numerical validation.
\end{abstract}

\maketitle

\tableofcontents

\section{Hard-core lattice gas}
In this Section we characterize the generalized profile $\gdisc_{\vb r}$
%e~c.f.~\cref{eq:gtilde}. 
introduced in Eq.~(2) of the Main Text for the lattice gas model.
To this end, we consider a gas of hard-core particles with density $\rhozero$, evolving on an infinite hypercubic lattice with spacing $\sigma$ and dimension $d$. 
Particles perform symmetric random walks with nearest-neighbour jumping rate $1/(2d\tau)$, except if the targed site is already filled. 
We denote by $\vb X$ the tracer position, and by $\rho_{\vb r}=\{0,1\}$ the occupation at site $\vb r$.
%, where $\vb r$ is evaluated in the frame of reference of the tracer. 
Even in the absence of a bias, we choose $\vu e_1$ as the reference direction, and we call $X_t\equiv \vb X\cdot \vu e_1$. We thus introduce the spatial profiles
\rev{
\begin{align}
    \prof_{\vb r} &\equiv \expval{\rho_{\vb X+\vb r}}-\rhozero, \label{eq:h} \\
    \gdisc_{\vb r} &\equiv \expval{\left(X_t-\expval{X_t} \right) \left( \rho_{\vb X+\vb r} -\expval{\rho_{\vb X+\vb r}} \right) }, \label{eq:gtilde}
\end{align}}
the former reducing to zero in the unbiased limit that we focus on here.

\subsection{Self-consistent equation for the stationary correlation profile $\gdisc\r$}

To compute $ \gdisc_{\vb r}$, we start
from the master equation~\cite{Illien2015_distribution}
\rev{
\begin{align}
    2d\tau \,  \partial_t P(\vb X, \rho,t) =&\, \sum_{\nu=1}^d \sum_{\vb r \neq \vb X, \vb X-\vu e_\nu} \left[ P(\vb X, \rho^{(\vb r,\bm \nu)},t)-P(\vb X, \rho,t)  \right ] \n\\
    &+\sum_{\mu}  \left[ (1-\rho_{\vb X}) P(\vb X-\vu e_\mu, \rho,t)-(1-\rho_{\vb X+\vu e_\mu}) P(\vb X, \rho,t)  \right ] .
    \label{eq:ME}
\end{align}
}
Above we indicated by $\rho=\{\rho\r\}$ a given configuration of the occupations, and by $\rho^{(\vb r,\bm \nu)}$ the configuration in which the occupations at sites $\vb r $ and $\vb r+\vu e_\nu$ have been exchanged. Note that the summation over $\mu$ runs on $ \mu \in \{\pm 1,\dots,\pm d\}$ (here and henceforth, unless stated otherwise).
In particular, the operator on the first line of \cref{eq:ME} encodes the jumps of the bath particles, whereas the operator on the second line corresponds to the jumps of the tracer particle.

The calculation proceeds as in the Supplemental Material of~\cite{Illien2018}, which we
specialize here to the unbiased case by 
setting $\prof_{\vb r}=0$ and applying the limits $p_\nu\to 1/(2d)$ and $\tau=\tau^*$.
We start by
averaging the spatial profiles in \cref{eq:h,eq:gtilde} over the master equation, upon
inserting
%and inserts 
the decoupling approximation
\rev{
\begin{align}
    &\expval*{\rho_{\vb X+\vb r}\rho_{\vb X+\vb r'}}\simeq \expval*{\rho_{\vb X+\vb r}}\expval*{\rho_{\vb X+\vb r'}},\\
    &\expval*{\delta X_t \,\rho_{\vb X+\vb r}\,\rho_{\vb X+\vb r'}}\simeq \expval*{\rho_{\vb X+\vb r}}\expval*{\delta X_t\,\rho_{\vb X+\vb r'}}+\expval*{\delta X_t\,\rho_{\vb X+\vb r}}\expval*{\rho_{\vb X+\vb r'}}.
\end{align}
}
This way 
find the equations satisfied by 
the cross-correlation functions $\gdisc_{\vb r}$ as~\cite{Illien2015_distribution}
\begin{equation}
    \begin{dcases}
        2d\tau\partial_t\gdisc_{\vb r}=\widetilde{L}\gdisc_{\vb r} \equiv \mathcal{L}_1(\vb r), 
        & \text{for } \vb r \notin \{\bm 0,\vu e_{\pm1},\dots,\vu e_{\pm d} \}, \\
        2d\tau\partial_t\gdisc_\nu  =  (\widetilde{L}+A_\nu)\gdisc_\nu \equiv \mathcal{L}_2(\nu), 
        & \text{for } \vb r=\vu e_\nu \text{ with } \nu \neq \pm 1,\\
        2d\tau\partial_t\gdisc_{1}  =  (\widetilde{L}+A_1)\gdisc_\nu
        -\sigma\rhozero(1-\rhozero)
        -\rhozero[\gdisc_1-\gdisc_{-1}], & \text{for } \vb r=\vu e_1,\\
        2d\tau\partial_t\gdisc_{-1}  =  (\widetilde{L}+A_{-1})\gdisc_{-1}
        -\sigma\rhozero(1-\rhozero)
        +\rhozero[\gdisc_1-\gdisc_{-1}], &
        \text{for } \vb r=\vu e_{-1}.
    \end{dcases}
    \label{eq:system-gr}
\end{equation}
Above we used the shorthand notation $\gdisc_\mu \equiv \gdisc_{\vb r =\vu e_\mu}$, and we introduced
the operator $\widetilde{L} \equiv\sum_\mu A_\mu\nabla_\mu$,
 where the function $A_\mu$ in Ref.~\cite{Illien2018} reduces in the symmetric limit to~\footnote{Unfortunately, the definition of $A_\mu$ given in Eq.~(8) of the Main Text of~\cite{Illien2018} differs from the one actually used in the calculation presented in the Supplemental Material. Here we adopt the latter for consistency.}
\begin{equation}
    A_\mu \to 2-\rhozero, \qquad A=\sum_\mu A_\mu \to 2d(2-\rhozero).
\end{equation}
The system of equations~\eqref{eq:system-gr} can be solved by introducing the auxiliary variable $\vb w=(w_1,\dots,w_d)$, and defining the generating function
\begin{equation}
    G(\vb w,t) \equiv \sum_{r_1,\dots,r_d=-\infty}^\infty \gdisc_{\vb r} \prod_{j=1}^d w_j^{r_j}.
\end{equation}
Multiplying 
all equations by $\prod_{j=1}^d {w_j}^{r_j}$, summing over all lattice sites and using the boundary conditions in \cref{eq:system-gr},
we find that $G(\vb w;t)$ satisfies the differential equation
\begin{equation}
    2d\tau \partial_t G(\vb w;t) = 
    %\left[\frac{A_1}{w_1}+{A_{-1}}{w_1}+A_2\sum_{\mu} w_{|\mu|}^{\sgn{\mu}}-\mathcal{A}\right] 
    (2-\rhozero)\left[ \frac{1}{w_1}+w_1  +\sum_{\mu} w_{|\mu|}^{\sgn{\mu}}-2d\right]
    G(\vb w;t) 
    %-\mathcal{L}_0
    -(2-\rhozero) \sum_{\mu}  \gdisc_\mu 
    +\sum_{\mu} w_{|\mu|}^{\sgn{\mu}} \left[\mathcal{L}_2(\mu)-\mathcal{L}_1(\vu e_\mu)\right]. \label{eq:EDP_G}
 \end{equation}
% where
%  \begin{equation}
%  \mathcal{L}_0= (2-\rhozero) \sum_{\mu}  g_\mu 
%  \end{equation}
In the stationary limit, this gives 
\begin{equation}
    G(\vb w) = \frac{J(\vb w)}{1-\lambda(\vb w)},
    \label{eq:G_stat}
\end{equation}
where 
\begin{equation}
    \lambda(\vb w) = \frac{1}{2d}\sum_{j=1}^d \left(w_j+\frac{1}{w_j} \right)
\end{equation}
is the structure factor associated to a symmetric random walk on a hypercubic lattice \cite{hughes1995random}, while
\begin{equation}
    J(\vb w) = (2-\rhozero)\sum_{ \mu}\left(\vb w_{\abs{\mu}}^{\T{sgn}(\mu)} -1\right)\gdisc_\mu -\rhozero \left[ \gdisc_1- \gdisc_{-1}-\sigma (1-\rhozero) \right]\left(w_1-\frac{1}{w_1} \right).
\end{equation}
%This specializes Eqs.~(S18,S19) in \cite{Illien2018} to the unbiased case.
The generating function found in \cref{eq:G_stat} can finally be inverted as
%by setting $w_j=\exp(iq_j)$ and taking the inverse Fourier transform as
\begin{equation}
    \gdisc\r = \int_\T{BZ} \dslash{q} \mathrm e^{-i\vb q\cdot \vb r} G(\{w_j= \mathrm e^{iq_j}\}_{j=1}^d),
\end{equation}
which yields the closed self-consistent equation
\begin{equation}
    \gdisc_{\vb r} = \frac{1}{2d} \int_\T{BZ} \dslash{q} \mathrm e^{-i\vb q\cdot \vb r}  G_d(\vb q) \lgraf  \sum_{ \mu} \left(\mathrm e^{ i\T{sgn}(\mu)q_\abs{\mu}  }-1\right) \gdisc_\mu 
    -\frac{2 i \rhozero}{2-\rhozero}\left[ \gdisc_1- \gdisc_{-1}-\sigma (1-\rhozero) \right] \sin(q_1)  
    \rgraf.
    \label{eq:g_r_implicit}
\end{equation}
Here each of the $d$ integrals is intended over the Brillouin zone $q_j\in[-\pi,\pi]$, and we introduced
\begin{equation}
    G_d(\vb q) \equiv \left[1-\frac{1}{d}\sum_{j=1}^d \cos(q_j) \right]^{-1}.
    \label{eq:rw_prop}
\end{equation}

\subsection{Solution of the self-consistent equation for $\gdisc_{\vb r}$}
\label{sec:g_discrete_sol}
Inspecting \cref{eq:g_r_implicit}, we deduce that an explicit expression for $\gdisc_{\vb r}$ can be found once we have fixed the values of the $\gdisc_\mu$, which in turn are given self-consistently by \cref{eq:g_r_implicit}. First, by symmetry we expect $\gdisc_\mu \equiv \gdisc_2$ for $\mu \in \{\pm 2, \dots, \pm d\}$, hence we only need to determine $\gdisc_1,\,\gdisc_{-1}$, and $\gdisc_2$. Moreover, in the absence of biases we physically expect $\gdisc_{-1}=-\gdisc_{1}$, which follows from the definition in \cref{eq:gtilde}. This way \cref{eq:g_r_implicit} reduces to
\begin{equation}
    \gdisc\r = 
    %g_1 \frac{2(1-\rhozero)}{(2-\rhozero)d} \cor I_d(\vb r) 
    \gdisc_1 \frac{2-3\rhozero}{(2-\rhozero)d} \cor I_d(\vb r) 
    + \frac{\sigma \rhozero(1-\rhozero)}{(2-\rhozero)d} \cor I_d(\vb r) + \frac{\gdisc_2}{d} \int_\T{BZ} \dslash{q} \mathrm e^{-i\vb q\cdot \vb r}  G_d(\vb q) \sum_{j=2}^d\left[ \cos(q_j)-1 \right],
    \label{eq:gr_self}
\end{equation}
where we have introduced for later convenience
\begin{equation}
    \cor I_d(\vb r) \equiv \int_\T{BZ} \dslash{q} G_d(\vb q) \sin(q_1) \sin (\vb q\cdot \vb r) = i \int_\T{BZ} \dslash{q} G_d(\vb q) \sin(q_1) \exp (-i \vb q\cdot \vb r).
    \label{eq:Id}
\end{equation}
We note in particular that 
%$\cor I_d(\vu e_1)=1$, while 
$\cor I_d(\vu e_j)=0$ if $j\neq 1$, so that by setting $\vb r=\vu e_2$ in \cref{eq:gr_self} we find
\begin{equation}
    \gdisc_2 = \gdisc_2 \frac{1}{d} \int_\T{BZ} \dslash{q} \mathrm \cos(q_2)  G_d(\vb q) \sum_{j=2}^d\left[ \cos(q_j)-1 \right].
\end{equation}
Numerical inspection shows that the integral on the r.h.s.~of this equation is in general different from one, which leads us to conclude that $\gdisc_2=0$. In Fig.~2(inset) in the Main Text, we further supported this claim by using Monte Carlo simulations of the system in $d=2$ (see \cref{app:simul_discrete} for further details).

Plugging $\gdisc_2=0$ in \cref{eq:gr_self} and setting $\vb r=\vu e_1$ now gives
\begin{equation}
    \gdisc_1 = 
    \gdisc_1 \frac{2-3\rhozero}{(2-\rhozero)d} \cor I_d(\vu e_1) 
    + \frac{\sigma \rhozero(1-\rhozero)}{(2-\rhozero)d} \cor I_d(\vu e_1),
\end{equation}
from which $\gdisc_1$ follows simply as
\begin{equation}
    \gdisc_1 = \frac{\sigma \rhozero (1-\rhozero) \cor I_d(\vu e_1)}{(2-\rhozero)d-(2-3\rhozero) \cor I_d(\vu e_1)}.
    \label{eq:g_1}
\end{equation}
Inserting $\gdisc_1$ found in \cref{eq:g_1} back into \cref{eq:gr_self}, together with $\gdisc_2=0$, finally gives
\begin{equation}
    \gdisc\r = \frac{ \sigma \bar{\rho} (1-\bar{\rho}) }{(2-\bar{\rho})d-(2-3\bar{\rho}) \cor I_d(\vu e_1)}\cor I_d(\vb r).
    \label{eq:gr_sol}
\end{equation}
This coincides with the expression reported in Eq.~(3) in the Main Text. In the next Sections we will inspect its large-distance behavior.

\subsection{Asymptotic behavior from singular-part analysis}
\label{sec:demery_method}
It is well known that the analytic structure of the Fourier transform of a function $f(r)$ determines its behavior for large $r \in \mathbb{R}$. Below we recall how the same ideas can be extended in higher dimensions, i.e.~for $r \in \mathbb{R}^d$. We will adopt the same method later on, for the continuous model analyzed in \cref{sec:large_distance}.

Consider a function $\hat f(q)$, with $q\in \mathbb{R}$. Basic theorems in Fourier analysis tell us that if $\hat f(q)$ is analytic, then its (anti-)Fourier transform $f(r)$ decays to zero rapidly --- in particular, faster than any polynomial $\propto r^{-n}$. Algebraic decays can thus only originate from singularities either in $\hat f(q)$ or in its derivatives: this is equivalent to requiring that not \textit{all} terms in the MacLaurin series of $\hat f(q)$ are regular. Generically, a singularity in $q=0$ will produce an algebraic tail at large distance $r$ (i.e.~$r^{-\alpha}$, with $\alpha\geq 1$), whereas a singularity in $q\neq 0$ can give rise to oscillations superimposed to an algebraic or exponential tail.
%($\propto\mathrm e^{\mathrm i\omega r} r^{-\alpha}$).

Now consider $\hat f(\vb q)$, with $\vb q=(q_1,q_2)\in \mathbb{R}^2$. If $\hat f$ is an analytic function of both $q_1$ and $q_2$, then it can be expanded in a Taylor series around $\vb q=\bm 0$ as 
\begin{equation}
    \hat f(\vb q) = \sum_{n,m} a_n b_m q_1^n q_2^m,
\end{equation}
and in particular these two sums commute. Equivalently,
\begin{equation}
    \lim_{q_1\to 0} \lim_{q_2\to 0} \hat f(\vb q) = \lim_{q_2\to 0} \lim_{q_1\to 0} \hat f(\vb q).
    \label{eq:limits}
\end{equation}
If, by contrast, the two limits in \cref{eq:limits} do \textit{not} commute, this gives an indication that $\hat f(\vb q)$ presents some kind of singularity for $\vb q=\bm 0$, and that its decay will be slow along certain directions. One then seeks to express
\begin{equation}
    \hat f(\vb q) = \hat f_\T{reg}(\vb q) + \hat f_\T{sing}(\vb q),
\end{equation}
by isolating the singular part of the function for $\vb q=\bm 0$ --- singular either in the sense that $\hat f_\T{sing}(\vb q)$ diverges by approaching $\vb q=\bm 0$, or else simply in the sense that the two limits in \cref{eq:limits} do not commute. Note that $\hat f_\T{sing}(\vb q)$ might as well be a simpler function than $\hat f(\vb q)$. Sometimes we can thus Fourier-invert exactly $\hat f_\T{sing}(\vb q)$, and exhibit explicitly the power-law decay of the corresponding $f_\T{sing}(\vb r)$ along a certain direction (including the prefactor). When this is the case, we can be sure that the regular part $\hat f_\T{reg}(\vb q)$ will not contribute to this asymptotic behavior, by the theorems recalled above.

Below we apply this method to estimate the large-distance behavior of $\gdisc\r$ derived in \cref{sec:g_discrete_sol}.

\subsection{Large-distance behavior of the correlation profile $\gdisc\r$}
We want to estimate the large-distance behavior of the profile $\gdisc_{\vb r}$ 
given in \cref{eq:gr_sol}, for $\vb r \parallel \vu{e}_1$.
To this end, we focus on the integrand function in \cref{eq:Id}, i.e.
\begin{equation}
    \hat f(\vb q) = G_d(\vb q) \sin(q_1) ,
\end{equation}
where $G_d(\vb q)$ is given in \cref{eq:rw_prop}. Calling $\vb q_\perp = (q_2,\dots,q_d)$, one can check that
\begin{align}
    \lim_{\vb q_\perp \to \bm 0} \hat f(\vb q) = \frac{2d}{q_1}, \qquad
    \lim_{q_1\to 0} \hat f(\vb q) = 0,
\end{align}
meaning that the order of the two limits cannot be exchanged, and thus that $\hat f(\vb q)$ is singular in $\vb q = \bm 0$ (in the sense discussed in \cref{sec:demery_method}). However, by introducing
\begin{equation}
    \hat f_\T{sing}(\vb q) \equiv \frac{2d q_1 }{\sum_{j=1}^d q_j^2}
\end{equation}
it is simple to check that $\hat f(\vb q)-\hat f_\T{sing}(\vb q)$ becomes regular in $\vb q\sim\bm 0$. The large-distance behavior of the Fourier transform $f(\vb r)$ of $\hat f(\vb q)$ is thus captured exactly by inverting instead $\hat f_\T{sing}(\vb q)$ : for $\vb r=x \vu{e}_1$ and $x\gg 1$, we get in particular
\begin{align}
    \cor I_d(x\vu{e}_1) &= i  \int_\T{BZ} \dslash{q}  \mathrm e^{ -iq_1 x} \hat f_\T{sing}(\vb q) =
    i 2 d\int_0^\infty \dd{\mu} \int_\T{BZ} \dslash{q}  \mathrm e^{ -\mu \sum_{j=1}^d q_j^2 -iq_1 x} q_1 \n \\
    &= i 2 d\int_0^\infty \dd{\mu} \left[\frac{ \T{Erf}(\pi \sqrt{\mu})}{2\sqrt{\pi\mu}} \right]^{d-1} \int_{-\pi}^\pi \frac{\dd{q_1}}{2\pi} \mathrm e^{-\mu q_1^2 -iq_1 x} \n\\
    &\sim \frac{ x^{1-d}d}{\pi^{d/2}}\int_0^\infty \dd{t} t^{d/2-1}\mathrm e^{-t}\left[ \T{Erf}\left(\frac{\pi x}{2\sqrt{t}}\right) \right]^{d-1} 
    \sim \frac{\Gamma (d/2)\, d}{\pi^{d/2}}x^{1-d},
\end{align}
where in the third line we extended the limits of integration over $q_1$ to $\pm \infty$ since $x$ is large, we computed the integral over $q_1$, and finally we changed variables as $t\equiv x^2/(4\mu)$.

The correctness of this asymptotic expansion can easily be checked numerically, as we did in Fig.~2 in the Main Text. Note that we would have obtained the same result by applying directly the stationary-phase approximation to $\hat f(\vb q)$ --- however, in this case extending the limits of integration over $q_j$ to $\pm \infty$ also for $j\neq 1$ would in principle be unjustified, since the large parameter $x$ only involves the direction $\vu e_1$.

%Finally, 
\rev{Additionally},
we note that similar estimates can be easily obtained along other spatial directions as well. For instance, one finds in $d=2$ and for $r\gg 1$ [compare with Fig.~2(inset) in the Main Text]
\begin{equation}
    \cor I_2(\vb r \parallel \vu \theta) \simeq \frac{2\cos \theta}{\pi r},
\end{equation}
where $\theta$ is the polar angle on the $2d$ plane.

\rev{Finally, a comment on the case $d=1$ is in order. 
First, we note that in $d=1$ we do not expect to find universality across hard-core and soft-core systems, as instead we unveiled in this work for $d>1$. Indeed, exclusion processes in $d=1$ are subject to the single-file constraint, which is well known to produce subdiffusive behavior in the motion of the tracer particle --- in sharp contrast with the diffusive behavior retained by a soft tracer even in $d=1$~\cite{Demery2014}.}

\rev{For the one-dimensional symmetric exclusion process (SEP), a spatial correlation profile analogous to $\gdisc\r$ has been recently discussed in Refs.~\cite{Poncet_2021,Grabsch2022}. This profile is known to attain at long times a scaling form as a function of $x/\sqrt{t}$ [see e.g.~Fig.~1(b) in Ref.~\cite{Poncet_2021}]. As a consequence, in an infinite system, such profile cannot truly reach a $t$-independent stationary state. Conversely, in any finite system correlations are expected to spread until a diffusive behavior is restored at long times.}

\rev{By contrast, the expression for $\gdisc\r$ given in Eq.~(3) of the Main Text has been derived under the \textit{assumption} that the profile becomes stationary. 
In fact, our prediction for $\gdisc\r$ given in \cref{eq:gr_sol,eq:Id} can be made explicit in $d=1$:
\begin{equation}
    \cor I_1(x>0) = \int_{-\pi}^\pi \frac{\dd{q}}{2\pi} \frac{\sin(qx)\sin(q)}{1-\cos(q)} = \int_{0}^{2\pi} \frac{\dd{q}}{2\pi} \frac{\sin(qx)\sin(q)}{2\sin^2(q/2)} = \int_{0}^{\pi} \frac{\dd{y}}{\pi} \frac{\sin(2xy)\cos(y)}{\sin(y)} = 1,
\end{equation}
where we set $y=q/2$ and used the exact result (3.612) in \cite{table}. By reflection symmetry, we deduce that in $d=1$ it is $\cor I_1(x)=\T{sgn}(x)$, so that
\begin{equation}
    \gdisc_{x=\vu{e}_1\cdot \vb r} = \frac{ \sigma \bar{\rho} (1-\bar{\rho}) }{(2-\bar{\rho})-(2-3\bar{\rho}) \cor I_1(\vu e_1)} \T{sgn}(x).
    \label{eq:SEP}
\end{equation}
We thus conclude that, using the decoupling approximation and assuming long-time stationarity, 
we recover the behavior $\gdisc_x\propto \T{sgn}(x)$ expected for a \textit{finite} one-dimensional SEP, where a stationary diffusive regime is eventually attained.}

\subsection{Details of the numerical simulation}
\label{app:simul_discrete}

The simulations in $d=2$ are performed on a periodic system of $L \times L$ sites, with $N=\bar\rho L^2$ particles at average density $\rho$. Initially, the particles are randomly placed on the lattice, with a uniform distribution. The dynamics is implemented as follows: (i) we first pick the time of the next event with rate $N/\tau$, with $1/\tau = 1/4$ the jump rate of one particle; (ii) we pick with equal probability one of the four possible jump directions (up, down, left, right); (iii) if the chosen destination site is empty, the jump is performed, otherwise it is rejected. We iterate this process until the wanted final time is reached.

During this process, we keep track of the position $\vb X_p(t)$ of each particle $p$ (including the winding through the periodic boundaries). At chosen times ($t = 1000$, $2000$, $3000$, $4000$ and $5000$), we compute, for each particle $p$, the occupations $\rho^{(p)}_{\vb r}(t)$ in the reference frame of $\vb X_p$, and the spatial profile $(\vb X_p(t) - \vb X_p(0) )\rho^{(p)}_{\vb r}(t)$. Finally, we sum over all the particles. We average over 800000 realizations of the simulation.

\section{Soft interacting Brownian particles}
Here we derive and characterize the correlation profile $\gcont(\vb x)$ introduced in Eq.~(10) of the Main Text, for the continuous interacting particles model. For completeness, we begin this Section by recalling the derivation of the Dean equations proposed in Refs.~\cite{Dean1996,Demery2014}, which we present here in a slightly generalized form, so as to possibly include the case in which the tracer is different (in terms of mobility and interaction potential) from the other bath particles. 

\subsection{Coarse-grained dynamics}
\label{sec:dean}
Let us consider a system of $(N+1)$ Brownian particles which evolve according to 
\begin{equation}
    \dot{\vb{X}}_i(t)=\delta_{i, 0} \vb{f}- \mu\i \sum_{j\neq i} \nabla_{\vb{X}_i} U_{ij}\left(\vb{X}_i(t)-\vb{X}_j(t)\right)+\bm\eta_i\left(t\right),
    \label{eq:langevin}
\end{equation}
where the Gaussian noises $\bm \eta_i$ have zero mean and
\begin{equation}
    \left\langle \bm \eta_i(t) \bm\eta^{\mathrm{T}}_j\left(t^{\prime}\right)\right\rangle=2T\mu\i \delta_{i j}\delta\left(t-t^{\prime}\right) \mathbb{1}.
\end{equation}
The particle with $i=0$ is the tracer, and can in general be different from the other bath particles, hence the mobility and the inter-particle interaction potentials in \cref{eq:langevin} read
\begin{equation}
    \mu\i = 
    \begin{dcases}
        \mu & i\geq 1,\\
        \mu_0 , & i=0,
    \end{dcases}
\end{equation}
and
\begin{equation}
    U_{ij}(\vb x) = 
    \begin{dcases}
        U(\vb x), & i,j\geq 1,\\
        U_0(\vb x), & i=0 \T{ or } j=0 ,
    \end{dcases}
    \label{eq:interaction_potential}
\end{equation}
respectively.

Following \cite{Dean1996,Demery2014}, we introduce the density of bath particles
\begin{equation}
    \rho(\vb x,t) = \sum_{i=1}^N \delta \left( \vb x-\vb X_i(t) \right),
    \label{eq:density}
\end{equation}
in terms of which the evolution equation of the tracer becomes
\begin{equation}
    \dot{\vb{X}}_0(t)= \vb{f}-\mu_0 \nabla_{\vb{X}_0} \int \dd{\vb y} U_0\left(\vb{X}_0(t)-\vb{y}\right)\rho(\vb y,t)+\bm\eta_0(t),
\end{equation}
while the bath density can be shown to follow the Dean equation
\begin{equation}
    \partial_t \rho(\vb x,t) = \mu\div \left[ \rho(\vb x,t) \grad \fdv{\mathcal F}{\rho(\vb x,t)} \right]+\div \left[ \rho^{\frac12}(\vb x,t) \bm \xi (\vb x,t) \right],
    \label{eq:dean}
\end{equation}
with
\begin{equation}
    \left\langle \bm \xi(\vb x,t) \bm\xi^{\mathrm{T}}\left(\vb x',t^{\prime}\right)\right\rangle=2T\mu\delta (\vb x-\vb x') \delta\left(t-t^{\prime}\right) \mathbb{1},
    \label{eq:xi_corr}
\end{equation}
and the pseudo free energy
\begin{equation}
    \mathcal F[\rho,\vb X_0] = T \int \dd{\vb x}\rho(\vb x) \log(\frac{ \rho(\vb x)}{\rho_0}) +  \frac 12 \int \dd{\vb x} \dd{\vb y} \rho(\vb x) U(\vb x-\vb y)\rho(\vb y) + \int \dd{\vb y}\rho(\vb y) U_0(\vb y-\vb X_0),
    \label{eq:pseudo-init}
\end{equation}
where $\rho_0$ is some uniform background density (e.g.~the average density of the system).
Interestingly, if $U_0(\vb x-\vb y)=U_0(\vb y-\vb x)$ --- that is, if the bath-particle interaction is \textit{reciprocal} --- then we may also rewrite (using the same free energy $\cor F$)
\begin{equation}
    \dot{\vb{X}}_0(t)= \vb{f}-\mu_0 \nabla_{\vb{X}_0} \mathcal F[\rho,\vb X_0]+\bm\eta_0(t).
    \label{eq:dean_tracer}
\end{equation}

\subsection{Comments on the expectation value of stationary observables}
\label{sec:stationary}
In the absence of biases, i.e.~for $\vb f\equiv 0$, we note that the system composed by the tracer $\vb X$ and the density $\rho$ of surrounding particles formally admits the joint stationary distribution
\begin{equation}
    \cor P[\rho, \vb X_0] \propto \mathrm e^{-\frac{1}{T}\cor F[\rho,\vb X_0]},
    \label{eq:false_stationary}
\end{equation}
where $\cor F$ is the pseudo free energy given in \cref{eq:pseudo-init}.
Crucially, however, an obvious translational (zero) mode may prevent the system from actually reaching this steady state. 
To get an intuition, consider a simpler system made of only two interacting Brownian particles $\vb X_1(t)$ and $\vb X_2(t)$, which can be conveniently analyzed by changing variables to their ``center of mass'' $\vb R=(\vb X_1+\vb X_2)/2$ and relative distance $\vb r=\vb X_2-\vb X_1$. Clearly, the variable $\vb R(t)$ performs free Brownian motion (irrespective of the specific form of the two-body interaction potential), which does not admit a stationary distribution for an unconfined system.

In general, using the stationary distribution in \cref{eq:false_stationary} to evaluate the equilibrium average of observables
that depend only on the relative distance(s) between particles \textit{does} render the correct result --- indeed, this is common practice in the statistical mechanics of equilibrium fluids~\cite{McDonald_book,mcquarrie2000statistical}, where the observables are the usual thermodynamic quantities.
By contrast, applying such procedure would render the wrong expectation value for observables that \textit{do} depend on this zero mode.

% As a consequence, stationary averages of observables that \textit{do} depend on this zero mode [such as the correlation profile $g(\vb x)$ that we address below] cannot be naively evaluated by using the stationary distribution in \cref{eq:false_stationary}. By contrast, applying such procedure would render the correct expectation value for observables that only depend on the relative distance(s)~\cite{mcquarrie2000statistical}. We will come back to this point later.

The latter is 
%unfortunately 
actually
the case for the spatial correlation profiles that we considered in the Main Text [see e.g.~Eq.~(10)]. To see why, let us use the stationary distribution~\cref{eq:false_stationary} to compute the cross correlation
\begin{align}
    \expval{ \vb X_0 \,\rho(\vb x+\vb X_0)} &\propto \int \cor D \rho \int \dd{\vb X_0}  \mathrm e^{-\frac{1}{T}\cor F[\rho,\vb X_0]}\, \vb X_0\, \rho(\vb x+\vb X_0) \n\\
    &= \int \cor D \rho' \int \dd{\vb X_0}   \mathrm e^{-\frac{1}{T}\cor F'[\rho',\vb X_0]} \,\vb X_0\, \rho'(\vb x) .
    \label{eq:avg_false}
\end{align}
In the first line we expressed the average as a path integral over all stationary configurations of the bath density $\rho(\vb x)$ and positions $\vb X_0$ of the tracer, while in the second line we applied the change of variables $\rho'(\vb x) = \rho(\vb x+\vb X_0)$, of unit Jacobian (indeed, this is simply a translation). In particular, in the absence of confinement the pseudo free energy transforms as
\begin{equation}
    \mathcal F'[\rho',\vb X_0] = T \int \dd{\vb x}\rho'(\vb x) \log (\frac{\rho'(\vb x)}{\rho_0}) +  \frac 12 \int \dd{\vb x} \dd{\vb y} \rho'(\vb x) U(\vb x-\vb y)\rho'(\vb y) + \int \dd{\vb y}\rho(\vb y) U_0(\vb y) = \mathcal F'[\rho'],
\end{equation}
thus losing its dependence on $\vb X_0$~\cite{Venturelli_2022}. As a consequence, the average in \cref{eq:avg_false} factorizes and we obtain
\begin{align}
    \expval{ \vb X_0 \,\rho(\vb x+\vb X_0)}\propto
    \left[\int \cor D \rho' \,  \mathrm  e^{-\frac{1}{T}\cor F'[\rho']} \rho'(\vb x) \right] \left( \int \dd{\vb X_0} \vb X_0 \right) =0,
\end{align}
where we used that $\expval{\vb X_0}=\bm 0$ in the unbiased case (i.e.~for $\vb f=\bm 0$). Clearly, this trivial zero is not the correct result, but rather stems from the translational invariance of the system.

\revv{Finally, this observation indicates that 
the tracer-bath correlation profiles cannot be directly expressed in terms of
the usual equilibrium correlation functions, predicted and measured e.g.~in simple liquids, such as pair correlations or the radial distribution function~\cite{McDonald_book,Evans1993,Evans1994,LeotedeCarvalho1994,Dijkstra2000,Hopkins2005,Stopper2019}.
%We conclude that 
Conversely,}
the correct way to access these stationary correlation profiles is to compute their averages dynamically, by using the coupled equations of motion~\eqref{eq:dean}~and~\eqref{eq:dean_tracer} for the tracer position and coarse-grained bath density --- and only eventually taking their long-time limit. We perform this calculation in the following Sections.

\subsection{Linearized Dean equation}
The derivation of the Dean equations presented in \cref{sec:dean} is exact.
However, \cref{eq:dean} is hard to solve as it stands, and thus we linearize it assuming small bath density fluctuations around a uniform background density~\cite{Demery2014}. To this end, we consider
\begin{equation}
    \rho(\vb x,t) = \rhozero+\rhozero^{\frac12}\phi(\vb x,t),
    \label{eq:fluctuation}
\end{equation}
where $\rhozero$ is the overall system density,
and plug it back into \cref{eq:dean}; we then discard small terms according to the approximation $\rhozero^{-\frac12}\phi \ll 1$. The result reads
\begin{align}
    \dot{\vb{X}}_0(t)&= \vb{f}-\rhozero^{-\frac12}\mu_0\nabla_{\vb{X}_0} \int \dd{y} V\left(\vb{X}_0(t)-\vb{y}\right)\phi(\vb y,t)+\bm\eta_0(t), \label{eq:tracer} \\
    \partial_t \phi(\vb x,t) &= \mu \div \left\lbrace \grad \left[ T  \phi(\vb x,t) + \int\dd{\vb y} u(\vb x-\vb y) \phi(\vb y,t)+  \rhozero^{-\frac12}  V(\vb x-\vb X_0(t))\right]+\bm \xi (\vb x,t) \right\rbrace, \label{eq:dean_linearized}
\end{align}
where we rescaled the interaction potentials as
\begin{equation}
    V(\vb x) = \rhozero\, U_0(\vb x), \qquad u(\vb x) = \rhozero\, U(\vb x).
\end{equation}
Although we will adapt to the common nomenclature by calling \cref{eq:dean_linearized} the \textit{linearized} Dean equation, we note in passing that we have not simply linearized \cref{eq:dean}. For instance, an extra term
\begin{equation}
    +\mu \rhozero^{-1}\div \phi(\vb x,t) \grad V(\vb x-\vb X_0)
\end{equation}
on the r.h.s.~of \cref{eq:dean_linearized} turns out to be suppressed by this approximation, even though it is linear in $\phi$.

Note that we can still write a pseudo free energy for the linearized system,
\begin{equation}
    \cor H[\phi, \vb X_0]\equiv  
    %T 
    \frac{T}{2}
    \int \dd{\vb x} \phi^2(\vb x) +  \frac 12 \int \dd{\vb x} \dd{\vb y} \phi(\vb x) u(\vb x-\vb y)\phi(\vb y) + \rhozero^{-\frac12}  \int \dd{\vb y}\phi(\vb y) V(\vb y-\vb X_0),
    \label{eq:F2}
\end{equation}
so that 
\begin{equation}
    \partial_t \phi(\vb x,t) = \mu \div \left[ \grad \fdv{\mathcal H}{\phi(\vb x,t)}+\bm \xi (\vb x,t) \right].
\end{equation}
As in \cref{sec:dean}, if $V(\vb x)=V(-\vb x)$ we can also rewrite
\begin{equation}
    \dot{\vb{X}}_0(t)= \vb{f}-\rhozero^{-\frac12} \mu_0 \nabla_{\vb{X}_0} \cor H[\phi,\vb X_0]+\bm\eta_0(t),
\end{equation}
and similar considerations apply.

To make progress, it is convenient to express the coupled equations for the tracer and the bath density in Fourier space, where they read
\begin{align}
    \dot{\vb{X}}_0(t)&= \vb{f}-h \mu_0 \int \dslash{q} i\vb q V_{\vb q}\phi_{\vb q}(t)\mathrm e^{i\vb q\cdot \vb X_0(t)}+\bm\eta_0(t), \label{eq:tracer_fourier} \\
    \partial_t \phi_{\vb q}(t) &= -\mu q^2(T+u_{\vb q}) \phi_{\vb q}(t)  -h\mu q^2 V_{\vb q}\mathrm e^{-i\vb q\cdot \vb X_0(t)} + \nu_{\vb q} (t) .\label{eq:field_fourier}
\end{align}
Here we called $h\equiv \rhozero^{-\frac12}$, we adopted the Fourier convention $f_{\vb q} = \int \dd[d]{\vb x} \exp(-i\vb q\cdot \vb x) f(\vb x) $, and we introduced the Gaussian noise $\nu\q(t)$ with zero mean and correlations
\begin{equation}
    \expval{\nu_{\vb q} (t) \nu_{\vb p} (t')} = 2\mu Tq^2\delta^d(\vb q+\vb p)\delta(t-t').
    \label{eq:nu_corr}
\end{equation}
In the following we will drop for simplicity the suffix zero from $\vb X_0$ and $\bm \eta_0$, as long as no ambiguity arises; we thus have
\begin{equation}
    \left\langle \bm \eta(t) \bm\eta^{\mathrm{T}}\left(t^{\prime}\right)\right\rangle=2\mu_0 T  \delta\left(t-t^{\prime}\right) \mathbb{1} .
    \label{eq:eta_corr}
\end{equation}
One can immediately derive by standard methods~\cite{Tauber,demerypath,Venturelli_2022} the long-time expectation values of
\begin{align}
    \expval{\phi_{\vb q}(t)\phi_{\vb p}(t)  }_0 &= \frac{T}{T+u_{\vb q}} \delta^d(\vb p+\vb q), \label{eq:field_corr}\\
    \expval{\mathrm e^{-i \vb q \cdot \vb X(t)}}_0 &= \exp[- (\mu_0 Tq^2+i \vb q \cdot \vb f)t], \label{eq:part_mgf}
\end{align}
where we indicated by $\expval{\bullet}_0$ the average taken over the non-interacting processes, obtained by setting $h=0$.

\subsection{Derivation of the stationary correlation profiles}
\label{sec:profiles}

We first introduce the average density profile in the frame of the tracer~\cite{Demery2014}
\begin{equation}
   \prof(\vb x,t) \equiv \expval{\phi(\vb x +\vb X(t),t)  } \qquad \to \qquad \prof\q(t) = \expval{\phi_{\vb q}(t)\mathrm e^{i \vb q \cdot \vb X(t)} } .
    \label{eq:hq_def}
\end{equation}
This has to be compared with \cref{eq:h} for the discrete case --- note that the uniform part $\rhozero$ of the density has already been subtracted in \cref{eq:fluctuation}. Similarly, we introduce in analogy with \cref{eq:gtilde} the generalized profile
\begin{equation}
    \gvecreal(\vb x,t) \equiv \expval{\vb X(t) \phi(\vb x +\vb X(t),t)  } \qquad \to \qquad\gvec
    (t) = \expval{\vb X(t) \phi_{\vb q}(t)\mathrm e^{i \vb q \cdot \vb X(t)} } .
    \label{eq:gq_def}
\end{equation}
Here we are interested in the stationary values attained by these quantities at long times, which we can compute (as shown below) up to their leading order in the coupling constant $h$ by using the Novikov theorem \cite{Novikov_1965,Venturelli_2022_2parts,Venturelli_2023}. We will assume $V(\vb x) = V(-\vb x)$ (real and even), so that $V_{\vb q}$ is also real and even, and we will set for simplicity $\mu=\mu_0$ and $\vb f\equiv 0$ (although the following derivation can be easily generalized).

\subsubsection{Average density profile}
\label{sec:avg}
We discuss the average density profile $\prof\q(t)$ first, since the knowledge of its stationary value will prove necessary to calculate also the stationary correlation profile $\gvec$, and to illustrate the strategy that we will use below to compute $\gvec$.

Using Stratonovich calculus [without loss of generality, since the white Gaussian noises $\bm \eta(t)$ and $\nu\q(t)$ in \cref{eq:tracer_fourier,eq:field_fourier} are additive], we first compute
\begin{align}
    \partial_t \prof\q(t) =&\, \expval{\dot \phi_{\vb q}(t)\mathrm e^{i \vb q \cdot \vb X(t)} } +i\vb q \cdot \expval{ \dot{\vb X}(t)\phi_{\vb q}(t)\mathrm e^{i \vb q \cdot \vb X(t)} } \n \\
    =&\, -\mu q^2(T+u_{\vb q})\prof\q(t)-h \mu q^2 V\q+ \expval{\nu\q(t) \mathrm e^{i \vb q \cdot \vb X(t)} }
    \n\\
    &+i\vb q \cdot\left[ \expval{ \bm\eta(t)\phi_{\vb q}(t)\mathrm e^{i \vb q \cdot \vb X(t)} }  -h  \mu \int \dslash{p} i\vb p V_{\vb p} \expval{\phi_{\vb q}(t)\phi_{\vb p}(t)\mathrm e^{i(\vb p+\vb q)\cdot \vb X(t)}} \right],\label{appeq:hq_steady}
\end{align}
where in the second line we used the equations of motion~\eqref{eq:tracer_fourier}~and~\eqref{eq:field_fourier}, with $\vb f=0$. We now compute the three expectation values that appear in \cref{appeq:hq_steady}, up to their lowest nontrivial order in the coupling $h$, which we assume to be small (we will comment in \cref{sec:validity} on the limits of validity of this perturbation theory). The last expectation value in \cref{appeq:hq_steady} gives simply
\begin{align}
    \expval{\phi_{\vb q}(t)\phi_{\vb p}(t)\mathrm e^{i(\vb p+\vb q)\cdot \vb X(t)}} &= \expval{\phi_{\vb q}(t)\phi_{\vb p}(t)}_0\expval{\mathrm e^{i(\vb p+\vb q)\cdot \vb X(t)}}_0+\order{h} \n\\
    &= \frac{T}{T+u_{\vb q}} \delta^d(\vb p+\vb q) \expval{\mathrm e^{i(\vb p+\vb q)\cdot \vb X(t)}}_0+\order{h} = \frac{T}{T+u_{\vb q}} \delta^d(\vb p+\vb q) +\order{h},
    \label{appeq:fullstructure}
\end{align}
where we used the stationary averages in \cref{eq:field_corr,eq:part_mgf} computed over the uncoupled stochastic processes.
To evaluate the other two expectation values in \cref{appeq:hq_steady}, which involve the product with a Gaussian noise, we first recall Novikov's theorem \cite{Novikov_1965,Luczka_2005}
\begin{equation}
    \expval{\zeta(t) F[\zeta]} = \int \dd{u} \expval{\zeta(t) \zeta(u)} \expval{\fdv{F[\zeta]}{\zeta(u)} } ,
    \label{eq:novikov}
\end{equation}
where $F[\zeta]$ is any functional of a Gaussian 
noise $\zeta$. For instance, using the correlator of the noise $\bm \eta(t)$ given in \cref{eq:eta_corr} we get
\begin{align}
    \expval{ \eta_\alpha(t)\phi_{\vb q}(t)\mathrm e^{i \vb q \cdot \vb X(t)} } = 2\mu T \left[ i\vb q\cdot \expval{ \phi_{\vb q}(t)\mathrm e^{i \vb q \cdot \vb X(t)} \fdv{\vb X(t)}{\eta_\alpha(t)} }
    +\expval{ \mathrm e^{i \vb q \cdot \vb X(t)}\fdv{\phi_{\vb q}(t)}{\eta_\alpha(t)} } \right].
    \label{appeq:expval_eta_phi_exp}
\end{align}
Concerning the first expectation value on the r.h.s.~of \cref{appeq:expval_eta_phi_exp}, from \cref{eq:tracer_fourier} it follows that
\begin{align}
     \fdv{ X_\beta(t)}{\eta_\alpha(u)} = \delta_{\alpha\beta} \Theta(t-u)
     -h \mu \int \dslash{q} iq_\beta V_{\vb q}\int^t_{t_0}\dd{t'} \sum_{\gamma=1}^d \int_{t_0}^{t'}\dd{s} \fdv{\left[\phi_{\vb q}(t')\mathrm e^{i\vb q\cdot \vb X(t')}\right]}{X_\gamma(s)}\fdv{ X_\gamma(s)}{\eta_\alpha(u)} ,
     \label{appeq:dx_deta}
\end{align}
where in the last term we used the chain rule for functional derivatives, with the integral over $s$ running up to $t'$ due to causality.
%where we used that $\fdv{\phi_{\vb q}(t')}{\eta_\alpha(u)}=\order{h}$.
Note that we have adopted so far the rules of standard calculus, corresponding to the choice of the Stratonovich convention for which $\Theta(0)=\frac12$. 
%This entails no loss of generality, because the two noises in \cref{eq:tracer_fourier,eq:field_fourier} are additive. 
By iterating \cref{appeq:dx_deta} we thus find
\begin{equation}
     \fdv{ X_\beta(t)}{\eta_\alpha(t)} = \frac12 \delta_{\alpha\beta}.
     \label{eq:fdv-exact}
\end{equation}
We stress that this last result is non-perturbative in $h$.
Similarly, we can use \cref{eq:field_fourier} to compute
\begin{align}
    \fdv{\phi_{\vb q}(t)}{\eta_\alpha(u)} &= h\mu q^2 V_{\vb q} i \vb q \cdot  \int_{t_0}^t \dd{t'} G\q(t-t') \mathrm e^{-i\vb q\cdot \vb X(t')} \fdv{\vb X(t')}{\eta_\alpha(u)} \n\\
    &= h \mu q^2 V_{\vb q} i q_\alpha  \int_{t_0}^t \dd{t'} G\q(t-t') \mathrm e^{-i\vb q\cdot \vb X(t')} \Theta(t'-u)\left[1+\order{h}\right],
\end{align}
where in the first line we introduced the propagator
\begin{equation}
    G\q(t) \equiv \exp[-\mu q^2(T+u_{\vb q})t ] \Theta(t),
    \label{eq:propagator}
\end{equation}
and in the second line we used \cref{appeq:dx_deta}. It thus follows clearly that [independently of the noise convention, and at all orders in $h$, due to \cref{eq:fdv-exact}]
\begin{equation}
    \fdv{\phi_{\vb q}(t)}{\eta_\alpha(t)} =0,
\end{equation}
and thus \cref{appeq:expval_eta_phi_exp} reduces to
\begin{equation}
    \expval{ \bm \eta(t)\phi_{\vb q}(t)\mathrm e^{i \vb q \cdot \vb X(t)} } = i\mu T \vb q \prof\q(t).
\end{equation}
Next, using the correlator of the noise $\nu\q(t)$ given in \cref{eq:nu_corr} we obtain from \cref{eq:novikov}
\begin{equation}
    \expval{\nu\q(t) \mathrm e^{i \vb q \cdot \vb X(t)} } = 2\mu Tq^2 i\vb q\cdot  \expval{ \mathrm e^{i \vb q \cdot \vb X(t)} \fdv{\vb X(t)}{\nu_{-\vb q}(t)} },
\end{equation}
where the minus sign in $\nu_{-\vb q}$ follows from the application of the functional derivative operator in Fourier space. However, by the same token as above one can easily deduce that 
\begin{equation}
    \fdv{\vb X(t)}{\nu_{-\vb q}(t)} = 0 \qquad \rightarrow \qquad \expval{\nu\q(t) \mathrm e^{i \vb q \cdot \vb X(t)} }=0.
    \label{appeq:expval_nu_exp}
\end{equation}
As a rule of thumb, we deduce that the noise $\bm\eta(t)$ can only influence (at equal times) a function of $\vb X(t)$, but not a function of $\phi\q(t)$ alone; similarly, the noise $\nu\q(t)$ can only influence a function of $\phi\q(t)$, but not a function of $\vb X(t)$ alone. Using \cref{appeq:expval_eta_phi_exp,appeq:expval_nu_exp,appeq:fullstructure} into \cref{appeq:hq_steady}, we find
at leading order for small $h$
\begin{equation}
    \partial_t \prof\q(t) =  -\mu q^2(2T+u_{\vb q})\prof\q(t)-h \mu q^2\left[ V\q+ \frac{V_{-\vb q}T}{T+u_{\vb q}}
    \right],
    \label{appeq:dt_hq}
\end{equation}
which rules the temporal evolution of $\prof\q(t)$ [starting from stationary initial conditions, which have been assumed in \cref{appeq:fullstructure} --- a different choice of initial conditions would otherwise enter the last term in \cref{appeq:dt_hq}].
Setting $\partial_t \prof\q(t) =0$,
together with the assumption $V\q= V_{-\vb q}$, finally renders the stationary average density profile
\begin{equation}
    \prof\q = -h\frac{V\q }{T+u\q} 
    +\order{h^2}.
\end{equation}
%result for the stationary $\psi\q$ reported in \cref{eq:hq}.

\subsubsection{Correlation profile $\gcont(x)$}
\label{sec:corr}
The calculation of $\gvec$ introduced in \cref{eq:gq_def} follows the same lines as above: writing 
\begin{align}
    \partial_t \gvec(t) = \expval{\dot{\vb X}(t)  \phi_{\vb q}(t)\mathrm e^{i \vb q \cdot \vb X(t)}} + \expval{\vb X (t) \dot \phi_{\vb q}(t)\mathrm e^{i \vb q \cdot \vb X(t)} } +i \expval{[\vb q \cdot \dot{\vb X}(t)] \vb X(t) \phi_{\vb q}(t)\mathrm e^{i \vb q \cdot \vb X(t)} } \label{appeq:gq_steady}
\end{align}
and using the equations of motion~\eqref{eq:tracer_fourier}~and~\eqref{eq:field_fourier} 
produces several expectation values, which can be dealt with by using the Novikov theorem, together with the rule of thumb stated above. 
First, note that the first expectation value on the right-hand side of \cref{appeq:gq_steady} has already been computed above [indeed, it appeared also in \cref{appeq:hq_steady}].
The only subtle point in the calculation of the second expectation value in \cref{appeq:gq_steady} is the treatment of the term
\begin{align}
    \expval{\phi\q(t)\phi\p(t) \vb X(t)\mathrm e^{i(\vb q+\vb p) \cdot \vb X(t) } } &=  \expval{\phi\q(t)\phi\p(t)}_0\expval{ \vb X(t) \mathrm e^{i(\vb q+\vb p) \cdot \vb X(t) }}_0 +\order{h} \n\\
    &=  \frac{T}{T+u_{\vb q}} \delta^d(\vb p+\vb q) \expval{\vb X(t)}+ \order{h} = \order{h},
    \label{eq:fullstructure_X}
\end{align}
where in the second step we noted that the two averages factorize at $\order{h^0}$, 
in the second line we used the correlation function given in \cref{eq:field_corr},
while in the final step we used that in the absence of a bias one has $\expval{\vb X}=0$.
Finally, the third expectation value in \cref{appeq:gq_steady} involves the correlation function
\begin{equation}
    \expval{\eta_j(t) X_i(t) \phi\q(t) \mathrm e^{i\vb q \cdot \vb X(t) }} = \delta_{ij} \mu T \prof\q + iq_j \mu T [\gvec]_i,
\end{equation}
which we computed again using the Novikov theorem. 
%After a tedious but straightforward calculation, c
Collecting all the various terms gives at leading order for small $h$
\begin{equation}
    \partial_t \gvec =  -\mu q^2(2T+u_{\vb q})\gvec +i\vb q \mu T \prof\q,
    \label{eq:gq_evolution}
\end{equation}
and since the left-hand-size is zero
%which gives 
in the stationary limit, we finally obtain
\begin{equation}
    \gvec = \frac{i \vb q T \prof\q}{q^2(2T+u\q)} +\order{h^2}.
\end{equation}

\subsubsection{Summary of the perturbative results}
In summary, we have obtained the leading perturbative order for small $h$ of the average spatial profiles as
\begin{align}
    \prof\q &= -h\frac{V\q }{T+u\q} 
    +\order{h^2}, 
    \label{eq:hq} \\
    \gvec &= -h\frac{V\q }{T+u\q}\cdot\frac{i \vb q T }{q^2 (2T+u\q)} +\order{h^2}. \label{eq:gq}
\end{align}
The component along $\vu e_1$ of the latter expression corresponds to the result reported in Eq.~(12) in the Main Text.

We remark that these expressions can be easily generalized to the case of a biased tracer, i.e.~with $\vb f\neq \bm 0$, with minor modifications of the calculation presented above. We thus omit the calculation and simply report the final result:
\begin{align}
    \prof\q &= -h\frac{V\q }{T+u\q} \cdot\frac{q^2 (2T+u\q)}{q^2 (2T+u\q)-i\vb q\cdot \vb f}+\order{h^2}, \label{eq:hq_bias} \\
    \gvec &= -h\frac{V\q }{T+u\q}\cdot\frac{i \vb q T [q^2 (2T+u\q)+i\vb q\cdot \vb f] }{[q^2 (2T+u\q)-i\vb q\cdot \vb f]^2} +\order{h^2}. \label{eq:gq_bias} 
\end{align}
We note that a similar expression for the average density profile $\prof\q$ in the presence of a bias had been first derived in Ref.~\cite{Demery2014}, by a distinct reasoning.  Indeed, one can alternatively recover \cref{eq:hq_bias}, in the stationary limit $t_0\to-\infty$, by solving explicitly for $\phi\q(t)$ in \cref{eq:field_fourier} and writing
    \begin{equation}
        \prof\q(t) = \expval{\phi_{\vb q}(t)\mathrm e^{i \vb q \cdot \vb X(t)} } = \int_{t_0}^t\dd{s} G\q(t-s)\left\lbrace \expval{\nu\q(s) \mathrm e^{i \vb q \cdot \vb X(t)} } -h\mu q^2V\q \expval{\mathrm e^{i\vb q\cdot \left[\vb X(t)-\vb X(s)\right]}}  \right\rbrace .
        \label{eq:psi_demery_route}
    \end{equation}
Unfortunately, however, above Eq.~(46) in Ref.~\cite{Demery2014} it was implicitly claimed that $\expval{\nu(\vb x+\vb X(t),s)}=0$,
% \begin{equation}
%     \expval{\nu(\vb x+\vb X(t),s)}=0 \qquad \rightarrow \qquad \expval{\nu\q(s)\mathrm e^{i\vb q \cdot \vb X(t)}}=0,
% \end{equation} 
whereas in fact this quantity is nonzero for $s<t$. Indeed, computing it via the Novikov theorem as above one finds
\begin{equation}
        \expval{\nu\q(s) \mathrm e^{i \vb q \cdot \vb X(t)} } = -2hT\mu^2 q^4V_{-\vb q} \int_s^t\dd{t'}G_{-\vb q}(t'-s) \expval{\mathrm e^{i\vb q\cdot \left[\vb X(t)-\vb X(t')\right]}}_0 +\order{h^2},
        \label{appeq:vs_demery}
    \end{equation}
    where the structure factor for (biased) Brownian motion reads in the stationary limit~\cite{demerypath,Venturelli_2022}
    \begin{equation}
        \expval{\mathrm e^{i\vb q\cdot \left[\vb X(t)-\vb X(t')\right]}}_0 = \mathrm e^{i\vb q\cdot \vb f (t-t')-\mu q^2 T \abs{t-t'}}.
        \label{eq:structure-factor}
    \end{equation}
Plugging this result back into \cref{eq:psi_demery_route} and computing the time integrals [using $G\q(t)$ introduced in \cref{eq:propagator} and sending $t_0\to -\infty$] correctly renders \cref{eq:hq} in the limit $t\to\infty$.
This explains the mismatch $T\mapsto 2T$ between Eq.~(49) in Ref.~\cite{Demery2014} and \cref{eq:hq_bias} computed here.
Besides, we stress that the method used in \cref{sec:avg} is more general, as it can be used to compute also higher-order correlation profiles (such as $\gvec$).

\subsubsection{Non-perturbativity of the average density profile in the unbiased case}
Here we prove that in the absence of a bias the average stationary profile $\prof\q$ found in \cref{eq:hq} is in fact non-perturbative in $h$. To see this, 
we recall that the stationary state of the system is formally represented by a canonical distribution with the pseudo free energy $\cor H[\phi,\vb X]$ given in \cref{eq:F2}. As we extensively discussed in \cref{sec:stationary}, observables that only depend on the relative distances between the particles can be averaged over such stationary distribution, which quickly provides their long-time limit. The average density profile [see \cref{eq:density,eq:fluctuation}]
\begin{equation}
    \phi(\vb x + \vb X_0(t),t) =h\left[ \sum_{i=1}^N \expval{\delta(\vb x + \vb X_0(t)-\vb X_i(t)} -\rhozero\right]
\end{equation}
evidently depends only on the relative distances, and so we can compute its stationary value in Fourier space using
\begin{equation}
    \prof\q = \expval{\phi\q \mathrm e^{i \vb q \cdot \vb X_0}} = \frac{1}{\cor Z}\int \cor D \phi \int \dd{\vb X_0} \mathrm  e^{-\frac{1}{T}\cor H[\phi,\vb X_0]} \phi\q \, \mathrm e^{i \vb q \cdot \vb X_0} .
\end{equation}
Above we temporarily reinstated the suffix $-0$ to the tracer position to avoid ambiguities, 
%where the free energy $\cor F_2$ was given in \cref{eq:F2}, and
%where 
and we introduced the partition function
\begin{align}
    \cor Z \equiv \int \cor D \phi \int \dd{\vb X_0}  \mathrm e^{-\frac{1}{T}\cor H[\rho,\vb X_0]},
    %\n\\
    %& = \int \dd{\vb X_0} \int \cor D \phi \, \exp{-\frac{1}{2T}\left[\int \dslash{q}\phi\q (T+u\q)\phi_{-\vb q} + h \int \dslash{q}  V\q \phi\q e^{i \vb q \cdot \vb X_0} \right]}.
\end{align}
where 
$\cor H[\phi,\vb X_0]$ was introduced in \cref{eq:F2}.
Note that the latter can be conveniently rewritten in Fourier space as
\begin{equation}
    \cor H[\phi,\vb X_0] = \frac12 \int \dslash{q}\phi\q (T+u\q)\phi_{-\vb q} + h \int \dslash{q}  V\q \phi\q \mathrm e^{i \vb q \cdot \vb X_0},
    \label{eq:H_fourier}
\end{equation}
which allows to write
\begin{equation}
    \expval{\phi\q \mathrm e^{i \vb q \cdot \vb X_0}} = -\frac{T}{h} \frac{\delta}{\delta V_{-\vb q} }\log \cor Z.
    \label{eq:derivative_profile}
\end{equation}
As in \cref{sec:stationary}, we now introduce $\varphi\q \equiv \phi\q \exp(i \vb q \cdot \vb X_0)$ and change variables in the functional integral (the Jacobian is 1, since this is merely a translation), upon which the partition function can be computed as
\begin{align}
    \cor Z &= \int \dd{\vb X_0} \int \cor D \varphi \, \exp{-\frac{1}{T}\left[\frac12\int  \dslash{q}\varphi\q (T+u\q)\varphi_{-\vb q} + h \int \dslash{q} V\q \varphi\q  \right]} \n\\
    &\propto  \exp[\frac{h^2}{2T} \int \dslash{q}  V\q (T+u\q)^{-1} V_{-\vb q}]. \label{appeq:genf}
\end{align}
Above, in the second line we evaluated the Gaussian functional integral over $\varphi$ and the integral over $\vb X_0$ up to a constant that is formally infinite, but structureless. Indeed, at equilibrium the statistical weight of $\vb X_0$ is uniform over all space, so that $\int \dd{\vb X_0}= \cor V$, where $\cor V$ is the volume of the system (the limit $\cor V\to\infty$ must eventually be taken, until which we assume periodic boundary conditions in order for the free energy to remain translational invariant).
Taking the functional derivative of $\cor Z$ as in \cref{eq:derivative_profile} then immediately renders the average profile
\begin{equation}
    \prof\q = -h\frac{V\q }{T+u\q},
\end{equation}
which coincides with the expression that we computed perturbatively in \cref{eq:hq}. 
This proves its validity beyond the leading order in perturbation theory.

Moreover, this procedure can be generalized to compute the long-time expectation value of other observables 
$\mathcal{O}(\varphi)$
% $\mathcal{O}(\varphi\q,\vb X_0)$ 
that are unaffected by the translational zero mode of the system. 
For instance, in \cref{appeq:fullstructure} we computed perturbatively
the two-point function
\begin{equation}
    \expval{\phi\q \phi\p \mathrm e^{i (\vb q+\vb p) \cdot \vb X_0}} = \expval{\varphi\q \varphi\p} =  -\frac{T^2}{h^2} \frac{\delta^2}{\delta V_{-\vb q}\, \delta V_{-\vb p} }\log \cor Z = \frac{T}{T+u_{\vb q}} \delta^d(\vb p+\vb q),
    \label{eq:derivative_corr}
\end{equation}
which also turns out to coincide with its leading order estimate [whereas~\cref{eq:derivative_corr} is valid for any $h$]. Note that this was the only step in \cref{sec:avg} in which we have resorted to perturbation theory.

By contrast, we have seen in \cref{sec:stationary} that the average correlation profile $\gvec$ is unfortunately affected by the translational zero mode of the system, so that applying the recipe above gives trivially zero. Similarly, we found no way of computing nonperturbatively the expectation value in \cref{eq:fullstructure_X}, which intervened in the derivation of $\gvec$. 

Finally, we stress that the path-integral method discussed in this last section cannot be applied in the presence of a bias, because then the canonical distribution no longer represents the correct stationary distribution of the system. By contrast, the perturbative calculation presented in \cref{sec:avg,sec:corr} can be extended to the biased case without difficulties.

\subsection{Large-distance behavior of the correlation profile $\gcont(x)$}
\label{sec:large_distance}

In the presence of a bias, it had been noted in Ref.~\cite{Demery2014} that the wake of $\prof(\vb x)$ decays for large (negative) $\vb x \parallel \vb f$ as [see Eq.~(57) 
%in Ref.~\cite{Demery2014}, 
\rev{therein,
where the quantity $\prof(\vb x)$ was called $\psi(\vb x)$}]
\begin{equation}
   \prof(x) \sim x^{-(1+d)/2},
\end{equation}
in surprising agreement with the exponent exhibited by the hard core lattice gas in dimension $d=2$ \cite{Benichou_2000}.

To test this analogy further, we now compare the large-$x$ behavior of $g_{x=\vu{e}_1\cdot \vb r} \sim x^{1-d}$, found in Eq.~(4) of the Main Text for the lattice model (and in the unbiased case), with that of $ \gcont(\vb x)= \vu e_1\cdot\gvecreal(\vb x)$, which follows from the Fourier inversion of $\gvec$ in \cref{eq:gq}:
\begin{equation}
    \gcont(x=\vb x \cdot \vu e_1 ) = -hT \int \dslash{q} \mathrm e^{iq_1 x_1} \frac{i q_1 \,V\q }{q^2 (T+u\q)(2T+u\q)}. 
    \label{eq:Fourier_inversion_gq}
\end{equation}
To proceed, we apply again the method presented in \cref{sec:demery_method}. 
We start by calling for brevity
\begin{equation}
    f(\vb q)= \frac{b(\vb q)iq_1}{a(\vb q)q^2}
\end{equation}
the integrand in \cref{eq:Fourier_inversion_gq}, with
\begin{equation}
    a(\vb q )\equiv -hTV\q, \qquad b(\vb q) \equiv (T+u\q)(2T+u\q).
\end{equation}
It is simple to check that\rev{, in dimension $d\geq 2$,}
\begin{align}
    f( \vb q) \xrightarrow[q_1\to 0]{} 0 , \qquad 
    f(\vb q) \xrightarrow[\vb q_\perp\to\bm 0]{} \frac{i b(q_1)/a(q_1)}{q_1}  ,
\end{align}
meaning that the two limits do not commute, and thus we deduce that the limit for $\vb q\to\bm 0$ of $f(\vb q)$ is singular. Let us then isolate the singular part of $f(\vb q)$ as
\begin{equation}
    f_\T{sing}(\vb q) = \frac{b(\bm 0) i q_1}{a(\bm 0) (q_\perp^2+q_1^2)}, 
\end{equation}
which admits a closed-form inverse Fourier transform (which can be obtained by complex integration) as
\begin{equation}
    \int \dslash{q} \mathrm e^{iq_1 x_1} f_\T{sing}(\vb q) = -\frac{b(\bm 0)}{a(\bm 0) \Omega_d}\T{sgn}(x) \abs{x}^{1-d} = \frac{hT V_{\vb q=\bm 0}  }{\Omega_d(T+u_{\vb q=\bm 0})(2T+u_{\vb q=\bm 0})} \T{sgn}(x) \abs{x}^{1-d}.
    \label{eq:g(x)_asimp}
\end{equation}
This is the result reported in Eq.~(13) in the Main Text.
Remarkably, we observe that $\gcont(x)$ exhibits the same algebraic decay exponent as in the discrete case. Additionally, \cref{eq:g(x)_asimp} correctly captures the spatial anti-symmetry of $\gcont(x)$, which follows from its definition~\eqref{eq:gq_def}.

\rev{Finally, a direct inspection of \cref{eq:Fourier_inversion_gq} shows that the asymptotic estimate in \cref{eq:g(x)_asimp} correctly holds also for $d=1$. This is interesting, since in this case the correlation profile $\gcont(x)$ does not decay to zero, but rather saturates to a finite value (we have confirmed this behavior using numerical simulations in $d=1$, see \cref{sec:simul-continuum}). Moreover, this is reminiscent of what we found in \cref{eq:SEP} for the SEP, whose realization in \textit{finite} systems is also characterized by a long-time diffusive regime, featuring a stationary (and structureless) correlation profile $\gcont(x)$.}

\subsubsection{Direct check with selected potentials}
As a further check, we have considered explicit examples of interaction potentials such as the Gaussian one, i.e.
\begin{equation}
    V(\vb x) = \frac{A}{\left( \sqrt{2\pi}\sigma \right)^d}\exp(-\frac{x^2}{2\sigma^2}) \qquad \rightarrow \qquad V\q = A \exp(-q^2\sigma^2/2),
    \label{eq:potential_gauss}
\end{equation}
and the Yukawa potential~\cite{Venturelli_2023}
\begin{equation}
    %V(\vb x) = \frac{A}{\Omega_d \Gamma(d)a^d} \exp(-\abs{\vb x}/a) \qquad \rightarrow \qquad 
    V\q = \frac{A}{1+q^2\sigma^2},
    \label{eq:potential_peak}
\end{equation}
corresponding e.g.~to $V(\vb x) \propto \exp(-\abs{\vb x}/\sigma)$ in $d=1$. In both cases the spatial normalization has been chosen such that $V_{\vb q=\vb 0}=A$, while $\sigma$ represents a generic microscopic cutoff scale.

To make progress, we assume $V\q,u\q$ to be rotationally invariant [hence to depend only on $q=\abs{\vb q}$, which is the case in \cref{eq:potential_peak,eq:potential_gauss}], and we step to spherical coordinates: we write $(\vb q \cdot \vb x)=qx\Psi(\Omega)$, where $\Psi(\Omega)$ is a generalized cosine in $d$ dimensions, and similarly $(\vb q \cdot \vu{e}_1)=q\Psi(\Omega)$. Then we use the angular integral \cite{Venturelli_2022}
\begin{equation}
        \int \frac{\dd{\Omega_d}}{(2\pi)^d} \Psi(\Omega) \mathrm e^{i qx \Psi(\Omega)}  = \frac{i J_{d/2}(qx) }{(2\pi)^{d/2} (qx)^{d/2-1} } 
\end{equation}
to rewrite
\begin{equation}
    \gcont(x=\vb x \cdot \vu e_1 ) = hT \frac{x^{1-d/2}}{(2\pi)^{d/2}} \int_0^\infty \dd{q} \frac{V_q \,q^{d/2-1} }{ (T+u_q)(2T+u_q)}J_{d/2}(qx) . 
\end{equation}
We may now write for large $x$
\begin{equation}
    \gcont(x=\vb x \cdot \vu e_1 ) \simeq \frac{hT  x^{1-d/2} }{(2\pi)^{d/2}(T+u_{\vb q=\bm 0})(2T+u_{\vb q=\bm 0})}\left[ \int_0^\infty \dd{q} J_{d/2}(q x) \, q^{d/2-1}  V_{q} \right]\left[1+\order{\frac{1}{x}}\right] .
    \label{eq:g1_asymptotic}
\end{equation}
The integral in square brakets can be evaluated explicitly, for the two potentials in \cref{eq:potential_peak,eq:potential_gauss},
by using 
the exact integrals
\begin{equation}
    \int_0^\infty \dd{q} J_{d/2}(q x) \, q^{d/2-1}  \mathrm e^{-q^2 \sigma^2/2} = 2^{\frac{d}{2}-1}x^{-\frac{d}{2}}\left[ \Gamma\left(\frac{d}{2} \right) - \Gamma\left(\frac{d}{2},\frac{x^2}{2\sigma^2} \right) \right],
\end{equation}
where $\Gamma(s,x)=\int_x^\infty \dd{t} t^{s-1}\mathrm e^{-t}$ is the incomplete Gamma function, and
\begin{equation}
    \int_0^\infty \dd{q} J_{d/2}(q x) \, q^{d/2-1}  \frac{1}{1+q^2\sigma^2} = 2^{\frac{d}{2}-1}\Gamma\left(\frac{d}{2} \right) x^{-\frac{d}{2}} -\sigma^{-\frac{d}{2}}K_{\frac{d}{2}}(x/\sigma)
\end{equation}
(valid for $d<7$ \cite{table}), where $K_\nu(x)$ is the modified Bessel function of the second kind. By expanding these special functions at leading order for large (positive) $x$, we recover the asymptotic estimate given in \cref{eq:g(x)_asimp}, as expected.

In particular, we note that this asymptotic result is independent of the microscopic cutoff scale $\sigma$, which confirms the irrelevance of the microscopic details encoded in the pairwise potential $V\q$ in determining the large-distance behavior of $\gcont(x)$.

\subsection{Comments on the validity of the perturbative expansion}
\label{sec:validity}
In order to obtain the stationary correlation profile given in \cref{eq:gq}, we have resorted to two approximations:
\begin{enumerate}[(i)]
    \item the assumption of small density fluctuations, see \cref{eq:fluctuation}, has allowed to linearize the Dean equations and eventually obtain \cref{eq:tracer,eq:dean_linearized};
    \item the assumption that the coupling terms proportional to $h=\rhozero^{-1/2}$ in \cref{eq:tracer_fourier,eq:field_fourier} are ``small'' justifies the perturbative expansion.
\end{enumerate}
Concerning the second among these approximations, in \ccite{Demery2014} it has been suggested that the perturbative expansion holds as long as 
\begin{equation}
    \frac{\rhozero^{1/2} U(\vb x)}{T} \ll 1
    \label{eq:claim}
\end{equation}
[see Eq.~(30) therein], which compares the typical magnitude of the interaction potential to the thermal energy. The authors concluded that the approximation works well in the limit of \textit{soft} interaction potentials (in the sense that particles can
overlap completely at a finite energy cost).

Although the intuition in \ccite{Demery2014} turns out to be totally correct, we note that neither the parameter 
$h$, nor the one in \cref{eq:claim} are dimensionless, meaning that they can hardly be judged ``small'' or ``large'' on their own.
One way to make this notion more precise is to consider the average of the pseudo free energy of the system given in \cref{eq:H_fourier}, i.e.
\begin{align}
    \expval{\cor H[\phi,\vb X_0]} &= \frac12 \int \dslash{q} \int \dslash{p} (T+u\q) \delta^d(\vb q+\vb p) \expval{\phi\q\phi\p} + h \int \dslash{q}  V\q \expval{\phi\q \mathrm e^{i \vb q \cdot \vb X}}\n\\
    &= \expval{\cor H_\phi} + \expval{\cor H_\text{int}},
    \label{eq:H_fourier_avg}
\end{align}
and compare the typical size of the interaction energy $\expval{\cor H_\text{int}}$ to that of the ``bare'' energy $\expval{\cor H_\phi}$. To do so, we recall that $\expval{\phi\q e^{i \vb q \cdot \vb X}}$ has already been computed in \cref{eq:hq}, hence we focus on the correlator
\begin{equation}
    \expval{\phi\q(t)\phi\p(t)} = \int_{t_0}^t \dd{t'} \int_{t_0}^t \dd{s'} G\q(t-t')G\p(t-s') \expval{j\q(t') j\p(s')},
    \label{eq:corr-general}
\end{equation}
which we aim to compute in the stationary limit $t_0\to -\infty$, up and including $\order{h^2}$. Here $G\q(t)$ is the propagator in \cref{eq:propagator} [which allowed to solve for $\phi\q(t)$ using its equation of motion~\eqref{eq:field_fourier}], while we introduced for brevity
\begin{equation}
    j\q(t) = \nu\q(t) -h\mu q^2 V_{\vb q}\mathrm e^{-i\vb q\cdot \vb X(t)}.
\end{equation}
Clearly, the term $\expval{\nu\q(t')\nu\p(s')}$ is already known from \cref{eq:nu_corr}, and it will account for the bare correlator given in \cref{eq:field_corr}. To compute the leading correction of $\order{h^2}$ to the correlator in \cref{eq:corr-general} one needs to inspect the cross terms
\begin{equation}
    \expval{\mathrm e^{-i\vb q\cdot \vb X(t') -i\vb p\cdot \vb X(s')  }}, \quad \expval{\nu\p(s') \mathrm e^{-i \vb q \cdot \vb X(t')} }, \quad \expval{\nu\q(t') \mathrm e^{-i \vb p \cdot \vb X(s')} }.
    \label{eq:expvalues}
\end{equation}
%that appear in \cref{eq:corr-general}. 
Note that in general we don't expect the dressed correlator in \cref{eq:corr-general} to be diagonal in the momenta, i.e.~$\propto \delta^d(\vb q+\vb p)$, as it happens instead for $h=0$ [see \cref{eq:field_corr}]. Fortunately, however, the presence of $\delta^d(\vb q+\vb p)$ in \cref{eq:H_fourier_avg} selects only $\vb p=-\vb q$, for which the expectation values in \cref{eq:expvalues} have already been calculated, up to their leading order for small $h$, in \cref{eq:structure-factor,appeq:vs_demery}. A long but straightforward calculation then renders in the stationary limit $t_0\to -\infty$
\begin{align}
    \expval{\phi\q\phi\p} &= \frac{T}{T+u\q} \delta^d(\vb q+\vb p) + h^2 C\t\qp +\order{h^4},\\
    C\t_{\vb q,-\vb q} &= \frac{\abs{V\q}^2}{(T+u\q)^2}.
\end{align}
Remarkably, this gives
\begin{align}
    \expval{\cor H_\phi} &= \frac12 \left( T+h^2 \int \dslash{q} \frac{\abs{V\q}^2}{T+u\q} \right),\\
    \expval{\cor H_\text{int}} &= -h^2 \int \dslash{q} \frac{\abs{V\q}^2}{T+u\q} ,
\end{align}
so that from \cref{eq:H_fourier_avg} we find
\begin{equation}
    \expval{\cor H} = \frac12 \left( T-h^2 \int \dslash{q} \frac{\abs{V\q}^2}{T+u\q} \right).
\end{equation}
The criterium of validity of the perturbative expansion can then be restated as
\begin{equation}
    h^2 \int \dslash{q} \frac{\abs{V\q}^2}{k_B T+u\q} \ll k_B T,
    \label{eq:criterium}
\end{equation}
where we have reinstated the Boltzmann constant $k_B$ for dimensional clarity.

To be concrete, we now set $V\q=u\q$, and assume for the interaction potential in real space $U(\vb x)$ the general rotationally invariant form
\begin{equation}
    U(\vb x) = \epsilon f(x/\sigma),
\end{equation}
where $ x=\abs{\vb x}$, $\epsilon$ is an energy scale, $\sigma$ is a microscopic length scale, and the function $f$ is dimensionless (see \cref{sec:simul-continuum} for explicit examples). In Fourier space, the corresponding rescaled interaction potential $u\q = \rhozero U\q $ then reads
\begin{equation}
    u\q = \rhozero\epsilon \, \sigma^d f(q\sigma).
\end{equation}
If we assume $u\q \ll k_B T$, then a sufficient condition for the criterium in \cref{eq:criterium} to be fulfilled is
\begin{equation}
    \frac{\epsilon}{k_B T} \ll \sigma^{-d} \left[\rhozero \int \dslash{q} f^2(q\sigma)  \right]^{-1/2}.
    \label{eq:criterium-2}
\end{equation}
Although other choices are possible that may still satisfy \cref{eq:criterium}, note that it is impossible for the opposite limit $u\q \gg k_B T$ to be fulfilled for all $q$, since $u\q$ must admit a Fourier transform and thus decay to zero for large $q$. Thus, \cref{eq:criterium-2} suggests that the ratio $\epsilon/(k_B T)$ of the typical interaction energy to the thermal energy of the system should be chosen small, in order for perturbation theory to hold. This essentially coincides with the \textit{soft potential} requirement put forward in Ref.~\cite{Demery2014}.

\subsection{Proof of Eq.~(9) of the Main Text}

Here we derive an evolution equation for the moment generating function $\Psi(\bm \lambda,t)= \ln\expval{\mathrm e^{\bm \lambda \cdot \vb X(t)}}$ of the tracer particle. To this end, we first use Stratonovich calculus to obtain 
\begin{equation}
    \partial_t \expval{\mathrm e^{\bm \lambda \cdot \vb X(t)}} = \bm \lambda \cdot \expval{\bm\eta (t)\, \mathrm e^{\bm \lambda \cdot \vb X(t)}} -h\mu_0 \bm \lambda \cdot \int \dslash{q} i \vb q V\q \expval{\mathrm e^{(\bm \lambda+i\vb q) \cdot \vb X(t)} \phi\q(t) },
    \label{eq:for-eq9}
\end{equation}
where we used the equation of motion~\eqref{eq:tracer_fourier} of the tracer (with $\vb f=\bm 0$).
The first expectation value on the r.h.s.~can be computed via the Novikov theorem as
\begin{align}
    \expval{ \eta_\alpha(t)\,\mathrm e^{\bm \lambda \cdot \vb X(t)} } = 2\mu T \bm\lambda \cdot \expval{ \mathrm e^{\bm \lambda \cdot \vb X(t)} \fdv{\vb X(t)}{\eta_\alpha(t)} }
     = \mu T \lambda_\alpha \expval{ \mathrm e^{\bm \lambda \cdot \vb X(t)} },
\end{align}
where in the last step we used the result obtained earlier in \cref{eq:fdv-exact}.
We thus recognize in this term essentially the Fourier transform of the Fokker-Planck equation for the non-interacting tracer. We stress, however, that this result applies to the interacting tracer, and is non-perturbative in $h$.

To obtain Eq.~(9) in the Main Text, we finally divide \cref{eq:for-eq9} by 
%$\Psi(\bm \lambda,t)$, 
$\expval{\mathrm e^{\bm \lambda \cdot \vb X(t)}}$,
and we recognize
\begin{equation}
    w\q(t) = \frac{\expval{\mathrm e^{(\bm \lambda+i\vb q) \cdot \vb X(t)} \phi\q(t) }}{\expval{ \mathrm e^{\bm \lambda \cdot \vb X(t)} }},
    \label{eq:wq-fin}
\end{equation}
i.e.~the Fourier transform of the generalized profile $w(\vb x,\bm \lambda,t) \equiv \expval{  \phi(\vb x +\vb X(t),t)\, \mathrm e^{ \bm \lambda \cdot \vb X(t)}} / \expval{\mathrm e^{ \bm \lambda \cdot \vb X(t)}}$ introduced in the Main Text under Eq.~(9).

We note that the leading-order correction to the stationary profiles computed in \cref{sec:profiles} turned out to be of $\order{h}$. Conversely, any correlation function involving $\vb X$ but not $\phi$ will only be corrected at $\order{h^2}$: the simplest way to prove this fact is to note that the system of equations~\eqref{eq:tracer_fourier}--\eqref{eq:field_fourier} is invariant under $\{h\mapsto -h,\phi \mapsto -\phi\}$~\cite{wellGauss,Venturelli_2022}. We thus consistently expect $w\q(t)$ in \cref{eq:wq-fin} to be at least of $\order{h}$ --- however, its computation goes beyond the scopes of the present work.

\revv{Finally, we stress that a relation analogous to Eq.~(9) in the Main Text, but involving the correlation profiles $\tilde w(\vb x,\bm \lambda,t) \equiv \expval{  \rho(\vb x +\vb X(t),t)\, \mathrm e^{ \bm \lambda \cdot \vb X(t)}} / \expval{\mathrm e^{ \bm \lambda \cdot \vb X(t)}}$, can be derived using the very same reasoning but \textit{before} linearizing the density field $\rho(\vb x,t)$ in the DK equation. This way one obtains the exact relation
\begin{align}
    \partial_t \Psi(\bm \lambda,t) = \lambda^2\mu T  - \mu  \bm \lambda \cdot \int \dslash{q} \mathrm i \vb q \, U\q  \tilde w\q(\bm \lambda,t),
\end{align}
which is the continuum counterpart of Eq.~(1)
in the Main Text
for the lattice gas case.}

\subsection{Details of the numerical simulation}
\label{sec:simul-continuum}
Brownian dynamics simulations 
%in $d=2$ 
were performed using the LAMMPS computational package \cite{Thompson2022}, and PyLammps, the wrapper Python class for LAMMPS.
The measurements in Fig.~3(a) in the Main Text were obtained by choosing Gaussian interaction potentials, i.e.
\begin{equation}
    U(r) = \epsilon \, \mathrm e^{-(r/\sigma)^2}
\end{equation}
(truncated at $2.5\sigma$),
and a periodic box
\revvv{in $d=2$}
with up to
$L=70$ simple cubic cells. The latter corresponds to placing initially one particle in the middle of each cell, so that the actual box size is $\tilde L=L \rhozero^{-1/d}$, where again $\rhozero$ is the system density.

After letting the system thermalize
for a time 
%$t_\text{term}=2\times 10^4$ (with time step $\Delta t=10^{-4}$),} 
$t_\text{term}=2000$ (with time step $\Delta t=0.001$), 
we measure the quantity
% \begin{align}
%     \expval{\vb X(t) \rho(\vb x +\vb X(t),t)  }_c &= \expval{\vb X(t) \rho(\vb x +\vb X(t),t)  }- \expval{\vb X(t)}\expval{\rho(\vb x +\vb X(t),t)  }\n\\
%     & = \sqrt{\rhozero}\expval{\vb X(t) \phi(\vb x +\vb X(t),t)  }
%     \label{eq:what-we-measure}
% \end{align}
\begin{align}
    \expval{\vb X(t) \rho(\vb x +\vb X(t),t)  }_c &= \sum_{i=1}^N \left[\vb X_i(t) -\vb X_i(0)\right]\rho(\vb x +\vb X_i(t),t) - 
    \sum_{i=1}^N \left[\vb X_i(t) -\vb X_i(0)\right] \sum_{j=1}^N \rho(\vb x +\vb X_j(t),t)
    %\expval{\vb X_i(t)}\expval{\rho(\vb x +\vb X_i(t),t)  }
    % \n\\
    % & \equiv \sqrt{\rhozero}\expval{\vb X(t) \phi(\vb x +\vb X(t),t)  },
    \label{eq:what-we-measure}
\end{align}
repeatedly over uncorrelated samples, and take their 
%temporal 
average.
Note that the symmetry according to which $\expval{\vb X(t)}=\bm 0$ is in general not realized exactly in a finite simulation, which is why we have to subtract the unconnected component. Comparing \cref{eq:what-we-measure} with the prediction for the stationary correlation profile $\sqrt{\rhozero}\expval{\vb X(t) \phi(\vb x +\vb X(t),t)  }$ in Eq.~\eqref{eq:Fourier_inversion_gq}, it appears that the measured quantity 
depends on the 
%dimensionless
combination $\theta\equiv \epsilon \rhozero/ T$, 
%and $\theta_2=\rhozero \sigma^d$, 
rather than on these parameters taken separately.
In turn, the value of $\theta$ controls both the spatial point at which $\gcont(x)$ crosses over to an algebraic behavior, and its amplitude in the asymptotic regime. The choice of parameters stated in the caption of Fig.~3 in the Main Text is dictated by the aim to amplify the signal-to-noise ratio.
Note that the thermalization time of the average correlation profile can be estimated a priori using \cref{eq:gq_evolution}, the slowest mode being the one corresponding to $q=2\pi/\tilde L$.
We averaged over approximately $10^6$ samples.

\begin{figure*}
\centering
\includegraphics[width=0.49\columnwidth]{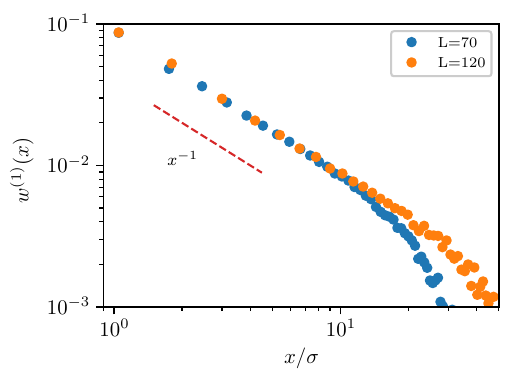}
\put(-195,45){(a)}
\includegraphics[width=0.49\columnwidth]{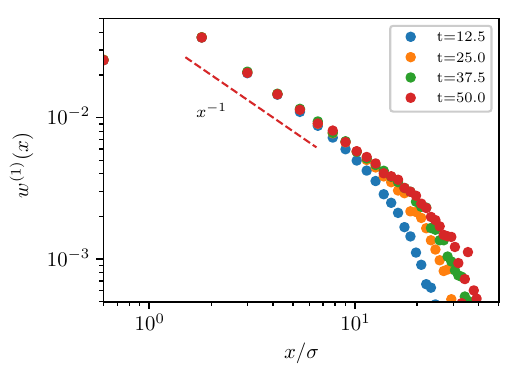}
\put(-195,45){(b)}
\caption{\revv{\textbf{(a)} Numerical simulation of a system of Brownian particles interacting via WCA potentials in $d=2$.
%and with $\bar \rho=0.25,\, T=1, \,\epsilon=0.1,\, \sigma=1$, and $\gamma=1$). 
The expected power-law behavior of $w^{(1)}(x)$ extends towards larger values of $x$ upon increasing the system size $L$.
\textbf{(b)}~Simulation of Brownian particles interacting via WCA potentials (in $d=2$, parameters as in the previous panel, with $L=120$). The expected power-law behavior of $w^{(1)}(x)$ extends towards larger values of $x$ upon increasing the total duration $t$ of the simulation.}
}
\label{fig:finite-size}
\end{figure*}

The simulations in Fig.~3(b) in the Main Text are instead obtained by using the Lennard-Jones potential
\begin{equation}
U_{\text{LJ}}(r) = 4\epsilon \left[ \left(\frac{\sigma}{r}\right)^{12} - \left(\frac{\sigma}{r}\right)^{6} \right],
\end{equation}
and its truncated version (Weeks-Chandler-Andersen)
\begin{equation}
U_{\text{WCA}}(r) = 
\begin{cases} 
4\epsilon \left[ \left(\frac{\sigma}{r}\right)^{12} - \left(\frac{\sigma}{r}\right)^{6} + \frac{1}{4} \right], & \text{for } r \leq 2^{1/6}\sigma, \\
0, & \text{for } r > 2^{1/6}\sigma, 
\end{cases}
\end{equation}
corresponding to retaining only the repulsive part of a LJ potential, shifted vertically so that the minimal potential energy is zero.
\revv{In order to verify the robustness of the large-distance algebraic behavior of $\gcont(x)$ (which in this case is not supported by an explicit analytical prediction), we considered larger}
\revvv{two-dimensional}
\revv{systems up to $L=120$ --- which require a longer thermalization time 
$t_\text{term}=2\times 10^4$ (with time step $\Delta t=10^{-4}$). 
In the simulations we chose 
% $\rhozero=0.5,\, T=1, \,\epsilon=0.1,\, \sigma=1$
$\rhozero=0.25,\, T=1, \,\epsilon=0.1,\, \sigma=1$
, and $\gamma=1$. In particular, the data presented in Fig.~\ref{fig:finite-size} clearly indicate that $\gcont(x)$ approaches the expected algebraic behavior upon increasing the size $L$ of the system, and the total duration of the simulation.}

\begin{figure}
\centering
\includegraphics[width=0.49\columnwidth]{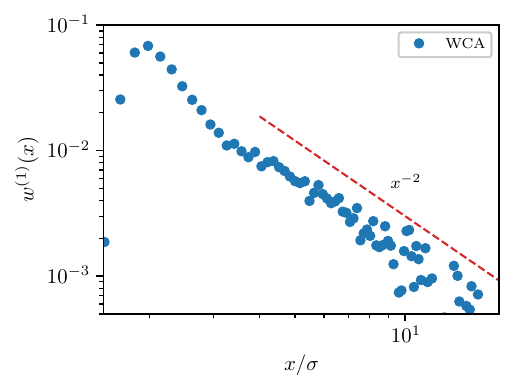}
\caption{\revvv{Tracer-bath correlation profile for a system of Brownian particles interacting via WCA potentials in dimension $d=3$. Preliminary numerical simulations point towards a power-law decay compatible with the exponent $x^{1-d}$ predicted in our work. Parameters: $L=40$, $\bar\rho=0.25$, $\epsilon=0.1$, all other parameters set to unity (data averaged over 1.5$\times 10^6$ trials).}
}
\label{fig:3d}
\end{figure}

\revvv{Finally, these very finite-size effects make it challenging to measure tracer-bath correlation profiles in higher-dimensional systems, since the number of particles to simulate grows as $L^d$. Besides, the function $\gcont(x)$ is expected to decay in $d=3$ as $\sim x^{-2}$, i.e.~more rapidly than in $d=2$, thus requiring even more trials to distinguish the signal from the thermal noise. Although more extensive computations would be in order, we have performed numerical simulations of particles interacting via WCA potentials for a relatively small system in $d=3$. Our preliminary results, reported in Fig.~\ref{fig:3d}, indeed seem to indicate a power-law behavior compatible with the expected exponent $x^{1-d}$ predicted in our work.}

\bibliography{references}